\documentclass[english]{article}
\usepackage[T1]{fontenc}
\usepackage[latin9]{inputenc}
\usepackage{geometry}
\usepackage{color}
\usepackage{babel}
\usepackage{float}
\usepackage{amsmath}
\usepackage{amssymb}
\usepackage{graphicx}
\usepackage[unicode=true,pdfusetitle,
 bookmarks=true,bookmarksnumbered=false,bookmarksopen=false,
 breaklinks=false,pdfborder={0 0 0},pdfborderstyle={},backref=false,colorlinks=true]
 {hyperref}

\makeatletter

\providecommand{\tabularnewline}{\\}
\newcommand{\lyxdot}{.}

\numberwithin{equation}{section}

\@ifundefined{date}{}{\date{}}
\usepackage{lmodern}
\usepackage[T1]{fontenc}
\usepackage{caption}

\usepackage{upgreek}

\@ifundefined{showcaptionsetup}{}{%
 \PassOptionsToPackage{caption=false}{subfig}}
\usepackage{subfig}
\makeatother

\begin{document}
\title{\vspace{1cm}
}
\title{Magnetic Levitation and Compression of Compact Tori \\
 }
\author{Carl Dunlea$^{1*}$, Stephen Howard$^{2}$, Wade Zawalski$^{2}$,
Kelly Epp$^{2}$, Alex Mossman$^{2}$, \\
 Chijin Xiao$^{1}$, Akira Hirose$^{1}$\\
}

\maketitle
$^{1}$University of Saskatchewan, 116 Science Pl, Saskatoon, SK S7N
5E2, Canada 

$^{2}$General Fusion, 106 - 3680 Bonneville Pl, Burnaby, BC V3N 4T5,
Canada

$^{*}$e-mail: cpd716@mail.usask.ca 
\begin{abstract}
The magnetic compression experiment at General Fusion was a repetitive
non-destructive test to study plasma physics  to Magnetic Target Fusion
compression. A compact torus (CT) is formed with a co-axial gun into
a containment region with an hour-glass shaped inner flux conserver,
and an insulating outer wall. External coil currents keep the CT off
the outer wall (radial levitation) and then rapidly compress it inwards.
The optimal external coil configuration greatly improved both the
levitated CT lifetime and the rate of shots with good flux conservation
during compression. As confirmed by spectrometer data, the improved
levitation field profile reduced plasma impurity levels by suppressing
the interaction between plasma and the insulating outer wall during
the formation process. Significant increases in magnetic field, density,
and ion temperature were routinely observed at magnetic compression
despite the prevalence of an instability, thought be an external kink,
at compression. Matching the decay rate of the levitation coil currents
to that of the internal CT currents resulted in a reduced level of
MHD activity associated with unintentional compression by the levitation
field, and a higher probability of long-lived CTs. An axisymmetric
finite element MHD code that conserves system energy, particle count,
angular momentum, and toroidal flux, was developed to study CT formation
into a levitation field and magnetic compression. An overview of the
principal experimental observations, and comparisons between simulated
and experimental diagnostics are presented.
\end{abstract}

\section{Introduction}

General Fusion is developing a magnetized target fusion (MTF) power
plant, based on the concept of compressing a compact torus (CT) plasma
to fusion conditions by the action of external pistons on a liquid
lead-lithium shell surrounding the CT \cite{Laberge,Laberge2}. Magnetic
confinement in self-organized compact toroids relies significantly
on internal plasma currents. The CT can take on a range of magnetic
profiles depending on boundary geometry and the amount of axial current
running on a central shaft. The final magnetic structure of the CT
can range from a spheromak up to spherical tokamak (ST) configuration
with a higher safety factor $q$ profile. For MTF applications, the
self-confining nature of the CT is essential because it enables CTs
to be injected into the flux conserver of an MTF compression system,
thereby physically separating and protecting the plasma formation
system from the extreme and repetitive forces of a full power compression
chamber. Understanding the thermal confinement physics associated
with CTs during compression is critical to the MTF path forward. To
study the plasma physics of CT compression, General Fusion has conducted
a sequence of subscale experiments, refered to as Plasma Compression
Small (PCS) tests \cite{PCS}. In a PCS test, which takes place outdoors
in a remote location, a CT is compressed by symmetrically collapsing
the outer flux conserver with the use of chemical explosives. PCS
tests are destructive, and therefore do not employ the full array
of diagnostics used in CT formation and characterization experiments
in the GF laboratory, and can only be executed every few months. However,
in parallel to this, in the lab, an upgraded duplicate of the PCS
plasma formation system called Super Magnetized Ring Test (SMRT) was
configured to operate as a magnetic compression experiment, and ran
from 2013 to 2016. It was designed as a repetitive, non-destructive
test to study CT compression, in support of the PCS tests. A CT, with
spheromak characteristics, is formed with a magnetized Marshall gun
into a containment region with an hour-glass shaped inner flux conserver,
and an insulating outer wall. Currents in external coils surrounding
the containment region produce a magnetic field which applies a radial
force on the plasma that \textquotedbl levitates\textquotedbl{} it
off the outer wall during CT formation and relaxation, and then rapidly
compresses it inwards. 

With external levitation/compression coils, it was relatively straightforward
to explore several different coil configurations. To accomplish fast
compression while supporting large mechanical forces, the basic design
of a stack of low-inductance single-turn coil plates was adopted (see
figures \ref{fig:Levitation-and-compression}, \ref{fig:Schematic-of-6-1}).
Reduction of error fields was explored via variation of the geometry
of current feed paths into each single turn plate, and also by departing
from this fast pulse design to try a levitation-only 25-turn wound
coil with higher inductance that had increased symmetry in order to
see the effect of error fields on formation stability. Overall coil
length and position was explored in a setup with a stack of six coil
plates and another with a stack of eleven coil plates. Drive circuit
parameters were also explored and we found optimum CT lifetime, and
repeatability of good shots, when the decay of current in the levitation
coil was tailored to match the typical decay time of the internal
plasma current. The magnetic compression experimental campaign shed
light on both fundamental plasma physics questions, as well as practical
engineering concerns. One example of a practical lesson was the observation
that interaction between plasma and the outer insulating wall during
the CT formation could generate high levels of plasma impurities and
radiative cooling under certain conditions. The wall material of high
purity alumina ceramic was noticably less contaminating than the quartz
wall that was tested. Analysis of the pulsed compression experiments
indicated fast MHD instabilities during compression, likely related
to the safety factor $q$ being too low. 

Along with the experiments described in, for example, \cite{RACE,marauder,Pi1,Pi2},
the experiment on which this work is based represents one of relatively
few studies in which magnetic compression has been attempted on CTs
produced by a magnetised Marshall gun. This is the first instance
of magnetic compression by poloidal field on a stationary spheromak
in a device without a co-axial railgun accelerator/compressor stage.
In the past, mostly in the 1970's and 1980's, there were several theoretical
and experimental studies looking at magnetic compression of conventional
tokamak plasmas \cite{Furth,TFTRconf,ATCconf,ATCpaper,Inoue,WatanabePhd,Kamada,Tuman-2Aconf},
inductively formed spheromaks \cite{ZeroD Adiabatic Comp Scaling,S1spheromak,S1_compression},
and field reversed configurations \cite{Rej,GotaTAE,Slough}.

Magnetic compression of spheromaks was the focus of the S-1 experiment
\cite{S1spheromak,S1_compression}, in which a spheromak, with pre-compression
major radius $R=50$cm, and pre-compression minor radius $a=25$cm,
was inductively formed and then magnetically compressed using toroidal
currents in coils located within the vacuum vessel. Pre-compression
S-1 spheromaks had toroidal plasma current $I_{p}\sim$200kA, corresponding
to peak poloidal field $B_{\theta}\sim$0.15T. A geometrical compression
factor of $C=\frac{R_{1}}{R}\sim$1.6$\sim\frac{a_{1}}{a}$, where
$R_{1}$ and $a_{1}$ denote the pre-compression major and minor radii,
while $R$ and $a$ denote the post compression radii, was achieved
on the S-1 device \cite{S1_compression}, leading the researchers
to classify the regime as constant aspect ratio $Type\,A$ \cite{Furth}
adiabatic compression. The availability of spheromak internal temperature,
density, and magnetic field point diagnostics allowed the researchers
to produce shot-averaged poloidal flux contours over the compression
cycle, and enabled determination of the compression scalings for density,
temperature and magnetic field, and comparison of the observed and
predicted scalings. Fine spatial resolution of the magnetic field
measurements allowed for experimental confirmation of several aspects
of basic theory. For example, relaxation to the Taylor state, a process
involving the anomalous conversion of poloidal to toroidal flux \cite{Woltjer,Taylor},
was observed during and after spheromak formation. Good spatial resolution
of field measurements also enabled experimental confirmation of the
approximate conservation of the individual fluxes, and of the absence
of flux conversion during compression. Electron and ion temperatures
were observed to increase at compression; peak $T_{e}$ rose from
\textasciitilde 40eV to \textasciitilde 100eV with compression,
and ion temperatures of up to 500eV were measured at compression.

The Adiabatic Toroidal Compressor (ATC) experiment \cite{ATCconf,ATCpaper}
($R=90$cm, $a=17$cm, pre-compression), which was operated in the
1970's, employed $Type\,B$ \cite{Furth} adiabatic compression, in
which $R$ scales in proportion to $C^{-1}$, and $a$ scales in proportion
to $C^{-\frac{1}{2}}$. As in standard tokamaks, the vacuum vessel
in enclosed by the toroidal field coils, but in the ATC, molybdenum
rail limiters guide the plasma inwards at compression, which is activated
by increasing toroidal current in the compression coils located outside
the (electrically resistive) vacuum vessel, from 2 to 10kA over 2ms.
Compression in $a$, $R$, $B_{\phi}$, $I_{p}$, $n_{e}$, and $T_{i}$
was observed with scalings consistent with predictions for $Type\,B$
compression. The ATC compression mechanism is similar to the \textquotedbl radial
magnetic pumping\textquotedbl{} scheme proposed in 1969 \cite{Artsimovich},
in which it was suggested that a tokamak plasma would be maintained
over several $B_{z}$ compression cycles, and that ions could be heated
further at each compression. In \cite{ATCpaper}, it was recommended
that high frequency magnetic compression on ATC would be technically
difficult, but that low frequency compression on a device, with dimensions
increased to five to ten times those of ATC, might lead to ignition. 

The scenario of attaining ohmic ignition through the combination of
an ultra-low-$q$ discharge and adiabatic magnetic compression was
explored in \cite{Inoue,WatanabePhd,Kamada}. It was envisaged that
with pre-compression conditions of $T_{1}=600$eV and $n_{1}=2\times10^{19}$m$^{-3}$,
that the ignition parameters $T=5.3$keV and $n=5\times10^{20}$m$^{-3}$
could, in principle, be achieved with $C=5$. 

TAE Technologies \cite{GotaTAE} and Helion Energy Inc. \cite{Slough}
have compressed FRCs to ion temperatures of several keV; a set of
independently triggered formation and acceleration coils are used
to form and merge two oppositely directed supersonic FRCs.

Merging-compression is a spherical tokamak (ST) plasma formation method
that involves the merging and magnetic reconnection of two plasma
rings, followed by inward radial magnetic compression of the resultant
single torus to form a spherical tokamak plasma configuration. The
initial tori are formed inductively around coils internal to the vacuum
vessel, and the compression coils are also internal, an approach with
some similarities to that developed on the S1 device. This ST plasma
formation method has the advantage of eliminating the need for a traditional
central solenoid - in an ST, space is limited in the central post
and is inadequate for solenoids capable of inducing toroidal plasma
currents in the MA range \cite{Gryaznevich}. The merging phase leads
to efficient transformation of magnetic to kinetic, then thermal energy
(up to 15MW of ion heating power was recorded on on MAST), and also
leads to a rapid increase of plasma current \cite{Gryaznevich}. The
merging compression ST formation method was first used on START \cite{Gryaznevich1,Sykes}
in 1991, and then in MAST \cite{Sykes1,Gryaznevich2}, and is currently
employed on the compact high field spherical tokamak ST40 at Tokamak
Energy Ltd. \cite{Gryaznevich}.

This paper is arranged as follows. Section \ref{sec:Machine-Overview}
is an overview of the magnetic compression device. Section \ref{part:levitation}
is focused on magnetic levitation of CTs. A description of various
external coil configurations experimented with is presented in section
\ref{subsec::Levitation coil configurations}. Principal results obtained
in the study of magnetically levitated CTs will be presented, discussed,
and compared with MHD-simulated diagnostics in section \ref{subsec:Levresultsoverview}.
A summary of the main findings from the study of levitated CTs, and
comparisons of CT performance in the principal configurations tested
constitutes section \ref{subsec:Main-points--}. Section \ref{sec:Magnetic-compression}
is focused on magnetic compression. The principal results from the
magnetic compression experiments will be shown, discussed, and compared
with simulation results in section \ref{sec:Overview_compresults}.
The mechanism behind the compressional instability that was routinely
observed is discussed in section \ref{subsec:Compressional-instability}.
A comparison of the performance of magnetically compressed CTs in
the principal configurations tested is presented in section \ref{sec:Comparison-of-compression}.
The main conclusions from the levitation and compression experiment
are outlined in section \ref{sec:Conclusion}. A method developed
to experimentally determine the outboard equatorial separatrix of
levitated CTs is described, with results, in appendix \ref{sec:rsep}.
In appendix \ref{subsec:SimDiagCompression-scalings}, the predicted
adiabatic compression scalings for various parameters are compared
with those obtained from MHD simulations.

\section{Machine Overview\label{sec:Machine-Overview}}

\begin{figure}[H]
\centering{}\includegraphics[scale=0.5]{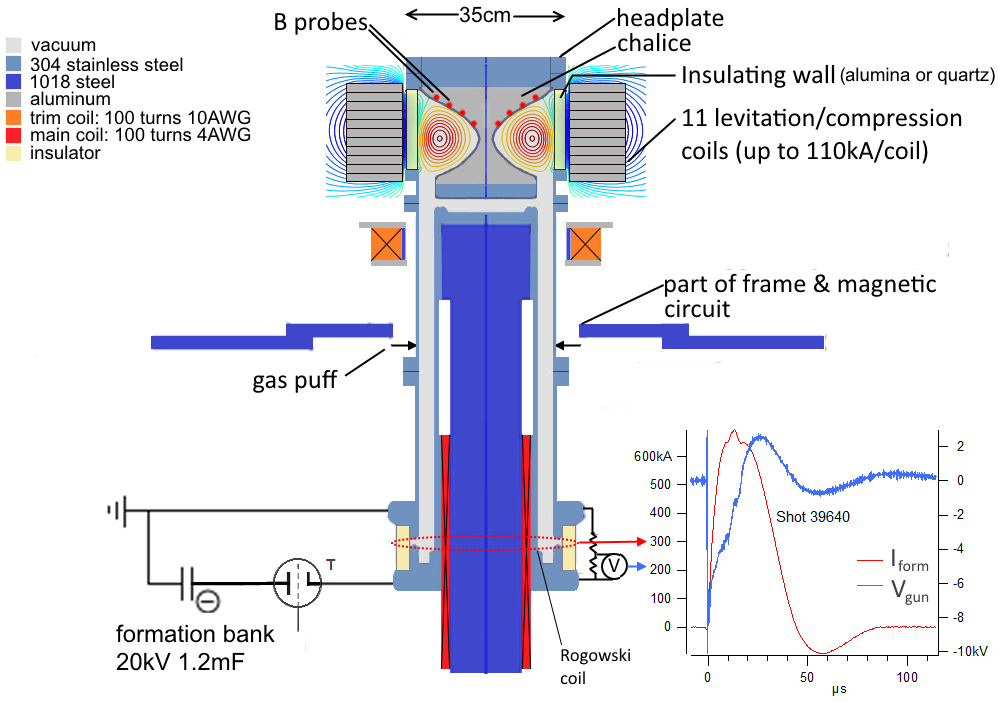}\caption{\label{fig:Machine-Schematic}Machine Schematic}
\end{figure}
The overall structure and experiment design of the SMRT magnetic compression
device can be seen in figure \ref{fig:Machine-Schematic}, including
a schematic of the formation circuit and with CT and levitation $\psi$
(poloidal flux) contours from an equilibrium model superimposed.  Measurements
of formation current $(I_{form}(t))$, and voltage across the formation
electrodes $(V_{gun}(t))$ are also indicated. Note that the principal
materials used in the machine construction, and some key components,
are indicated by the color-key at the top left of the figure. Apart
from the inclusion of the levitation/compression coils and the insulating
tube around the CT containment region, the machine is identical to
the standard pre - 2016 General Fusion $\mbox{MRT}$ (Magnetized Ring
Test) plasma injectors, which had an aluminum outer flux conserver
in place of the insulating tube. The sequence of machine operation
is as follows:\vspace{1cm}
\begin{table}[H]

\begin{centering}
\begin{tabular}[t]{lll}
(1) & $t\sim-3s$  & Main coil is energised with steady state ($\sim4\mbox{s}$ duration)
current ($I_{main}$)\tabularnewline
(2) & $t=t_{gas}\sim-400\upmu$s  & Gas is injected into vacuum \tabularnewline
(3) & $t=t_{lev}\sim-400\upmu\mbox{s}\rightarrow-40\upmu\mbox{s}$  & Levitation banks, charged to voltage $V_{lev}$, are triggered\tabularnewline
(4) & $t=0$s  & Formation banks, charged to voltage $V_{form}$, are triggered\tabularnewline
(5) & $t=t_{comp}\sim40\upmu\mbox{s}\rightarrow150\upmu\mbox{s}$  & Compression banks, charged to voltage $V_{comp}$, are triggered\tabularnewline
\end{tabular}\caption{\label{tab:Sequence-of-machine}Sequence of machine operation}
\par\end{centering}
\end{table}
\vspace{1cm}
\begin{figure}[H]
\subfloat{\raggedright{}\includegraphics[scale=0.5]{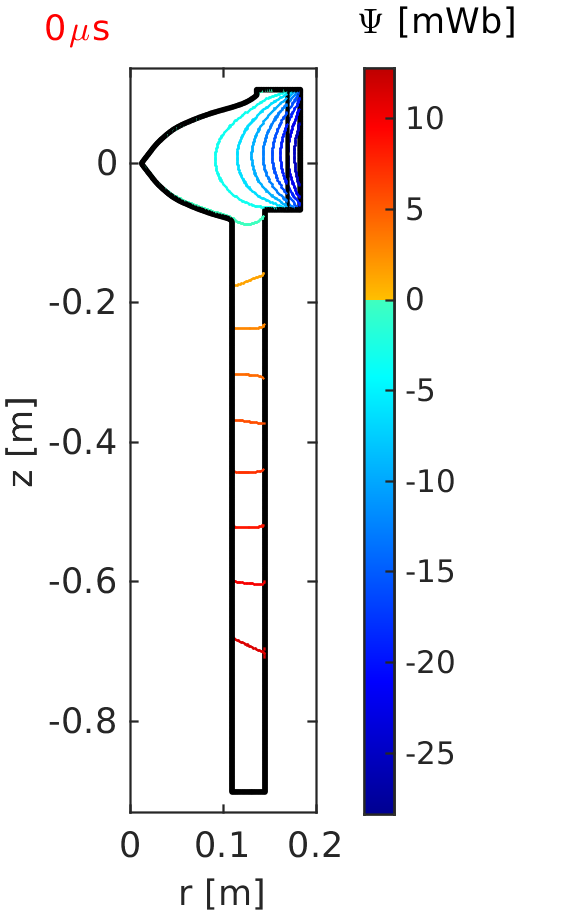}}\hfill{}\subfloat{\raggedright{}\includegraphics[scale=0.5]{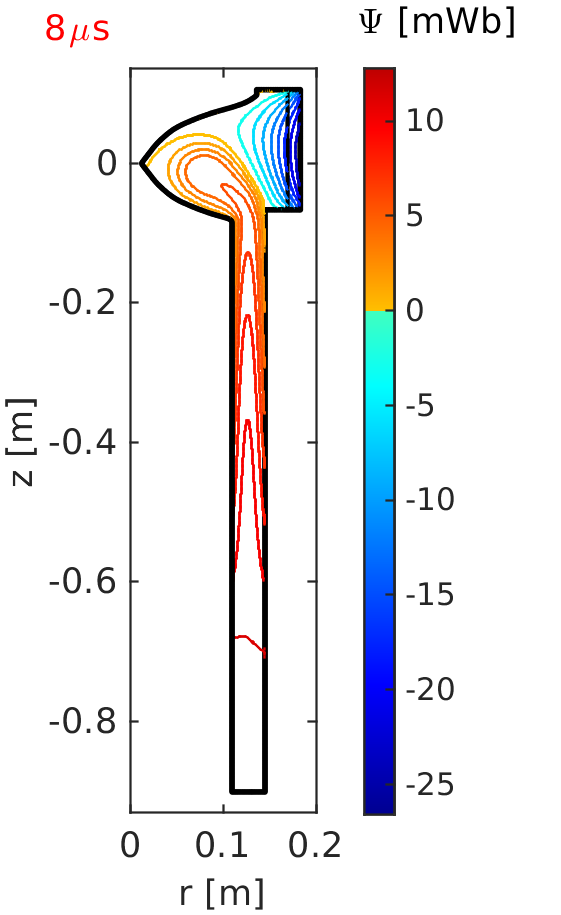}}\hfill{}\subfloat{\raggedright{}\includegraphics[scale=0.5]{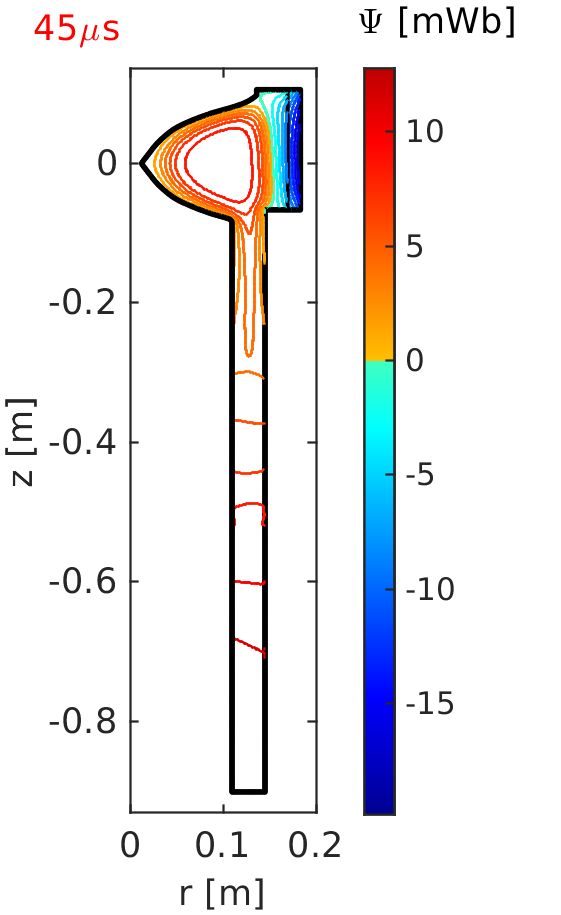}}

\subfloat{\raggedright{}\includegraphics[scale=0.5]{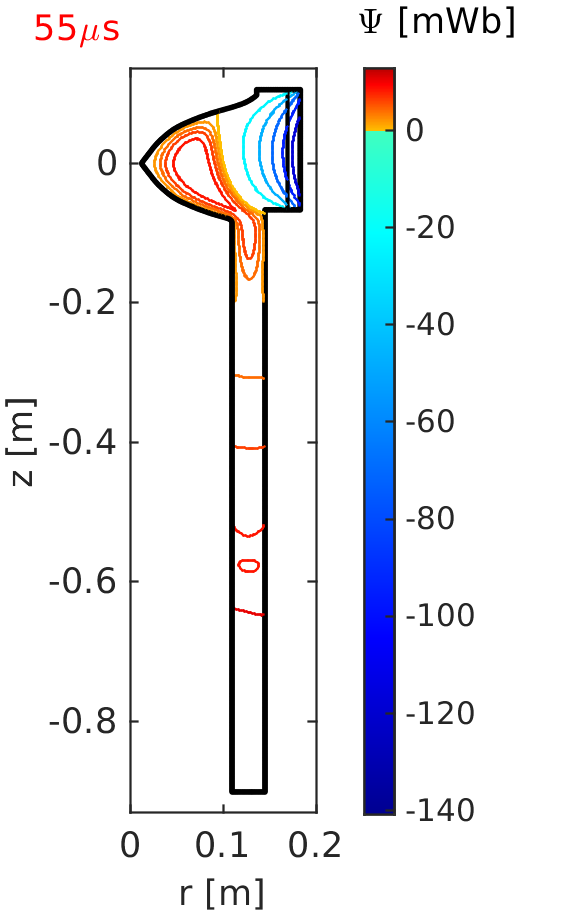}}\hfill{}\subfloat{\raggedright{}\includegraphics[scale=0.5]{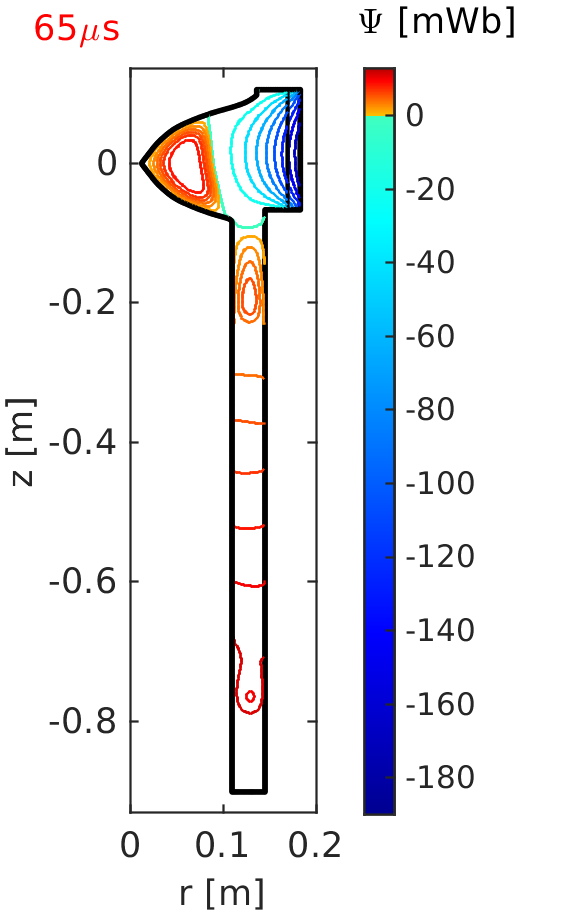}}\hfill{}\subfloat{\raggedright{}\includegraphics[scale=0.5]{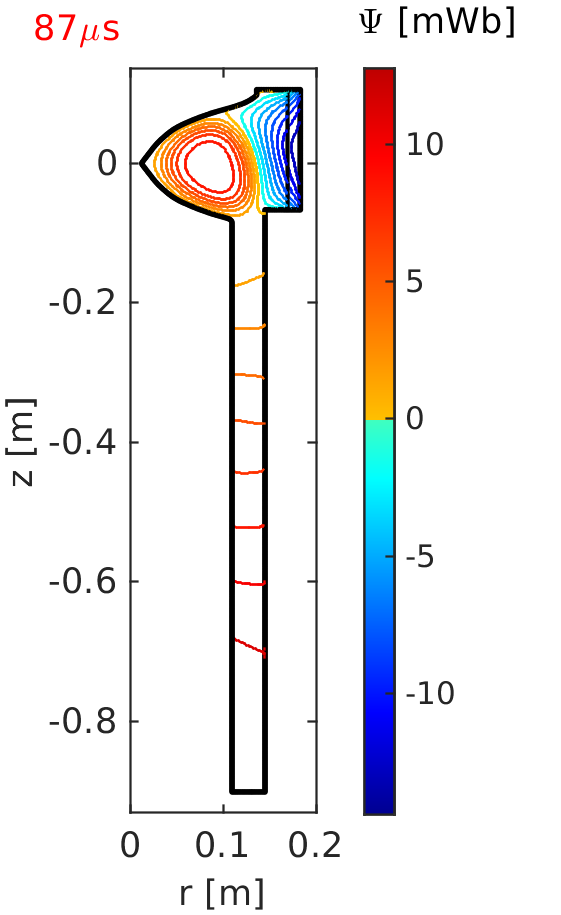}}

\caption{\label{fig:MHDform}$\psi$ contours from an MHD simulation of CT
formation into levitation field, and magnetic compression, with $t_{comp}=45\upmu\mbox{s}$}
\end{figure}
To better illustrate the sequence of operation, $\psi$ contours from
an MHD simulation of the magnetic compression experiment are shown
in figure \ref{fig:MHDform}. As described separately in \cite{SIMpaper,thesis-1},
an energy, particle, and toroidal flux conserving finite element axisymmetric
MHD code was developed to study CT formation into a levitation field,
and magnetic compression. The Braginskii MHD equations with anisotropic
heat conduction were implemented. To simulate plasma / insulating
wall interaction, we couple the vacuum field solution in the insulating
region to the full MHD solution in the remainder of the domain. A
plasma-neutral interaction model including ionization, recombination,
charge-exchange reactions, and a neutral particle source, was implemented
in order to study the effect of neutral gas in the gun on simulated
formation \cite{Neut_paper}. In figure \ref{fig:MHDform}, note that
$\psi$ contours represent poloidal field lines, and that the vertical
black line at the top-right of the figures at $r\sim17\mbox{cm}$
represents the inner radius of the insulating wall. Vacuum field only
is solved for to the right of the line, and the plasma dynamics are
solved for in the remaining solution domain to the left of the line.
The inner radius of the stack of eleven levitation/compression coils
(which are not depicted here) is located at the outer edge of the
solution domain, at $r\sim18\mbox{cm}$. Simulation times are notated
in red at the top left of the figures. Note that the colorbar scaling
changes over time; $\max(\psi)$ decreases slowly over time as the
CT decays, while min$(\psi)$ increases as the levitation current
in the external coils decays, and then drops off rapidly as the compression
current in the external coils is increased, starting at $t_{comp}=45\upmu\mbox{s}$
in this simulation. At time $t=0,$ the stuffing field ($\psi>0$)
due to currents in the main coil fills the vacuum below the containment
region, and has soaked well into all materials around the gun, while
the levitation field fills the containment region. Simulated CT formation
is initiated with the addition of toroidal flux below the gas puff
valves located at $z=-0.43$m; initial intra-electrode radial formation
current is assumed to flow at the z-coordinate of the valves. As described
in detail in \cite{SIMpaper,thesis-1}, toroidal flux addition is
scaled over time in proportion to $\int_{0}^{t}V_{gun}(t')\,dt'$.
Open field lines that are resistively pinned to the electrodes, and
partially frozen into the conducting plasma, have been advected by
the $\mathbf{J}_{r}\times\mathbf{B}_{\phi}$ force into the containment
region by $t=8\upmu\mbox{s}$ ($\mathbf{J}_{r}$ is the radial formation
current density across the plasma between the electrodes, and $\mathbf{B}_{\phi}$
is the toroidal field due to the axial formation current in the electrodes).
By $45\upmu\mbox{s}$, open field lines have reconnected at the entrance
to the containment region to form closed CT flux surfaces. At these
early times, open field lines remain in place surrounding the CT.
Compression starts at $45\upmu\mbox{s}$ and peak compression is at
$65\upmu\mbox{s}$. The CT expands again between $65\upmu\mbox{s}$
and $87\upmu\mbox{s}$ as the compression current in the external
levitation/compression coils decreases. Note that at $55\upmu$s,
magnetic compression causes closed CT poloidal field lines that extend
down the gun to be pinched off at the gun entrance, where they reconnect
to form a second smaller CT. Field lines that remain open surrounding
the main CT are then also reconnectively pinched off, forming additional
closed field lines around the main CT, while the newly reconnected
open field lines below the main CT act like a slingshot that advects
the smaller CT down the gun, as can be seen at $65\upmu\mbox{s}$.\\
\\
\\
A pulse-width modulation system was used for current control in the
main coil circuit. The working gas was typically $\mbox{He},\,\mbox{H}_{2},$
or $\mbox{D}_{2}$, with valve plenum pressure $\sim30\mbox{psi}\,(\mbox{gauge})$,
and optimal vacuum pressure $\sim1\times10^{-8}\,\mbox{Torr}$. The
formation capacitor bank ($240\mbox{kJ},\,\mbox{bank consisting of twenty four }50\upmu\mbox{F},\,20\mbox{kV}$
capacitors in parallel) drives up to 1MA of current, with a half period
of $50\upmu\mbox{s}$ (see figure \ref{fig:Machine-Schematic} inset).
The original machine configuration had six levitation/compression
coils, with each coil having its own levitation and compression circuit.
The $120\mbox{kJ}$ levitation bank consisted of two $50\upmu\mbox{F},\,20\mbox{kV}$
capacitors in parallel for each coil, and there were four of these
capacitors in parallel for each coil for the $240\mbox{kJ}$ compression
bank. 
\begin{figure}[H]
\centering{}\includegraphics[scale=0.15]{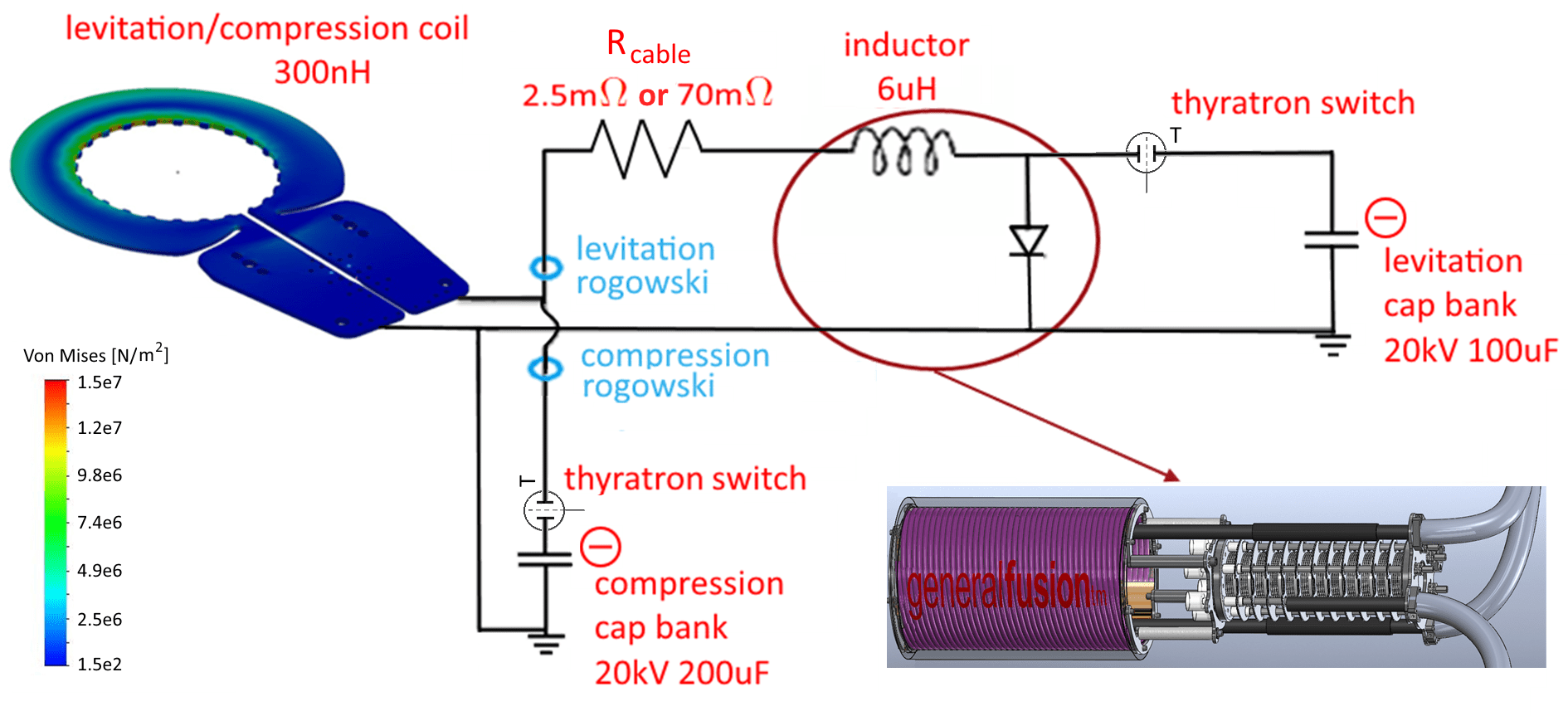}\caption{\label{fig:Levitation-and-compression}Levitation and compression
circuit }
\end{figure}
The circuit for one of the single-turn levitation and compression
coils is depicted in figure \ref{fig:Levitation-and-compression}.
Each coil (or coil-pair in the case of the configuration with 11 coils)
had a separate identical circuit. Unlike the crowbarred levitation
currents, the compression currents are allowed to ring with the capacitor
discharge. Typical levitation and compression current waveforms are
shown in figures \ref{fig:Poloidal-field-for} and \ref{fig:39735BpMagnetic-compression-shot}
(right axes). 
\begin{figure}[H]
\begin{centering}
\subfloat[\label{fig:Machine-headplate-schematic}Machine headplate schematic ]{\begin{centering}
\includegraphics[scale=0.3]{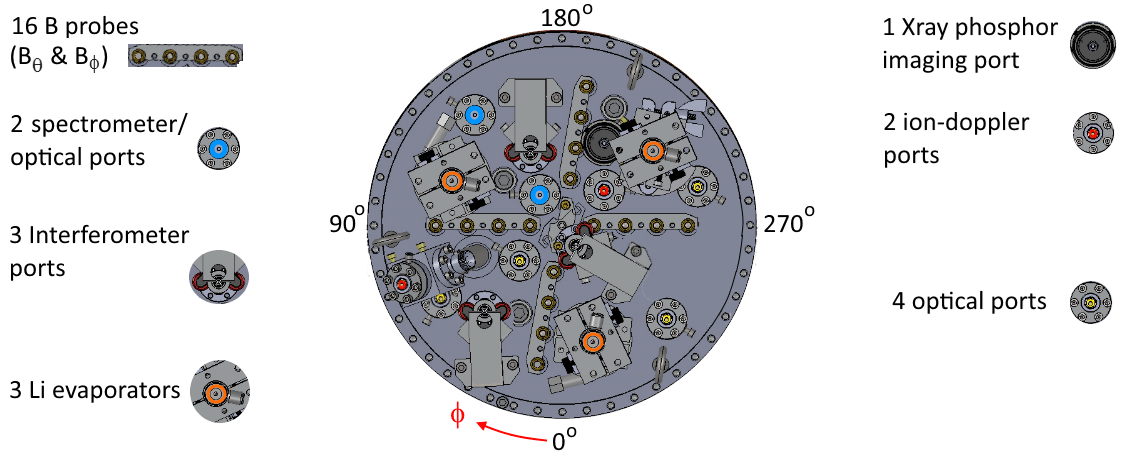}
\par\end{centering}
\centering{}}
\par\end{centering}
\begin{centering}
\subfloat[\label{fig:Chalice} Inner flux conserver view ]{\begin{centering}
\includegraphics[scale=0.3]{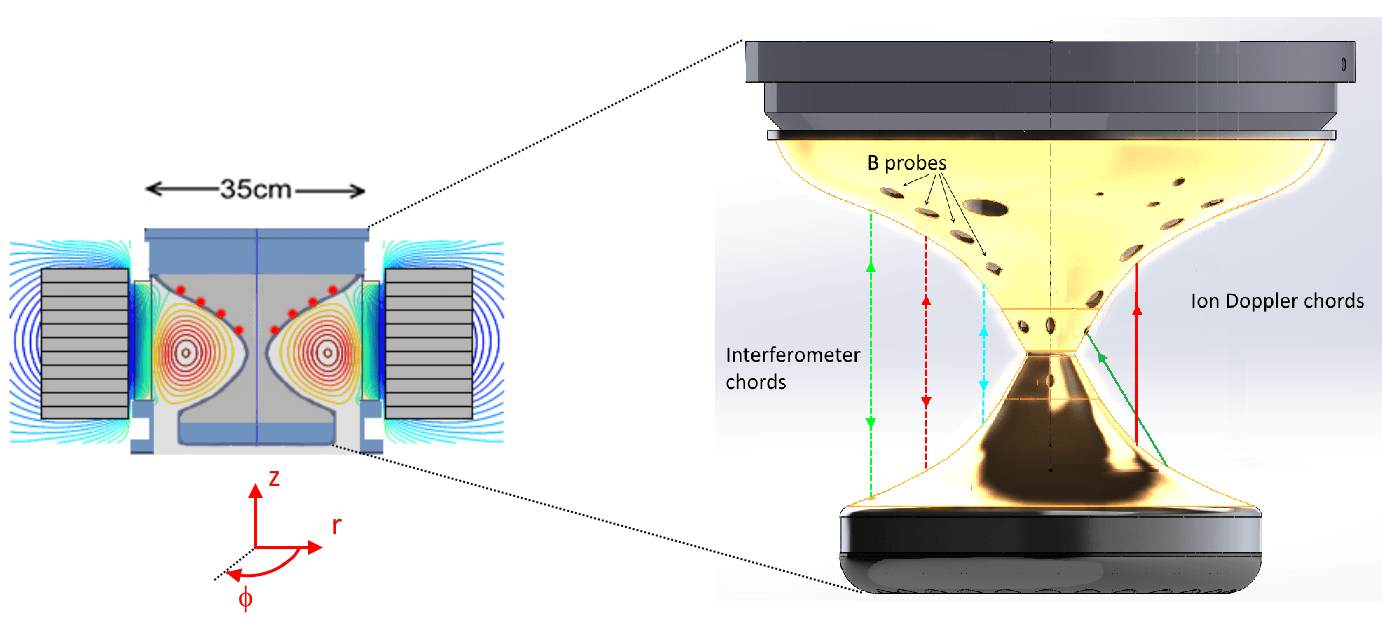}
\par\end{centering}
\centering{}}\caption{Diagnostics overview\label{fig:Diagnostics-overview}}
\par\end{centering}
\end{figure}
A schematic of the machine headplate (top view) indicating the principal
diagnostics and lithium evaporator ports is shown in figure \ref{fig:Diagnostics-overview}(a).
 Lithium coating on the surfaces of the containment region was applied
routinely as a gettering agent, and resulted in levitated CT lifetime
improvements of up to 70\%, depending on the insulating wall material,
as outlined in section \ref{subsec:Main-points--}. Figure \ref{fig:Diagnostics-overview}(b)
depicts the tungsten-coated aluminum inner flux-conserver, indicating
the locations of some of B-probe ports and the lines-of-sight for
the ion Doppler and interferometer diagnostics. For ease of depiction,
the ion Doppler/interferometer chords are shown to be located on the
same poloidal plane. Line-averaged electron density was obtained along
chords at $r=35\mbox{mm},\,r=65\mbox{mm}$ and $r=95\mbox{mm}$ using
dual $1310$nm and $1550$nm He-Ne laser interferometers. Dual wavelength
Michelson-type interferometers were used to enable compensation for
errors due to machine vibration during a shot. Note that the plasma-traversing
beam crosses through the plasma twice. Retroreflectors positioned
in the base of the inner flux conserver reflect the beam back up through
the plasma. The reference beam is directed along a path of equal length
in ambient air. An indication of ion temperature, along the vertical
chord at $r=45\mbox{mm}$ and the diagonal chord with its upper point
at $r\sim25\mbox{mm}$, was found from Doppler broadening of line
radiation from singly ionized Helium (He II line at 468.5nm). Visible
light emission is recorded by two survey spectrometers which have
variable exposure durations, and by six fiber-coupled photodiodes
that record time-histories of total optical emission. 

Two magnetic probes, for recording poloidal and toroidal field, were
located at the closed ends of each of sixteen thin-walled stainless
steel tubes embedded in axially directed holes in the inner flux conserver.
The $r,\,\phi$ coordinates of the probes, where $r=0$ is defined
as being at the machine axis, are:\hfill{} \\
\\
\begin{table}[H]
\centering{}{\footnotesize{}}%
\begin{tabular}{|c|c|c|c|c|c|c|c|c|c|c|c|c|c|c|c|c|}
\hline 
{\footnotesize{}$\mathbf{r}\,[\mbox{\mbox{mm}}]$} & {\footnotesize{}$26$} & {\footnotesize{}$26$} & {\footnotesize{}$39$} & {\footnotesize{}$39$} & {\footnotesize{}$52$} & {\footnotesize{}$52$} & {\footnotesize{}$64$} & {\footnotesize{}$64$} & {\footnotesize{}$77$} & {\footnotesize{}$77$} & {\footnotesize{}$90$} & {\footnotesize{}$90$} & {\footnotesize{}$103$} & {\footnotesize{}$103$} & {\footnotesize{}$116$} & {\footnotesize{}$116$}\tabularnewline
\hline 
{\footnotesize{}$\phi\,${[}deg.{]}} & {\footnotesize{}$90$} & {\footnotesize{}$270$} & {\footnotesize{}$10$} & {\footnotesize{}$190$} & {\footnotesize{}$90$} & {\footnotesize{}$270$} & {\footnotesize{}$10$} & {\footnotesize{}$190$} & {\footnotesize{}$90$} & {\footnotesize{}$270$} & {\footnotesize{}$10$} & {\footnotesize{}$190$} & {\footnotesize{}$90$} & {\footnotesize{}$270$} & {\footnotesize{}$10$} & {\footnotesize{}$190$}\tabularnewline
\hline 
\end{tabular}\caption{\label{tab: coordinates-ofBprobes}$r,\,\phi$ coordinates of magnetic
probes}
\end{table}

\section{\label{part:levitation}Magnetic Levitation}

\subsection{\label{subsec::Levitation coil configurations}Overview of external
coil configurations}

\begin{figure}[H]
\begin{raggedright}
\subfloat[]{\centering{}\includegraphics[scale=0.5]{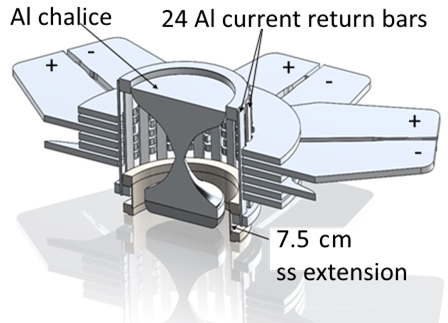}}\hfill{}\subfloat[]{\centering{}\includegraphics[scale=0.22]{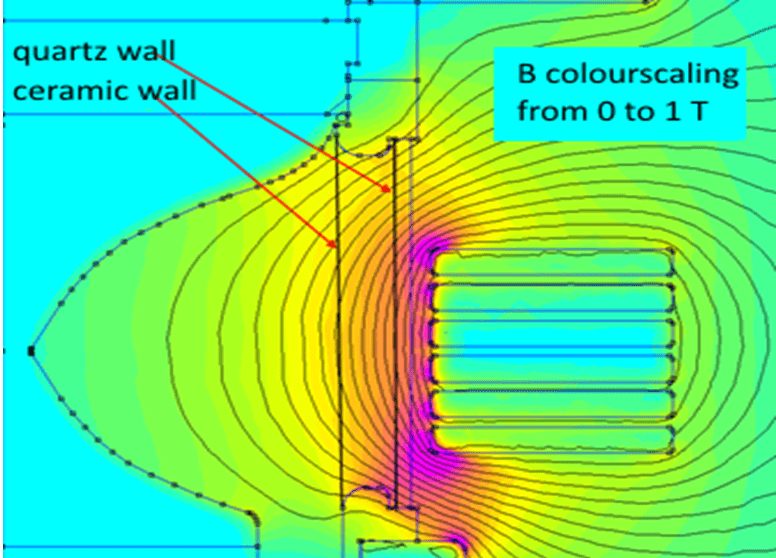}} 
\par\end{raggedright}
\centering{}\caption{\label{fig:Schematic-of-6-1}Schematic of six coils (a) and FEMM model
of levitation field (b)}
\end{figure}
 Figure \ref{fig:Schematic-of-6-1}(a) indicates, for the original
configuration with six coils, the inner flux conserver, coils,  stainless
steel extension, and aluminum return-current bars that carry axial
current outside the insulating wall. The inner radii of the original
ceramic (alumina - $\mbox{Al}{}_{2}\mbox{O}_{3}$) wall, and of the
quartz (silica - $\mbox{SiO\ensuremath{_{2}}}$) wall that was tested
later, are shown in \ref{fig:Schematic-of-6-1}(b). This is an output
plot from the open-source FEMM (Finite Element Method Magnetics) program\cite{FEMM},
with $30\mbox{kA}$ per coil and an input current frequency of $800$Hz.
Contours of $\psi_{lev}$, poloidal levitation flux, are shown, with
the plot colour-scaling being proportional to $|B|.$ FEMM models
alternating currents as sinusoidal waveforms in time, so we chose
$|t_{lev}|=300\upmu$s to be the quarter- period, giving a frequency
of $800$Hz. In the configuration with six coils, it was found that
CT lifetimes could be increased by $\sim10\%$ by firing the levitation
capacitors at $t_{lev}\sim-300\upmu\mbox{s}$, well before firing
the formation capacitors. This allows the levitation field to soak
into the stainless steel above and below the wall, resulting in line-tying
(field-line pinning) - magnetic field that is allowed to soak into
the steel can only be displaced on the resistive timescale of the
metal, which is longer than the time it takes for the CT to bubble-in
to the containment region. Note that the principal materials used
in the construction of the magnetic compression machine are indicated
in figure \ref{fig:Machine-Schematic}. As confirmed by MHD simulations
\cite{thesis-1}, this line-tying effect is thought to have reduced
plasma-wall interaction and CT impurity inventory by making it a little
harder for magnetised plasma entering the confinement region to push
aside the levitation field. FEMM models were used to produce boundary
conditions for $\psi$, pertaining to the peak values of toroidal
currents in the main, levitation, and compression coils at the relevant
frequencies, for MHD and equilibrium simulations. For MHD simulations,
boundary conditions for $\psi_{lev}(\mathbf{r},t)$ and $\psi_{comp}(\mathbf{r},t)$
are scaled over time according to the experimentally measured waveforms
for $I_{lev}(t)$ and $I_{comp}(t)$. 

The $7.5\mbox{cm}$ high stainless steel extension indicated in figure
\ref{fig:Schematic-of-6-1}(a) was an addition to the original configuration
that also helped reduce the problem of plasma-wall interaction. In
 the original design without the extension, the ceramic insulating
outer wall extended down an additional 7.5cm. With the original levitation
field profile from six coils, and without the extension, levitated
CTs were short-lived, up to $\sim100\upmu$s as determined from the
poloidal B-probes embedded in the aluminum inner flux conserver at
$r=52\mbox{mm}$ (see figures \ref{fig:Machine-Schematic} and \ref{fig:Diagnostics-overview}(a)),
compared with over $300\upmu$s for non-levitated CTs produced in
MRT injectors with an aluminum outer flux conserver. CT lifetime was
increased, up to $\sim170\upmu$s, with the addition of the steel
extension. The extension mitigated the problems of sputtering of steel
at the alumina/steel lower interface, and of plasma interaction with
the insulating wall during the formation process. 

An insulating wall with larger internal radius was tested (original
alumina tube with $r_{in}=144\mbox{mm}$ was replaced with a quartz
tube with $r_{in}=170\mbox{mm}$). The resistive part of $\dot{\psi}$
is $\dot{\psi}_{\eta}=\eta\Delta^{^{*}}\psi$, where $\Delta^{^{*}}$
is the elliptic Laplacian-type operator used in the Grad-Shafranov
equation, and $\eta\,[\mbox{m}^{2}/\mbox{s}]$ is the magnetic diffusivity,
so CT lifetime should scale approximately with $l^{2}$, where $l$
is the characteristic length scale associated with the CT. The radius
of the inboard wall at the inner flux conserver waist at $z=0$ is
$r_{w}\sim20\mbox{mm}$. The minor CT radius for a given $r_{in}$
would be approximately $a\sim\frac{r_{in}-r_{w}}{2}$, so assuming
that $l\sim a$, we have $\frac{l_{quartz}^{2}}{l_{ceramic}^{2}}\sim1.5$.
From this rough estimate, for a given $\psi_{CT}$, an increase from
$\sim170\upmu$s to $\sim260\upmu$s was expected with the transition
to the larger radius quartz tube. However lifetime decreased noticeably
($\sim170\upmu$s to $\sim150\upmu$s) with the transition, so that
in terms of CT lifetime, plasma interaction with quartz was almost
$twice$ as bad as with alumina. The quartz wall led to more plasma
impurities (see section \ref{subsec:Main-points--}), and consequent
further radiative cooling, and therefore an increased rate of resistive
decay. 

In the 6-coil configuration, the longest-lived CTs were achieved at
generally low settings for $V_{form}$, $I_{main}$, and $V_{lev}$
(note that for optimal settings, these parameters, defined in table
\ref{tab:Sequence-of-machine}, scale with one another), resulting
in low-flux CTs. For example, $V_{form}\mbox{ and }I_{main}$ would
typically have been $12\mbox{kV}$ and $45\mbox{A}$ compared with
$16\mbox{kV}$ and $70\mbox{A}$ for best performance on standard
$\mbox{MRT}$ machines. Note that $V_{form}=16$kV corresponds to
 a peak formation current of $I_{form}\sim700$kA, while $I_{main}=70$A
corresponds to a gun flux of around 12mWb. Increasing these parameters
on the magnetic compression injector in the 6-coil configuration led
to increased impurity levels and degraded lifetime further.

After CT formation and relaxation, it was usual, with the standard
$\mbox{MRT}$ machines from around 2013 onwards, to observe magnetic
fluctuations with toroidal mode number $n=2$ on the measured $B_{\theta}$
signals, as determined by phase analysis of the $B_{\theta}$ signals
from probes located at the same radius $180^{o}$ apart toroidally
(see table \ref{tab: coordinates-ofBprobes}). These fluctuations
are evidence of coherent CT toroidal rotation, and were absent on
shots taken on the magnetic compression device. There was concern
that rotation could be impeded by mode-locking caused by toroidal
asymmetry in the levitation field, introduced by the gaps in toroidal
levitation current associated with the single-turn coils. A set of
six new coils (coil outline is depicted in figure \ref{fig:Levitation-and-compression}),
which reduced the original field error by a factor of $\sim10$, was
manufactured. Also, a 25-turn, high inductance ($160\upmu$H) coil
was experimented with - this reduced the original field error by a
factor of $\sim100.$ Due to its long $150\upmu$s current rise time,
and inadequate structural resistance against $\mathbf{J\times}\mathbf{B}$
forces, the 25-turn coil could be used only for CT levitation (in
connection with a single levitation circuit), and not for compression.
The coil was made with a height that extended all the way along the
insulating wall, closing the gaps outboard of the wall that were present
above and below the 6-coil stack. It was thought that the presence
of the gaps facilitated the process by which magnetised plasma entering
the confinement region at formation can push aside the levitation
field and interact with the insulating wall, sputtering impurities
into the plasma. 

It turned out that levitation field asymmetry associated with the
original single-turn coils was not a problem at the level of performance
achieved. At the settings for low-flux CTs, no improvement in CT lifetime
or symmetry was seen with either the new set of discrete coils or
the 25-turn coil, and there was no additional evidence of CT rotation,
nor evidence that a mode-locking issue had been alleviated. Movement
of filamentary structures observed with X-ray phosphor imaging indicated
the likelihood of CT rotation, but couldn't confirm it. Coherent CT
rotation was confirmed later in the experiment; $n=1$ fluctuations
regularly appeared on the $B_{\theta}$ traces when additional toroidal
field was included with the use of $80\mbox{kA}$ crowbarred formation
current. The $n=2$ fluctuations observed on $B_{\theta}$ signals
with standard $\mbox{MRT}$ machines may be connected with internal
reconnection events that occurred upon exceeding a threshold in CT
temperature. The first appearance of the $n=2$ fluctuations on $\mbox{MRT}$
machines was in 2013, when titanium gettering was first experimented
with. Back in 2013, titanium gettering led to CT lifetime increases
of up to $30\%$, to $300\upmu\mbox{s}$, and an increase in electron
temperature, as determined with Thomson scattering, from $\sim20\rightarrow80\mbox{eV}$
near the CT core.

As a result of the modification of the levitation field profile, the
25-turn coil allowed for the production of high flux levitated CTs,
with a corresponding improvement in lifetime. At $V_{form}=16\mbox{kV}$
and $I_{main}=70\mbox{A}$, best CT lifetimes with the quartz insulating
wall improved $\sim80\%$, from $\sim150\upmu$s (low flux CTs with
6 coils) to $\sim270\upmu$s. With a single levitation circuit, the
25-turn coil also facilitated the optimization of circuit parameters,
and it was found that CT lifetime and repeatability of good shots
(long CT lifetime) could be improved by adding resistance to the circuit
in order to match the decay rate of the levitation current to that
of the CT current. 

The 11-coil configuration consisted of 5 coil pairs and one single
coil, and approximately reproduced the levitation field profile of
the 25-turn coil, allowing for formation and compression of higher-flux
CTs with correspondingly increased lifetimes. Each pair was assembled
using one of the original coils, clamped together in parallel with
one of the newer coils that were designed to increase toroidal symmetry
in the levitation/compression field. The remaining newer coil was
included on its own, positioned 3rd from the bottom of the 11-coil
stack, to further increase the field at the top and bottom of the
wall.

\begin{figure}[H]
\begin{raggedright}
\subfloat[eleven coils on machine]{\raggedleft{}\includegraphics[scale=0.245]{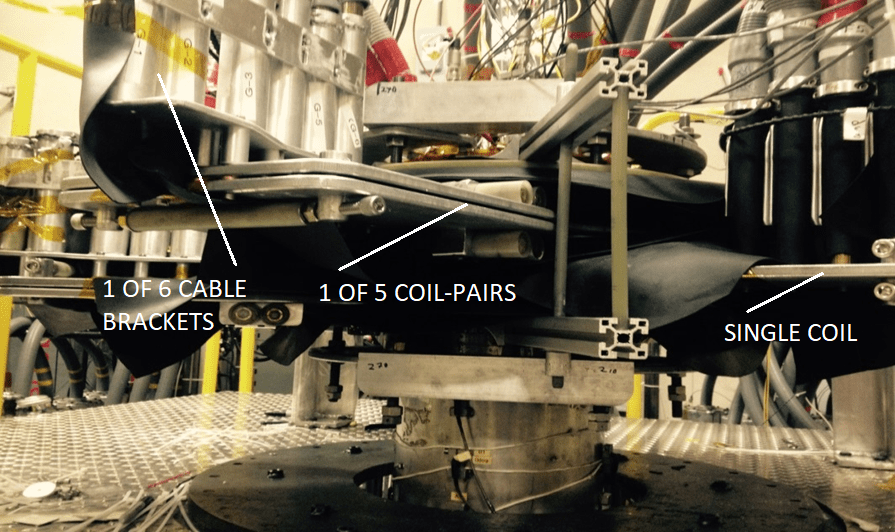}}\hfill{}\subfloat[FEMM model - 11 coils]{\includegraphics[scale=0.5]{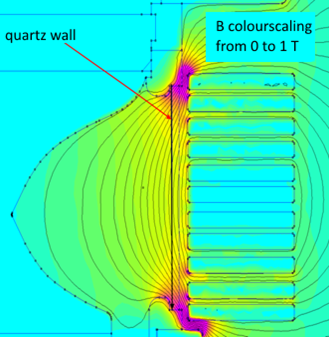}} 
\par\end{raggedright}
\centering{}\caption{\label{fig:11-coil-configuration}11-coil configuration}
\end{figure}
 The 11-coil stack installed on the machine is shown in figure \ref{fig:11-coil-configuration}(a)
- the single coil is visible on the lower right.  Each coil/coil-pair
is connected to its own levitation circuit via the two outer co-axial
cables in the cable connecting bracket attached to the coil/coil-pair.
One of the six brackets can be seen in the upper left foreground.
Each of the inner four co-axial cables in each bracket links individually
to a $52\upmu\mbox{F},\,20\mbox{kV}$ compression capacitor and thyratron
switch. Figure \ref{fig:11-coil-configuration}(b) shows a FEMM output
plot of the levitation field for the 11-coil setup with $16\mbox{kA}$
per coil and a solution frequency of $4$kHz, corresponding to the
experimentally determined optimal delay of $|t_{lev}|=50\upmu$s between
the firing of the levitation and formation capacitor banks. Note that
the strategy used with the 6-coil configuration of increasing $|t_{lev}|$
to reduce plasma/wall interaction was not required with the 11-coil
configuration. 

\subsection{Overview of results and comparison with simulations - CT levitation\label{subsec:Levresultsoverview}}

\subsubsection{Magnetic field measurements\label{subsec:Magnetic-field-mesurements}}

\begin{figure}[H]
\subfloat[$B_{\theta}$, $2.5\mbox{m}\Omega$ cables]{\raggedright{}\includegraphics[width=8cm,height=5cm]{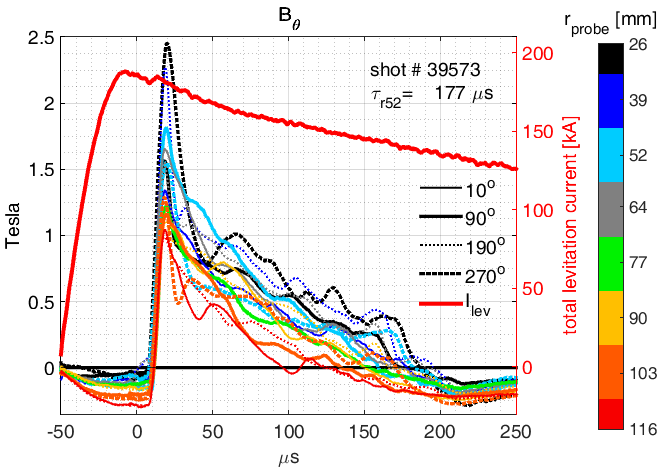}}\hfill{}\subfloat[$B_{\theta}$, $70\mbox{m}\Omega$ cables]{\raggedleft{}\includegraphics[width=8cm,height=5cm]{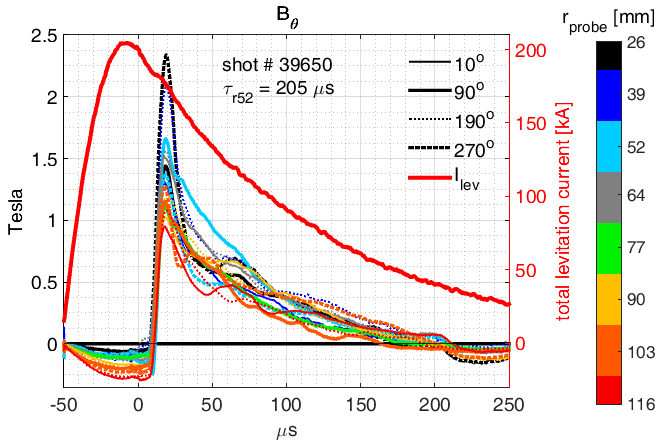}}

\subfloat[$B_{\phi}$, $2.5\mbox{m}\Omega$ cables ]{\raggedright{}\includegraphics[width=8cm,height=5cm]{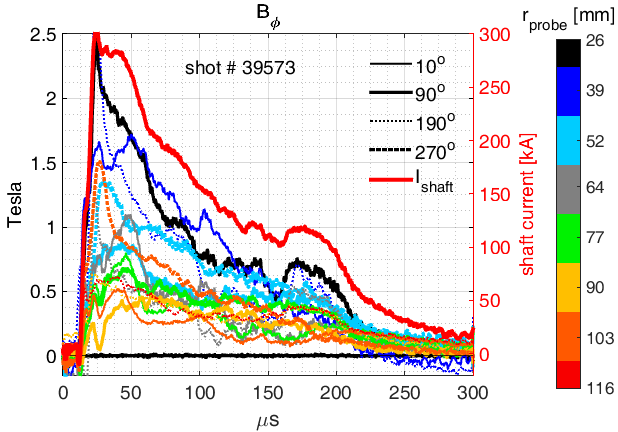}}\hfill{}\subfloat[$B_{\phi}$, $70\mbox{m}\Omega$ cables]{\raggedleft{}\includegraphics[width=8cm,height=5cm]{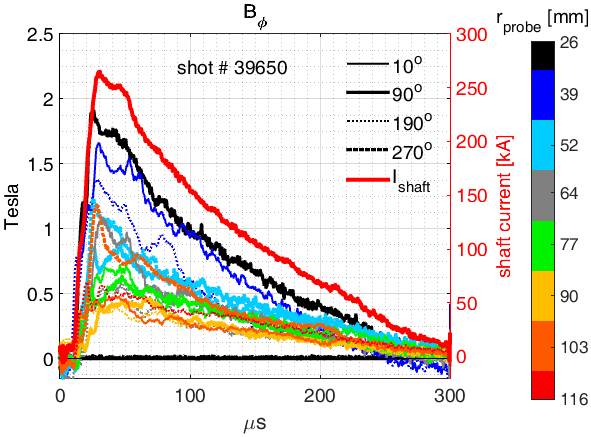}}

\caption{\label{fig:Poloidal-field-for}$B_{\theta}$ (figures (a), (b)) and
$B_{\phi}$ ((c), (d)) for levitated CT with eleven coils. Note that
$I_{lev}$ and $I_{shaft}$ are also indicated on the plots of $B_{\theta}$
and $B_{\phi}$ respectively.  }
\end{figure}
Measurements of $B_{\theta}$ and $B_{\phi}$ for two shots taken
in the 11-coil configuration, with different cable resistances (denoted
as $R_{cable}$ in the levitation/compression circuit schematic in
figure \ref{fig:Levitation-and-compression}), is shown in figure
\ref{fig:Poloidal-field-for}.  As indicated in table \ref{tab: coordinates-ofBprobes},
there are sixteen magnetic probe heads located in the inner flux conserver
- eight of these are located at four different radii ($r=39,\,64,\,90$,
and 116mm) at toroidal angle $\phi=10^{o}$ and $\phi=190^{o}$, and
there are an additional eight probes at ($r=26,\,52,\,77$, and 103mm)
at $\phi=90^{o}$ and $\phi=270^{o}$. Magnetic probe signals are
colored by the radial coordinates of the probe locations, with toroidal
coordinates of the probe locations denoted by linestyle, as denoted
in the plot legends. CT lifetime is gauged using the $\tau_{r52}$
metric (indicated in figures \ref{fig:Poloidal-field-for}(a) and
(b)), which is the time at which the average of the poloidal field
measured at the two probes at $r=$52mm crosses zero. Note that $B_{\theta}$
is the field component parallel to the inner flux conserver surface
in the poloidal plane. Total levitation current, measured with Rogowski
coils, is also indicated in figures \ref{fig:Poloidal-field-for}(a)
and (b) (thick red lines, right axes) for the two shots. $t_{lev}=-50\upmu$s
for these shots, so with a current rise time of $\sim40\upmu$s in
the levitation coils, the poloidal levitation field measured at the
probes reaches its maximum negative value at $t\sim-10\upmu$s. Formation
capacitors are fired at $t=0$s, and (referring to figure \ref{fig:MHDform})
it takes $\sim10-20\upmu$s for the gun (stuffing) flux to be advected
up to the probe locations. The stuffing field has opposite polarity
to the levitation field. Over the next several tens of $\upmu$s,
during and after reconnection of poloidal field to form closed flux
surfaces, the CT undergoes Taylor relaxation, involving conversions
between poloidal and toroidal fluxes. The CT shrinks and is displaced
inwards by the levitation field as the CT currents and fields decay
resistively. As the CT decays, starting at the outer probes and progressing
inwards towards the inner probes, the CT field measured at the probes
is once again replaced by the levitation field. After $\sim200\upmu$s
(figures \ref{fig:Poloidal-field-for}(a) and (b)), the field measured
at all the B probes is the levitation field. Note that, when levitation
field is being measured at the probes, that $|B_{\theta}|$ is larger
at the outer probes, due to the $1/(r_{coil}-r_{probe})$ scaling
of levitation field with levitation current in the external coils.
On the other hand, when CT field is being measured at the probes,
$B_{\theta}$ is larger at the inner probes, due to the $1/r_{probe}^{2}$
scaling of CT field with CT flux - poloidal field lines are bunched
together progressively more at smaller radii. 

The shots referred to in figures \ref{fig:Poloidal-field-for}(a)
and (b) were both at $V_{form}=16\mbox{kV}$, $I_{main}=70A$, additional
settings included $V_{lev}=13.8\mbox{kV}$ for shot  39573, and $V_{lev}=16\mbox{kV}$
for shot  39650. To achieve approximately the same levitation current,
$V_{lev}$ was increased for shot  39650 to compensate for the additional
cable resistance. It can be seen how the decay rate of $I_{lev}$
approximately matches that of the CT toroidal current (as determined
by positive $B_{\theta}$) with a $70\mbox{m}\Omega$ cable replacing
the original pair of $5\mbox{m}\Omega$ cables in parallel ($i.e.,$
total $2.5\mbox{m}\Omega$) between the main holding inductor and
coil in each levitation circuit. A much higher rate of \textquotedbl good\textquotedbl{}
shots, smoother decays of $B_{\theta}$ and $B_{\phi}$ (less apparent
MHD activity), and a lifetime increase generally of around $10-20\%$,
was observed with the $70\mbox{m}\Omega$ cables in place. With the
low resistance $5\mbox{m}\Omega$ cables, the CT is decaying far faster
than the levitation field, so that it is being compressed (without
firing the compression banks) by the levitation field more and more
as $\psi_{CT}$ decreases. By matching the levitation field decay
rate to that of the CT currents, the CT is allowed to retain the size
that it would have if it was being held in place by field due to eddy
currents induced in an outboard flux-conserver, instead of by an outboard
levitation field. It can be seen how the CT is being displaced from
larger radii much faster in shot  39573, compared with shot  39650,
in which decay rate matching was implemented. In shot  39573, the
CT has been displaced inwards beyond the probe at $116\mbox{mm}$
by $\sim120\upmu$s, when the $B_{\theta}$ signal from the 116mm
probe goes negative, but this displacement is delayed until $\sim170\upmu$s,
in shot  39650. As described in appendix \ref{sec:rsep}, a method
was developed to experimentally measure the outboard CT separatrix
using data from a set of eight magnetic probes located on the insulating
wall. This method allows for time-resolved determination of the outboard
CT separatrix at eight equally spaced toroidal angles, and comparison
of this between configurations. 

Referring again to figures \ref{fig:Poloidal-field-for}(a) and (b),
note that in shot  39573, CT poloidal field at the inner probes collapses
rapidly to zero at $\sim170\upmu$s, whereas the decay is much smoother
in shot  39650 - this sudden collapse is due to the compressional
instability that will be discussed in section \ref{sec:Magnetic-compression}.

With reference to figures \ref{fig:Poloidal-field-for}(c) and (d),
$B_{\phi}$, the toroidal field measured at the probes, is due to
poloidal shaft current in the inner flux conserver. Shaft current
is induced to flow in conducting material surrounding the CT as the
system tries to conserve the toroidal flux introduced at CT formation,
and continues to decay away resistively for several tens of microseconds
after the CT currents have decayed. The current path includes the
inner flux conserver walls, aluminum bars (indicated in figure \ref{fig:Schematic-of-6-1}(a)),
and a path through ambient plasma in the gap below the CT between
the bottom of the inner flux conserver and the outer electrode. $I_{shaft}$
(thick red traces in figures \ref{fig:Poloidal-field-for}(c) and
(d)) is calculated from $B_{\phi}$ using Ampere's law. As the CT
shrinks due to compression, increasing proportions of poloidal shaft
current can divert from the initial paths in the aluminum bars, and
flow through the ambient plasma outboard of the CT. This will be clarified
in section \ref{sec:Magnetic-compression}. Shaft current increases
when it flows along the reduced inductance path through the ambient
plasma. There is evidence in figure \ref{fig:Poloidal-field-for}(c)
of mild magnetic compression by the levitation field starting at around
$150\upmu$s on shot 39573 (with the low resistance cables). This
is evident from the overall rise in shaft current at $\sim150\upmu$s,
and from the rise $B_{\phi}$ at the probes, particularly at the $(26\mbox{mm},\:90^{\circ})$
and $(52\mbox{mm},\:190^{\circ})$ probes. This unintentional compression
is absent with the implementation of decay-rate matching, as seen
in figure \ref{fig:Poloidal-field-for}(d) for shot  39650.\\
\\
\begin{figure}[H]
\subfloat[$2.5\mbox{m}\Omega$ cables]{\raggedright{}\includegraphics[width=7cm,height=5cm]{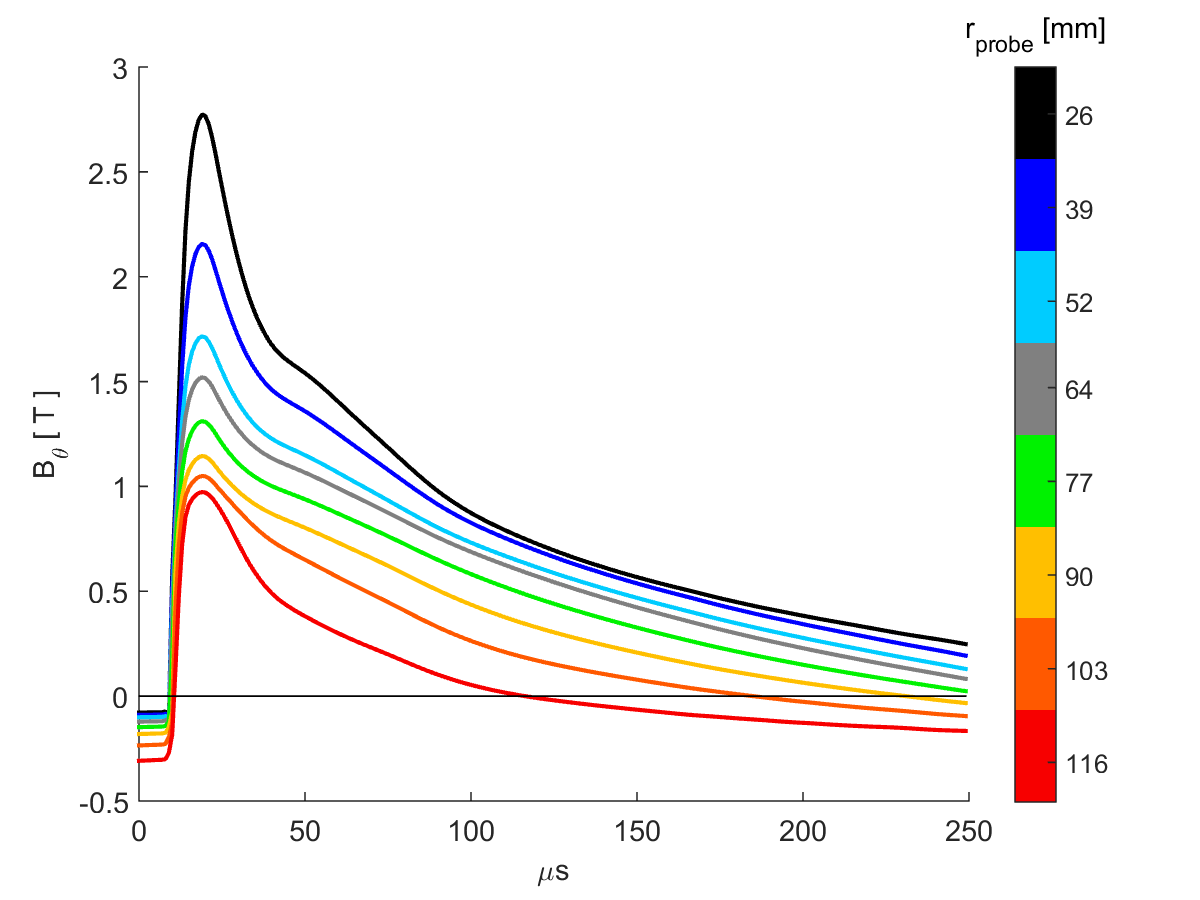}}\hfill{}\subfloat[$70\mbox{m}\Omega$ cables]{\raggedleft{}\includegraphics[width=7cm,height=5cm]{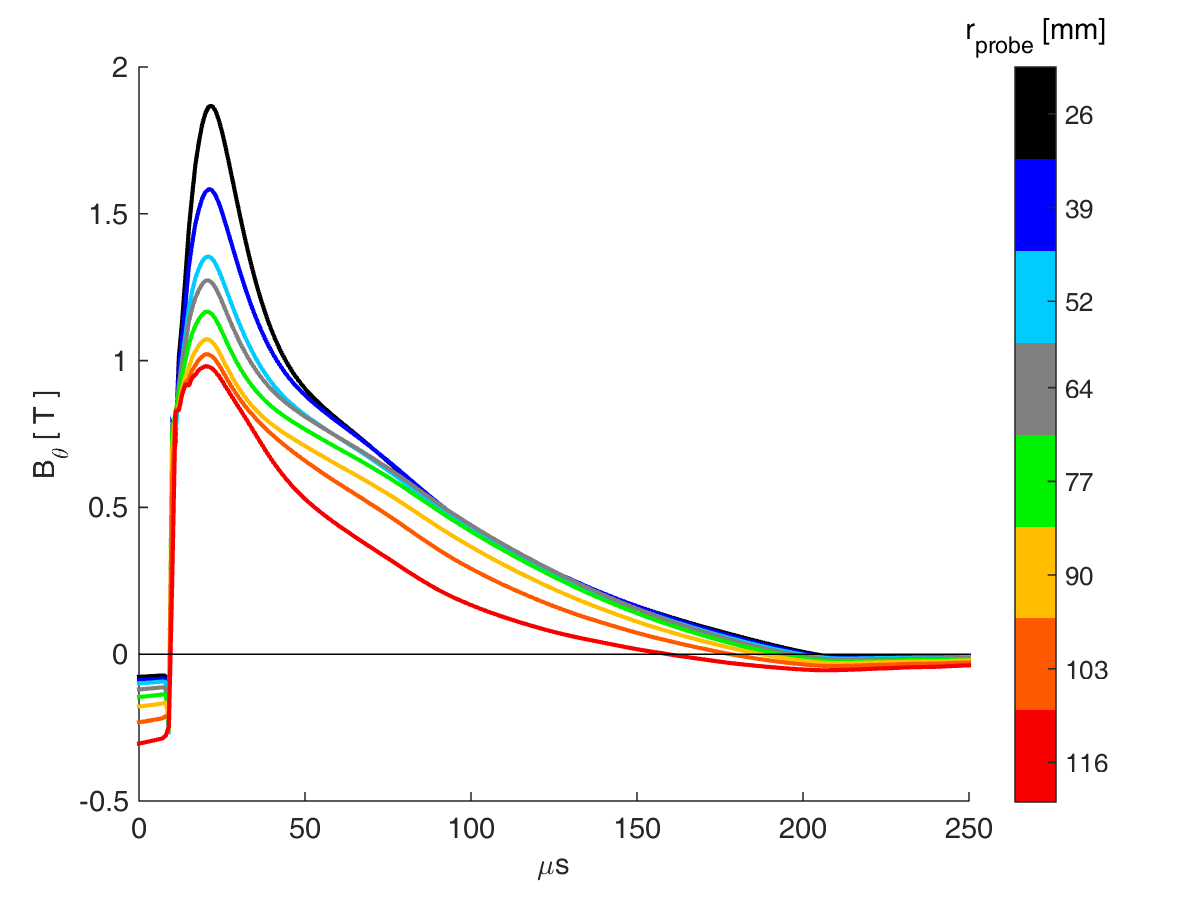}}

\caption{\label{fig:BpolSIM_res}Simulated $B_{\theta}$ for levitated CT with
eleven coils }
\end{figure}
Figures \ref{fig:BpolSIM_res}(a) and (b) show $B_{\theta}$ recorded
at the probe locations from MHD simulations in which the boundary
conditions pertaining to the levitation field were evolved over time
according to the experimentally measured waveforms for $I_{lev}$,
which depend on the resistance of the cables in the levitation circuit,
as indicated in figure \ref{fig:Poloidal-field-for}(a) and (b) (right
axes). Comparing the $B_{\theta}$ traces in figures \ref{fig:Poloidal-field-for}(a)
and \ref{fig:BpolSIM_res}(a), it can be seen how the comparison is
qualitatively good up until around $150\upmu$s, when the compressional
instability, which is not captured by the 2D MHD dynamics, causes
the CT to be extinguished rapidly. The comparison in figures \ref{fig:Poloidal-field-for}(b)
and \ref{fig:BpolSIM_res}(b) remains good at all times, as the compressional
instability did not arise in the case with decay rate matching. 

\begin{figure}[H]
\subfloat[Measured $B_{\phi}$ ($70\mbox{m}\Omega$ cables). $I_{shaft}$ is
also indicated. ]{\raggedright{}\includegraphics[width=8.6cm,height=5cm]{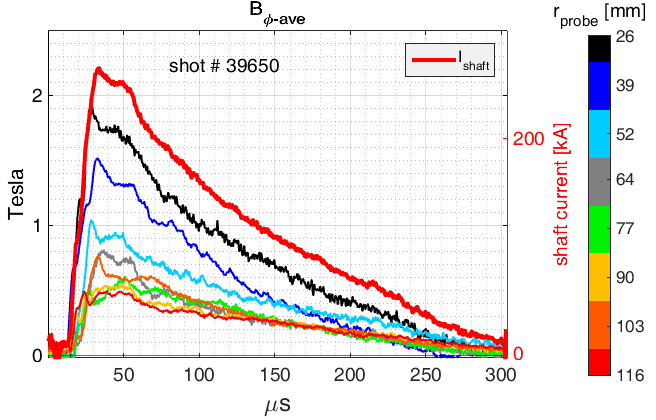}}\hfill{}\subfloat[Simulated $B_{\phi}$ ($70\mbox{m}\Omega$ cables) ]{\raggedleft{}\includegraphics[width=7.5cm,height=5cm]{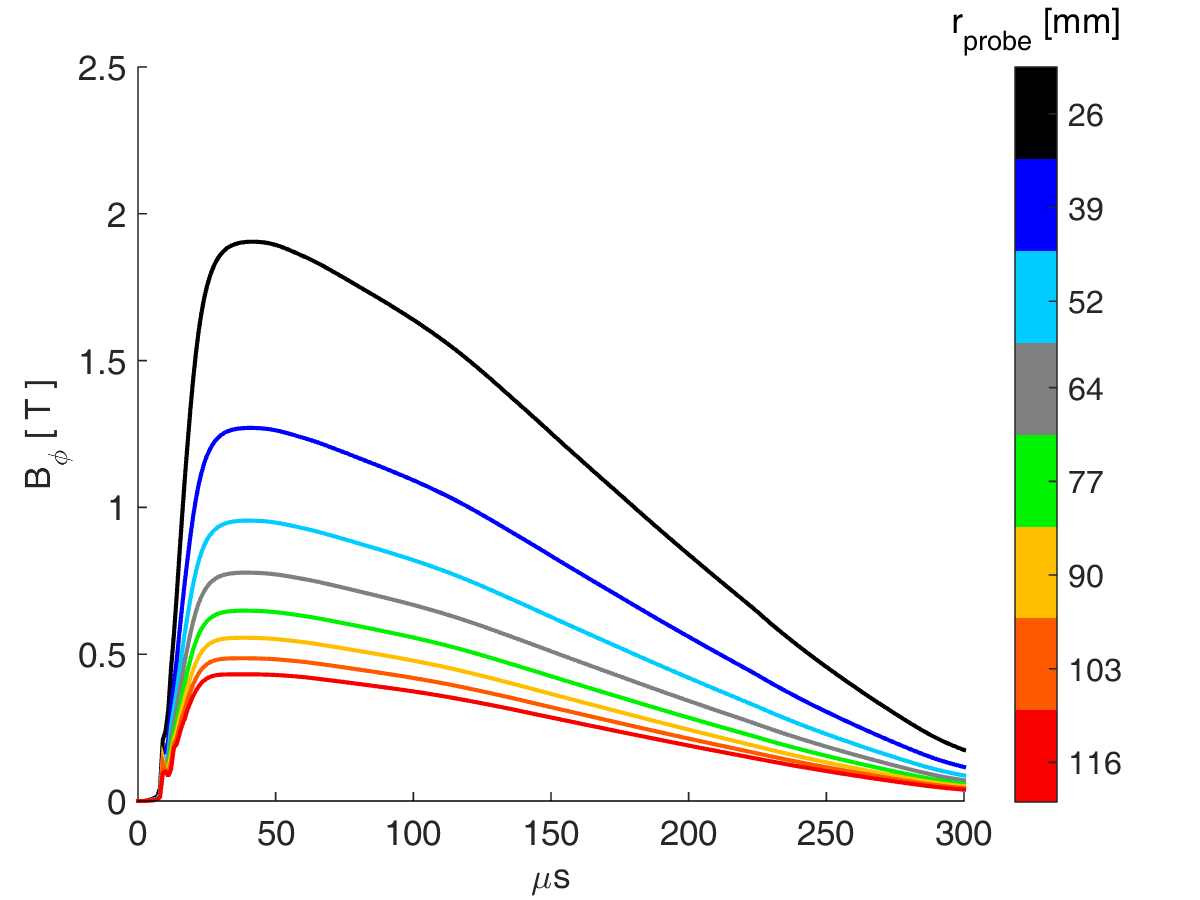}}

\caption{\label{fig:Bphi_exp_sim_comparison}Comparison of measured (a) and
simulated (b) $B_{\phi}$ (levitation - 11 coils). }
\end{figure}
Experimentally measured $B_{\phi}$ for shot 39650, taken in the eleven
coil configuration with $70\mbox{m}\Omega$ cable resistance, is shown
in figure \ref{fig:Bphi_exp_sim_comparison}(a).  For ease of comparison,
the toroidal averages of the toroidal field traces measured at the
two probes 180$^{o}$ apart at each of the eight radii, at which the
magnetic probes in inner flux conserver are located, are shown here.
With $70\mbox{m}\Omega$ cable resistance, the compressional instability
that was routinely observed on levitation-only shots with the $2.5\mbox{m}\Omega$
cable resistance is not observed, so the simulated toroidal field,
shown in figure \ref{fig:Bphi_exp_sim_comparison}(b), is a good match
to the experimental measurements. Shaft current is not used as an
input to the simulation, rather it arises naturally in simulations
as a consequence of induced wall-to-wall currents that act to conserve
total toroidal flux (see \cite{SIMpaper,thesis-1} for details).

\subsubsection{\label{subsec:Ion-temperature-and}Ion temperature and electron density
measurements}

\begin{figure}[H]
\subfloat[]{\raggedright{}\includegraphics[width=7cm,height=4.3cm]{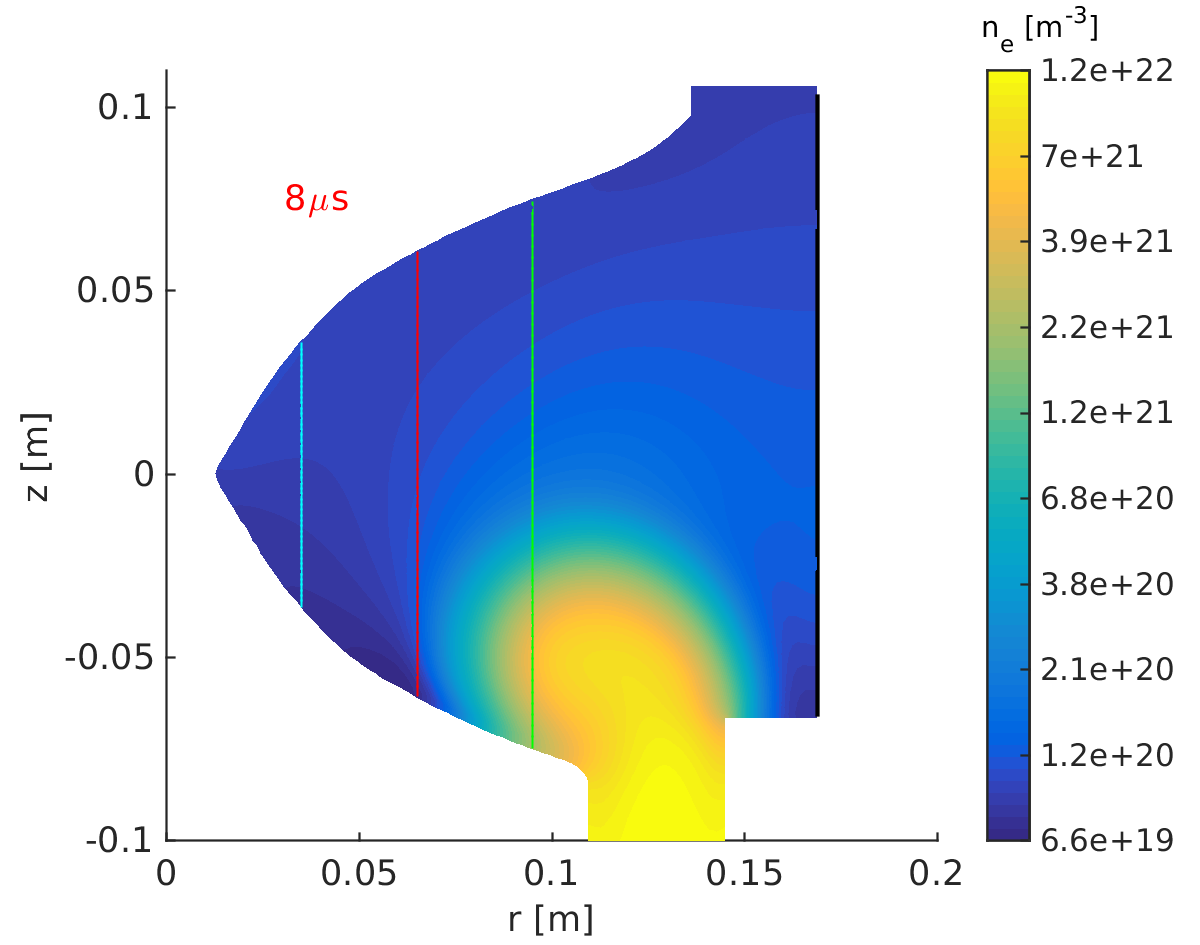}}\hfill{}\subfloat[]{\raggedleft{}\includegraphics[width=7cm,height=4.3cm]{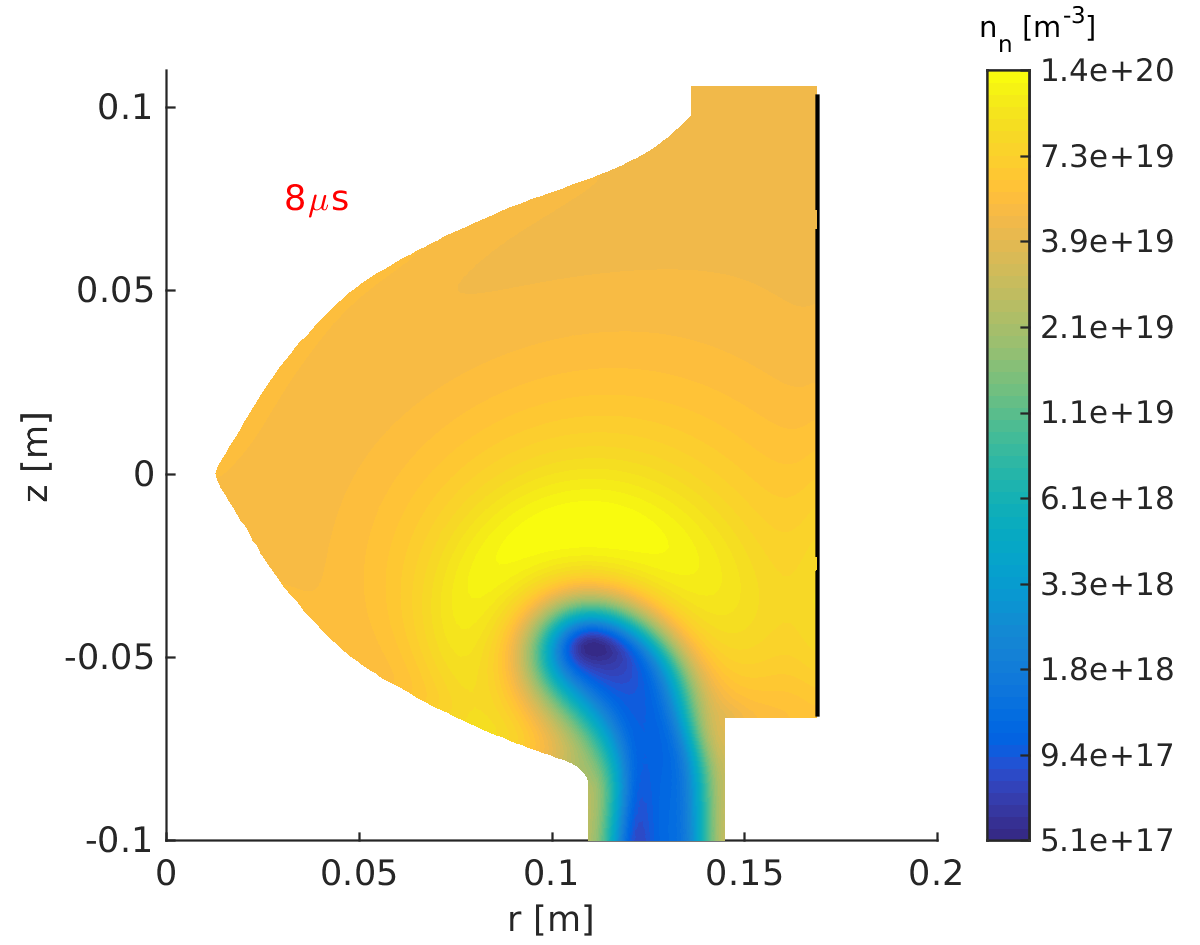}}

\subfloat[]{\raggedright{}\includegraphics[width=7cm,height=4.3cm]{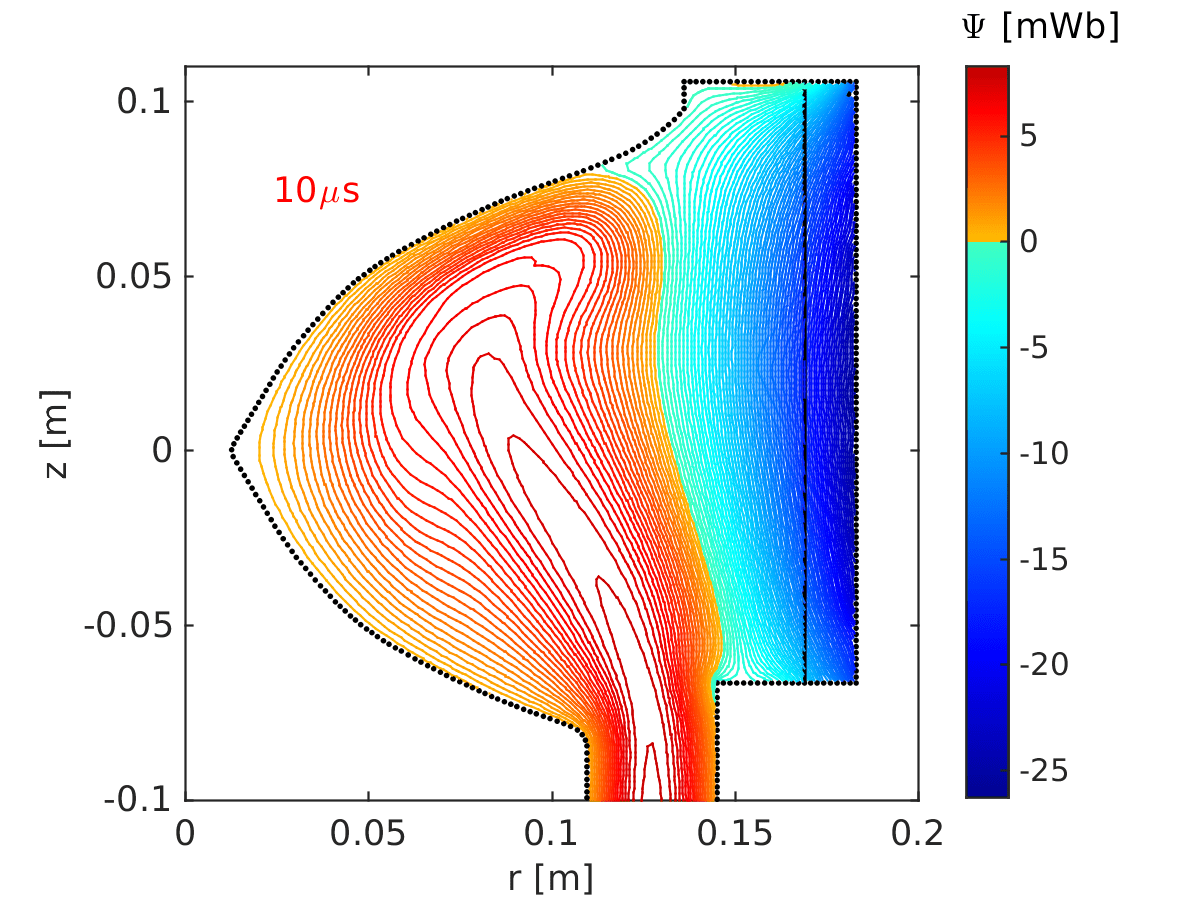}}\hfill{}\subfloat[]{\raggedleft{}\includegraphics[width=7cm,height=4.3cm]{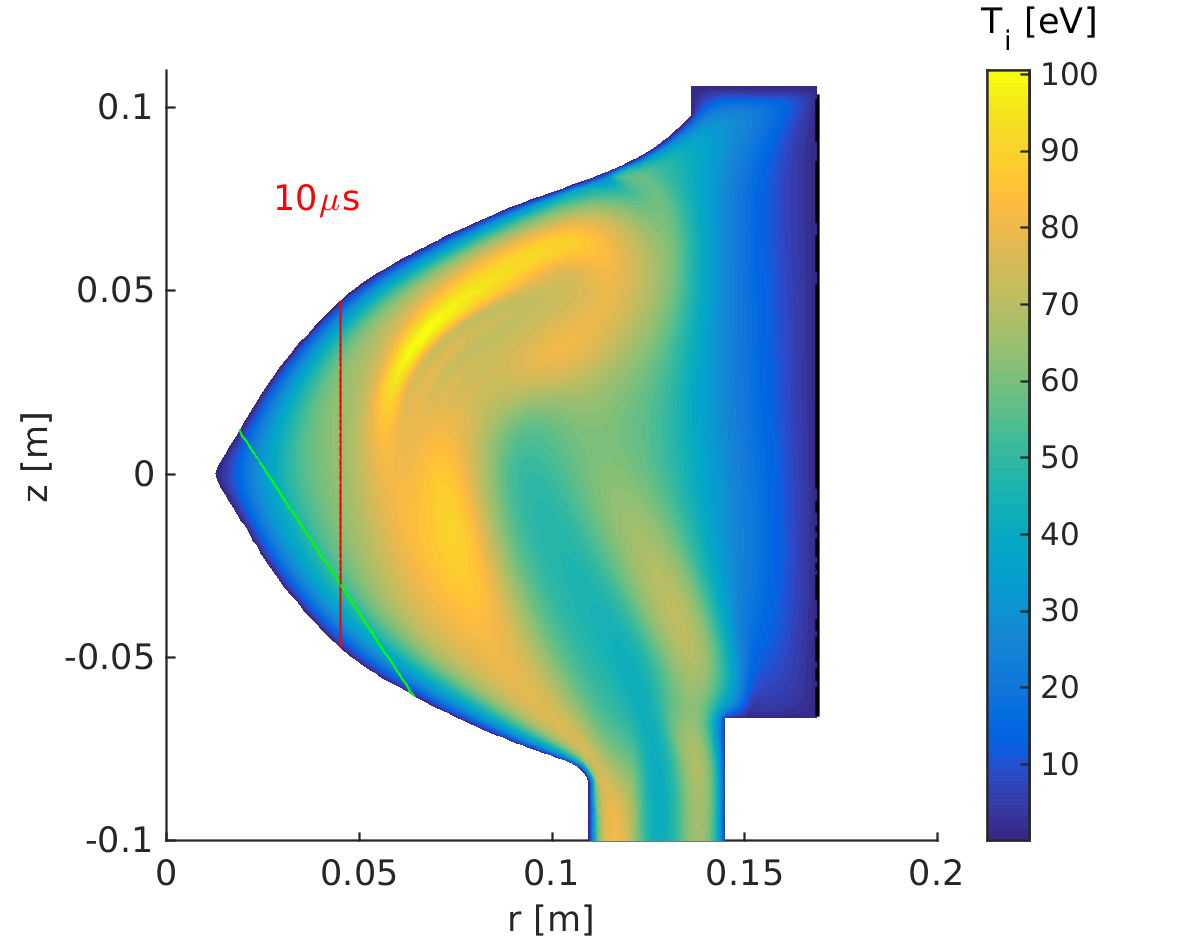}}

\subfloat[]{\raggedright{}\includegraphics[width=7cm,height=4.3cm]{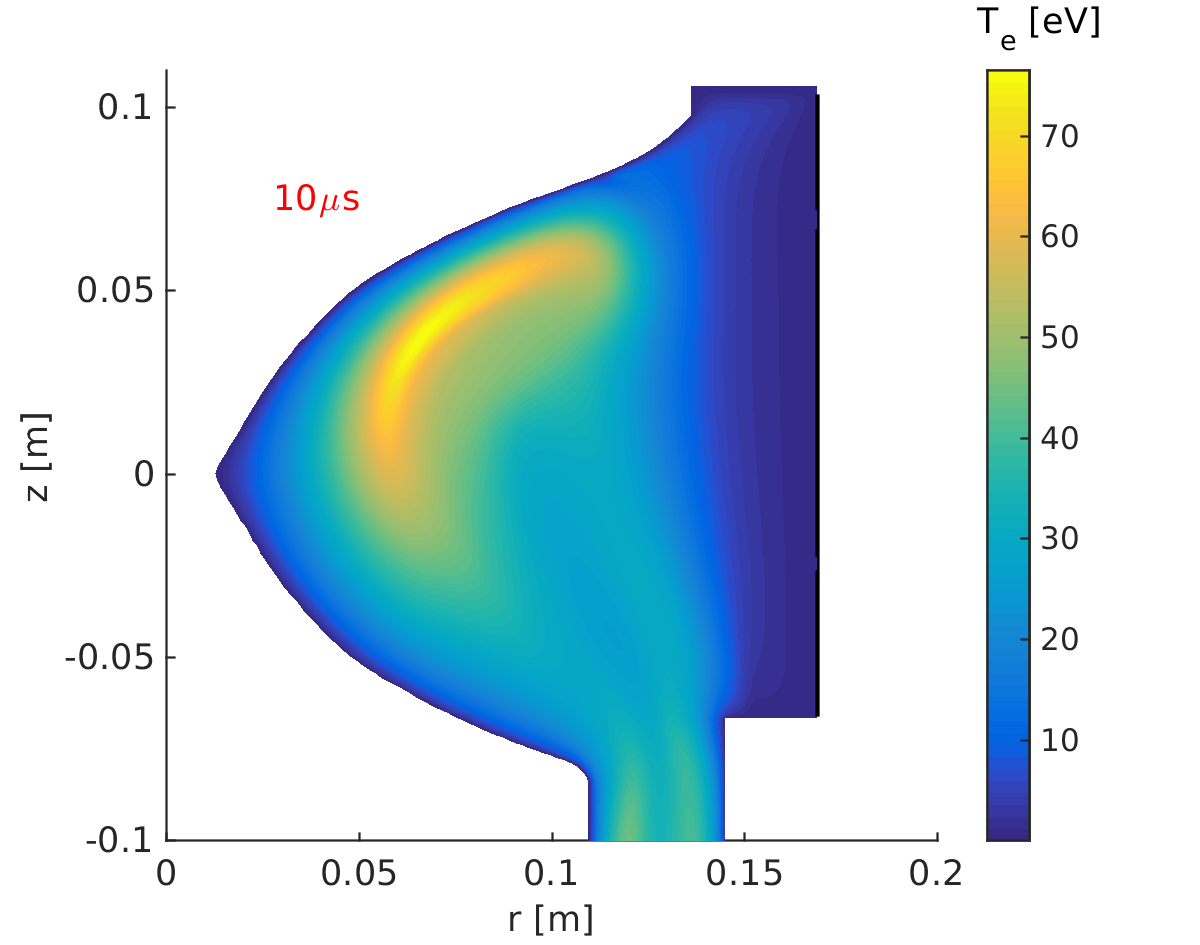}}\hfill{}\subfloat[]{\raggedleft{}\includegraphics[width=7cm,height=4.3cm]{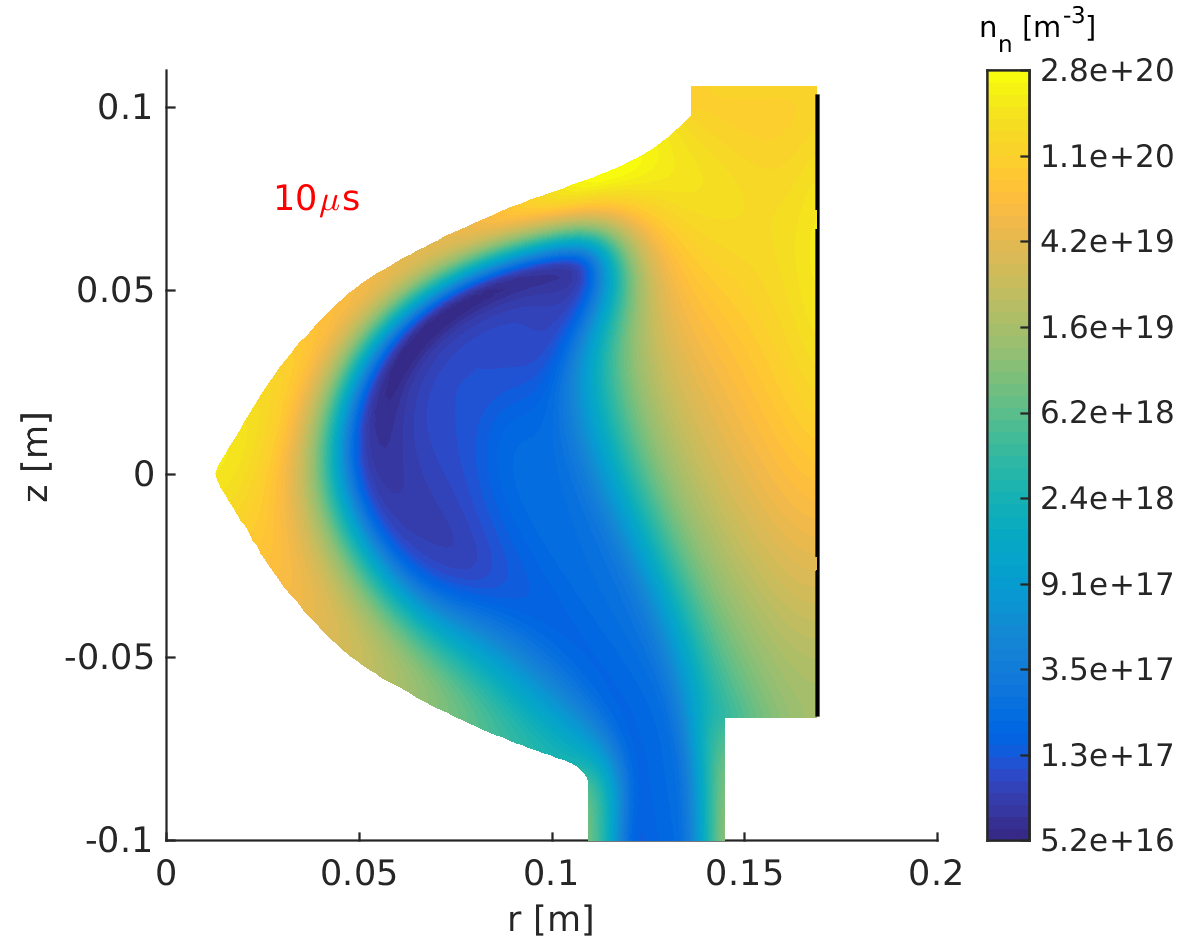}}

\subfloat[]{\raggedright{}\includegraphics[width=7cm,height=4.3cm]{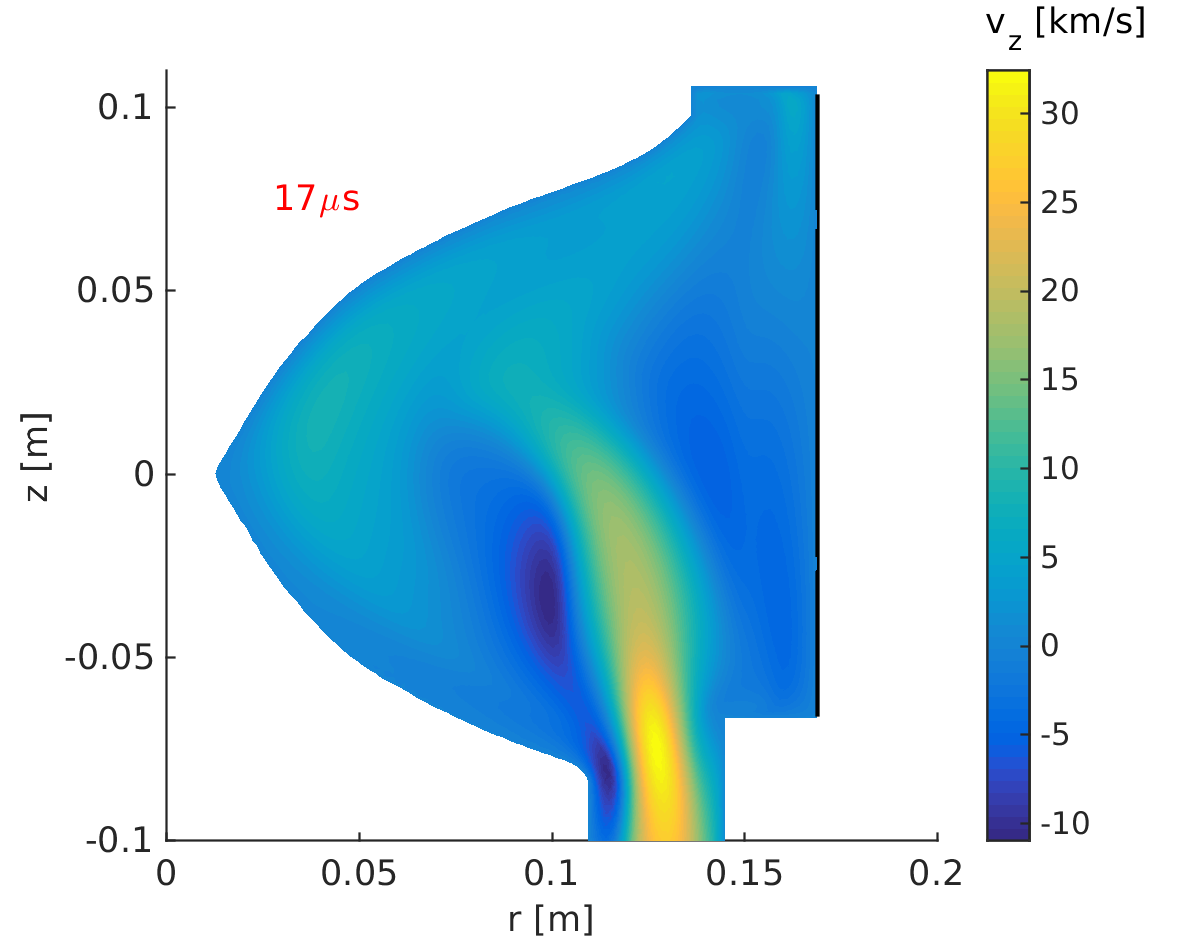}}\hfill{}\subfloat[]{\raggedleft{}\includegraphics[width=7cm,height=4.3cm]{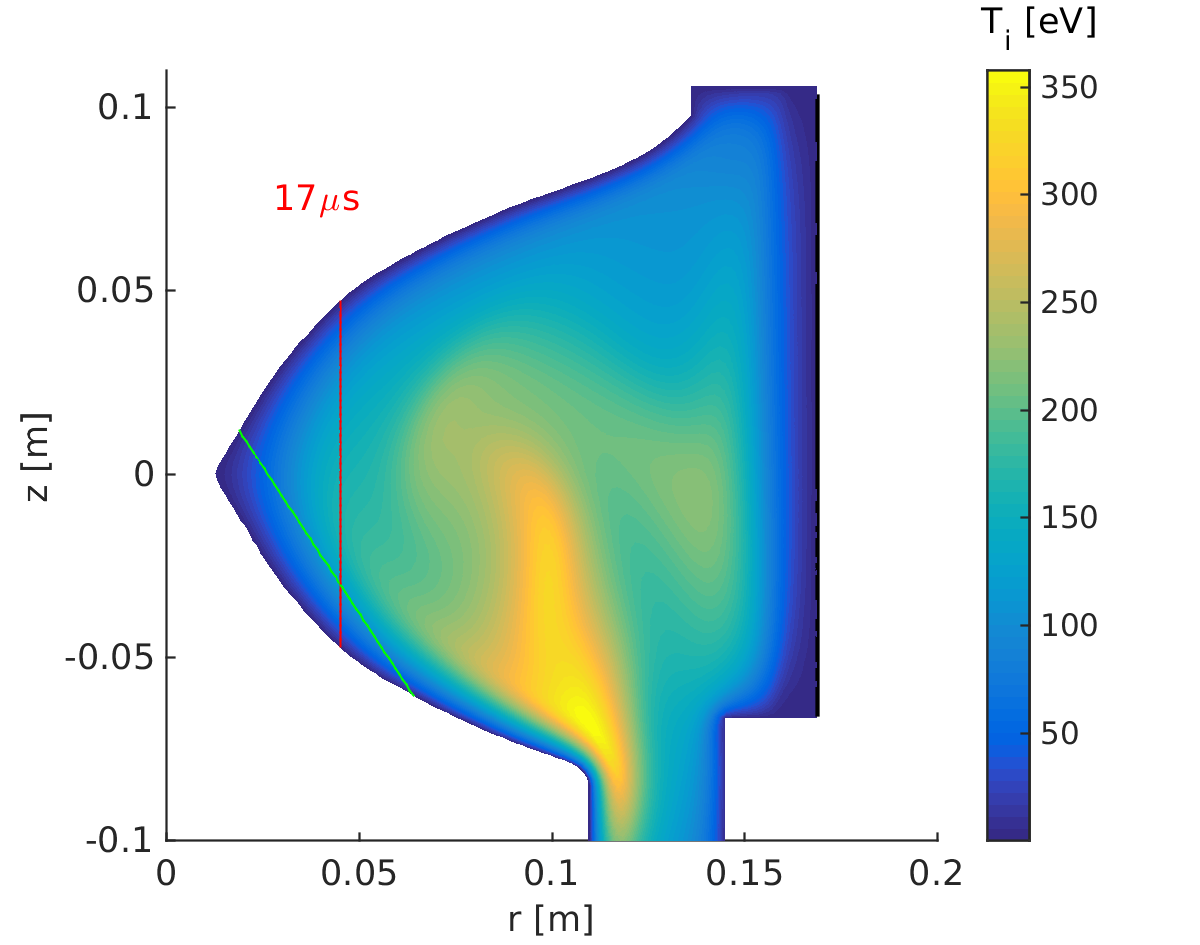}}

\caption{Evolution of various fields from MHD simulations\label{fig:MHDfields_T_n}:
$n_{e}$ at 8$\upmu$s (a), $n_{n}$ at 8$\upmu$s (b), $\psi$ at
10$\upmu$s (c), $T_{i}$ at 10$\upmu$s (d), $T_{e}$ at 10$\upmu$s
(e), $n_{n}$ at 10$\upmu$s (f), $v_{z}$ at 17$\upmu$s (g), and
$T_{i}$ at 17$\upmu$s (h). }
\end{figure}
\begin{figure}[H]
\subfloat[]{\raggedright{}\includegraphics[width=7cm,height=5cm]{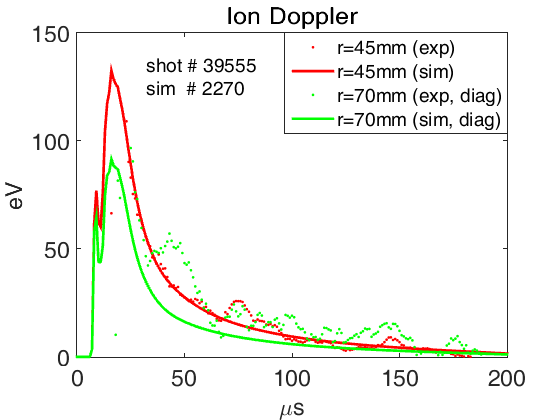}}\hfill{}\subfloat[]{\raggedleft{}\includegraphics[width=7cm,height=5cm]{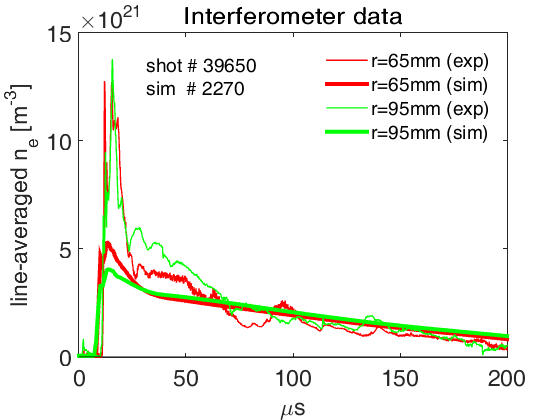}}

\caption{Comparison of measured and simulated $T_{i}$ (figure (a)) and $n_{e}$(b)\label{fig:comparisonT_n}}
\end{figure}
Profiles of electron density $n_{e}$ and neutral particle density
$n_{n}$ from an MHD simulation (\# 2270) \cite{SIMpaper,thesis-1,Neut_paper}
at early time ($t=8\upmu$s) during the formation of the CT are shown
in figures \ref{fig:MHDfields_T_n}(a) and (b).  At this time, plasma
has been advected up the Marshall gun and is entering (bubbling into)
the CT containment region, along with partially frozen-in open stuffing
field that is resistively pinned to the inner and outer electrodes
further down the gun at $z<-0.1$m, and is displacing the levitation
field in the containment region. The vertical blue, red and green
chords in figure \ref{fig:MHDfields_T_n}(a) represent the lines of
sight of the interferometer measurements ($cf.$ figure \ref{fig:Diagnostics-overview}(b)).
Initial plasma fluid density (note we use a single fluid MHD, but
partition the energy equation into ion and electron components) is
concentrated around the gas valves down the gun at $z=-0.43$m. The
initial neutral fluid density distribution, also centred around the
gas valves, extends further than the initial plasma fluid density
distribution, so that a front of neutral fluid precedes the plasma
as it is advected into the containment region. 

Figure \ref{fig:MHDfields_T_n}(c) shows contours of $\psi$ from
the same simulation at $t=10\upmu$s. Levitation field continues to
be displaced in the containment region. Figures \ref{fig:MHDfields_T_n}(d),
(e) and (f) show contours of $T_{i},\,T_{e}$ and $n_{n}$ at the
same time. Thermal diffusion is anisotropic in this simulation, with
constant coefficients $\chi_{\parallel}^{i}=5000,\,\chi_{\parallel}^{e}=16000,\,\chi_{\perp}^{i}=120,\,\chi_{\perp}^{e}=240\,[\mbox{m}^{2}/\mbox{s}]$.
In general, upper bounds on the parallel ion and electron thermal
diffusion coefficients are imposed by the minimum practical timestep
- an explicit timestepping scheme is used. The perpendicular coefficients
are chosen so as to match the decay rate of CT currents and fields
to experimentally indicated rates - radiative cooling due to the presence
of high $Z$ impurities in the plasma is not modelled directly - instead,
enhanced perpendicular thermal diffusion is used as a proxy for this
cooling. It can be seen how temperature is equilibriating along field
lines even at these early times, and, as a consequence of ionization,
neutral density is reduced at regions of high electron temperature.

As indicated in figures \ref{fig:MHDfields_T_n}(g) and (h), high
plasma-fluid velocities during the simulated formation process, largely
due to rapid upward advection, and due to jets associated with magnetic
reconnection of CT polodal field near the entrance to the containment
region, lead to significant levels of ion viscous heating. The chords
along which simulated ion Doppler measurements are taken, for comparison
with experimental measurements (see figure \ref{fig:Diagnostics-overview}(b)),
can be seen in figure \ref{fig:MHDfields_T_n}(h).

Figures \ref{fig:comparisonT_n}(a) and (b) indicate the agreement
between experimentally measured and simulated ion temperature and
electron density. These simulated diagnostics are the corresponding
line-averaged quantities along the chords indicated in figures \ref{fig:MHDfields_T_n}(h)
and (a) respectively. Note that the diagonal green coloured chord
indicated in figure \ref{fig:MHDfields_T_n}(h) has its lower point
at $r=70$mm. With reference to data presented in \cite{Kunze,Pittman},
a maximum error in the temperature measurement (He II line at 468.5nm)
due to density broadening has been evaluated as around 13eV for the
peak density of 1.2$\times10^{22}$m$^{-3}$, and the error falls
off in proportion to $n_{e}^{0.83}$. The interferometer looking along
the chord at $r=35$mm (figure \ref{fig:Diagnostics-overview}(b))
was not working for this shot, so this measurement has not been included
here.

\subsection{Main points - CT levitation\label{subsec:Main-points--}}

\begin{figure}[H]
\centering{}\includegraphics[scale=0.5]{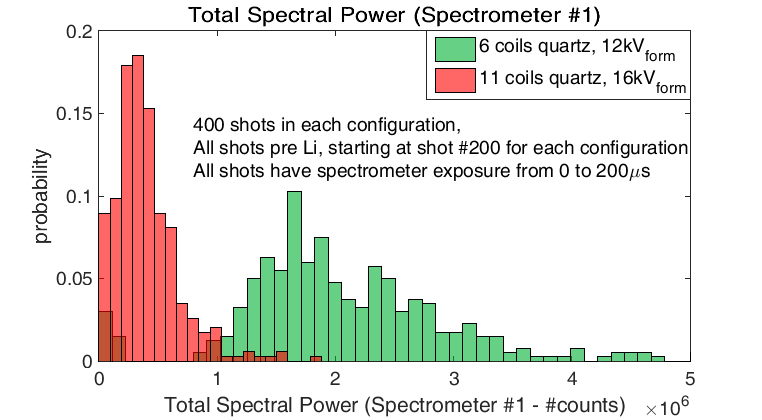}\caption{\label{fig:11coils_spectrometerdata}Spectrometer data, indicating
reduced plasma impurity concentration with the 11-coil configuration}
\end{figure}
Normalised histograms comparing total spectral power recorded with
spectrometer 1 (the spectrometer located at larger radius, depicted
in figure \ref{fig:Diagnostics-overview}(a)) for the 6-coil and 11-coil
configurations with the quartz wall, are shown in figure \ref{fig:11coils_spectrometerdata}.
 Data from 400 pre-lithium shots for each configuration, all with
spectrometer exposure from $0$ to $200\upmu$s, is included. The
validity of the data was verified by comparing the total spectral
power recorded with the measured intensity of plasma optical emission
at the same location as the spectrometer (the spectrometers shared
ports with optical feedthroughs), and finding a good correlation.
Even at increased formation voltage, total spectral power is around
four times lower with eleven coils. This is particularly unusual because
on a given configuration, it's expected that higher formation current
leads to increased ablation of electrode material and consequently
increased impurity levels and total spectral power. The 11-coil setup
reduced impurities and the associated energy losses due to line radiation
because it reduced the level of interaction between plasma and the
outer insulating wall during the bubble-in process, and the benefit
from the reduction of plasma-insulating wall interaction was more
significant than any impurity level augmentation caused by increased
plasma-electrode interaction.
\begin{figure}[H]
\begin{raggedright}
\subfloat[6-coil configuration]{\raggedleft{}\includegraphics[width=7cm,height=5cm]{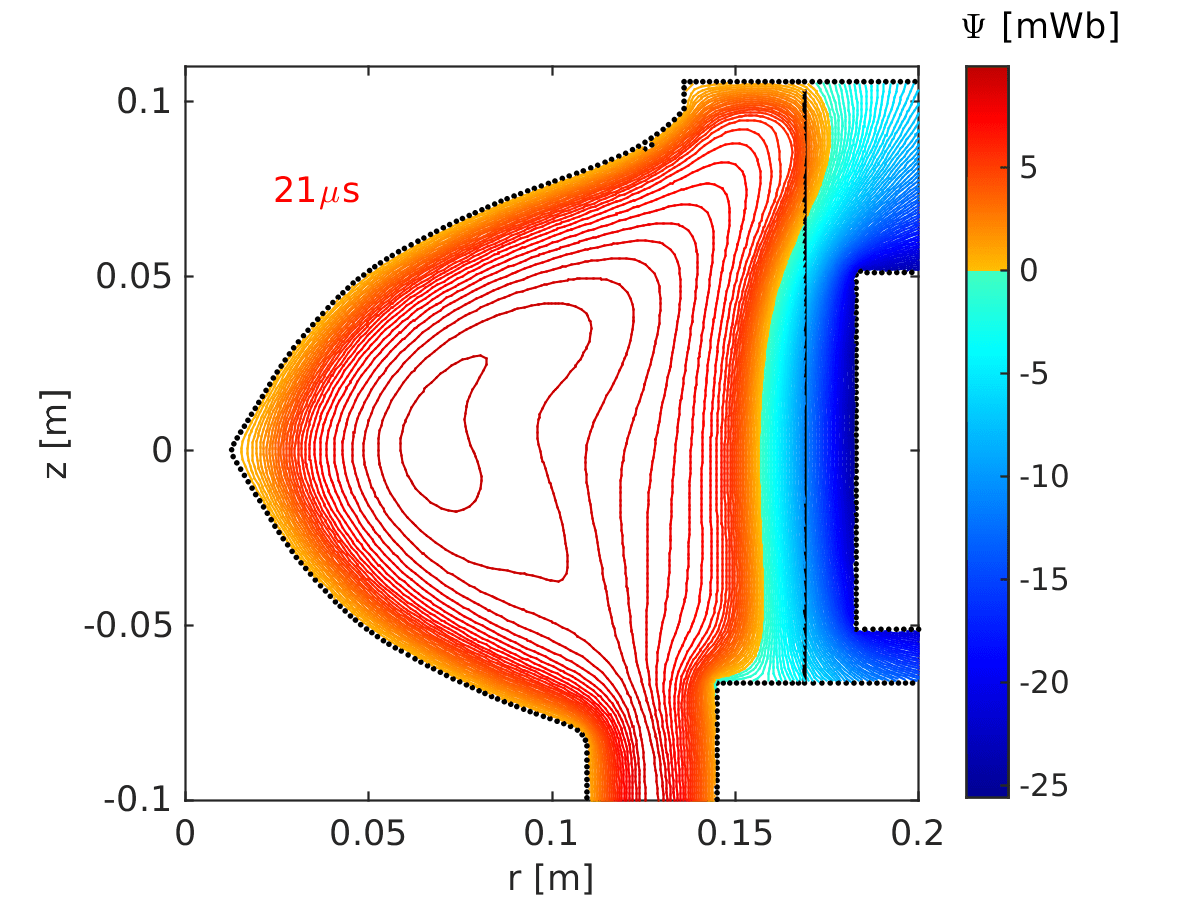}}\hfill{}\subfloat[11-coil configuration]{\raggedleft{}\includegraphics[width=7cm,height=5cm]{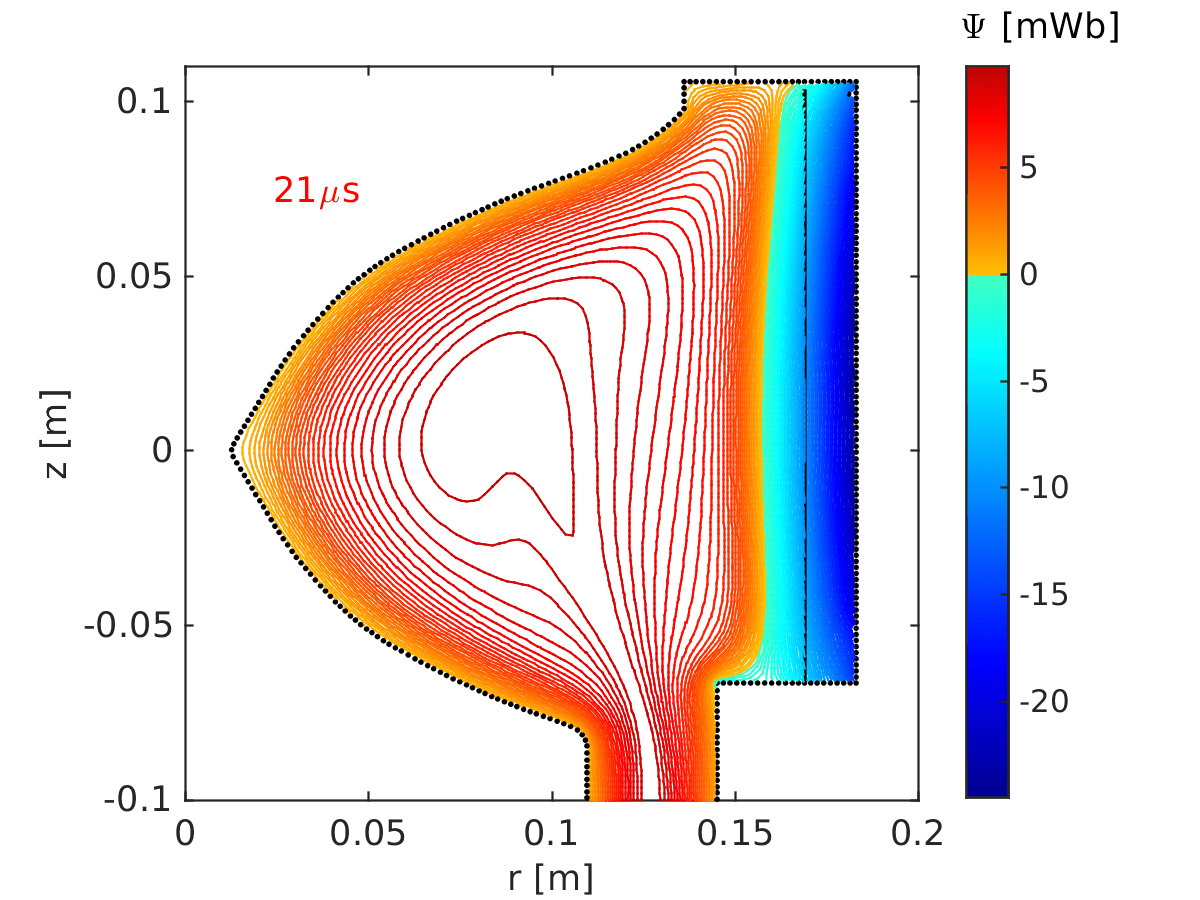}} 
\par\end{raggedright}
\centering{}\caption{\label{fig:plasmaWall}Simulated plasma-wall interaction. Note how
poloidal field penetrates the insulating wall during the bubble-in
process in the six coil configuration (a). }
\end{figure}
MHD simulations confirm the reduction of plasma-wall interaction with
the eleven coil configuration, as indicated in figure \ref{fig:plasmaWall}.
 In figure \ref{fig:plasmaWall}(a), the stack of six coils is partly
located in the blank rectangle on the right, centered around $z=0\mbox{cm}$,
and extends off further to the right (not shown). The region above,
below, and just to the left of the coil-stack represents the air around
the stack. The vertical black line at $r=17\mbox{cm}$ represents
the inner radius of the insulating wall, and the outer radius of the
insulating wall at $r=17.7\mbox{cm}$ is not indicated. In figure
\ref{fig:plasmaWall}(b), the stack of eleven coils extends all the
way from the top to the bottom of the insulating wall, and the inner
radius of the coil stack is the same as that for the six coil stack.
In both cases, as described in \cite{SIMpaper,thesis-1}, only $\psi$,
which determines the vacuum poloidal field, is evaluated in the insulating
region to the right of the inner radius of the insulating wall. The
solution for $\psi$ is coupled to the full MHD solution in the remainder
of the domain. To maintain toroidal flux conservation, boundary conditions
for $f$, which has a finite constant value in the insulating wall
and is zero outside the current-carrying aluminum bars depicted in
figure \ref{fig:Schematic-of-6-1}(a), are evaluated and applied to
the part of the boundary representing the inner radius of the insulating
wall. Both simulations have boundary conditions for $\psi_{main}$
and $\psi_{lev}$ from FEMM models, pertaining to $I_{main}=70$A,
and with the total levitation current such that $\psi_{lev}$ is approximately
the same for each configuration. Figure \ref{fig:plasmaWall}(a) indicates
how poloidal field penetrates the insulating wall during the bubble-in
process in the six coil configuration. In practice, ions streaming
along and gyro-rotating around the field lines would then sputter
insulating material into the plasma, leading to impurity radiation
and radiative cooling, with consequent increased resistivity and reduced
CT magnetic lifetimes. 

\begin{figure}[H]
\centering{}\includegraphics[height=4cm]{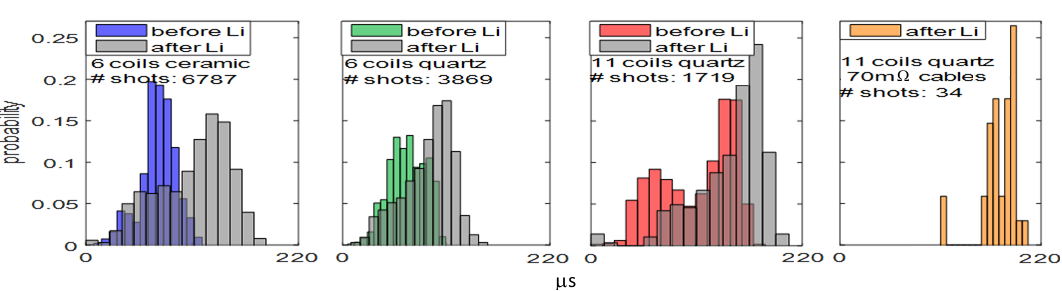}\caption{\label{fig:Effect-of-lithium}Effect of lithium gettering on levitated
CT lifetimes for different wall materials and coil configurations}
\end{figure}
As indicated in the normalised histograms in figure \ref{fig:Effect-of-lithium},
pre-lithium CT lifetimes were longer with the ceramic wall despite
the smaller volume. Lithium gettering was very effective on the ceramic
wall ($\sim70\%$ lifetime increase), but not so effective on quartz
($\sim30\%$ lifetime increase). Lifetime increased significantly
with the 11-coil configuration. The \textquotedbl double-Gaussian\textquotedbl{}
shape of the (before Li) distribution for eleven coils may be due
to the $\sim35$\% of shots taken in that configuration in suboptimal
machine-parameter space ($i.e.,$ values of $V_{form},\,V_{lev},\:I_{main}$,
and $t_{gas}$) that were rapidly explored in the last days of the
experiment in new configurations such as without levitation inductors,
with additional crowbarred sustain current ($\sim80\mbox{kA}$ addition
formation current with a decay time of $\sim1\mbox{ms}$), and with
passive or open-circuited levitation/compression coils. Note that
of the $>10000$ shots from which data is taken for this levitated
CT lifetime comparison, only 34 shots in the best of the configurations
tested - eleven coils with $70\mbox{m}\Omega$ cables - are shown
because the 11-coil configuration was explored rapidly in the days
before the experiment was decommissioned. The repeatability of good
shots was significantly improved in that configuration.

\begin{figure}[H]
\begin{raggedright}
\subfloat[six coils, ceramic wall]{\centering{}\includegraphics[width=8cm,height=5cm]{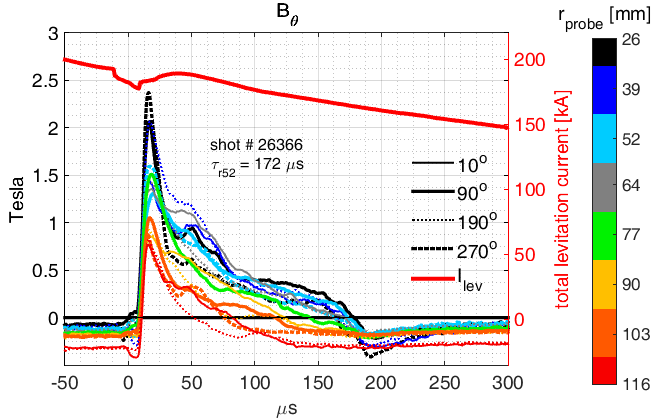}}\hfill{}\subfloat[six coils, quartz wall]{\centering{}\includegraphics[width=8cm,height=5cm]{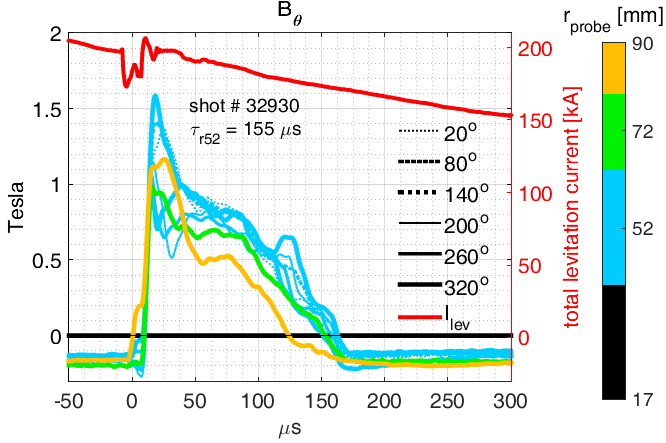}}
\par\end{raggedright}
\begin{raggedright}
\subfloat[25 turn coil, quartz wall]{\centering{}\includegraphics[width=8cm,height=5cm]{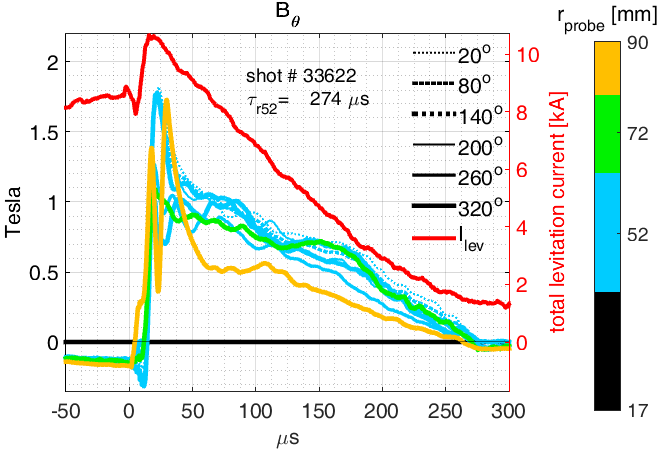}}\hfill{}\subfloat[11 coils, quartz wall]{\centering{}\includegraphics[width=8cm,height=5cm]{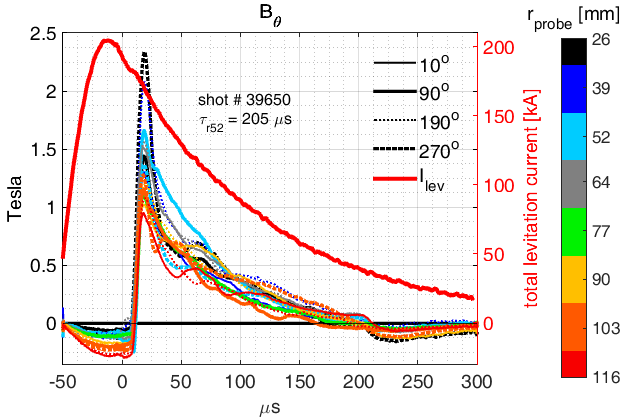}}
\par\end{raggedright}
\begin{raggedright}
\subfloat[no coils, stainless steel wall]{\centering{}\includegraphics[width=8cm,height=5cm]{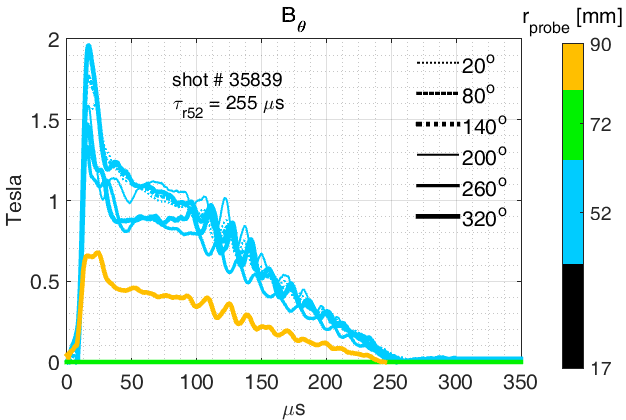}}\hfill{}\subfloat[no coils, aluminum wall]{\centering{}\includegraphics[width=8cm,height=5cm]{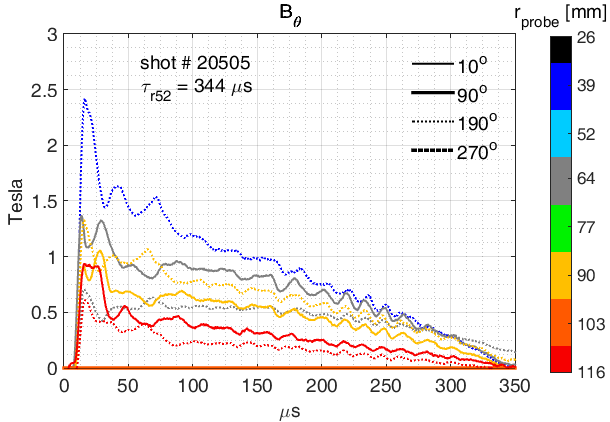}}
\par\end{raggedright}
\centering{}\caption{\label{fig:-for-four4configs}Measured $B_{\theta}$ for six configurations}
\end{figure}
Poloidal field traces for the six principal configurations tested
are shown in figure \ref{fig:-for-four4configs}.  Note that the magnetic
probes are located at radial and azimuthal coordinates different to
those listed in table \ref{tab: coordinates-ofBprobes} for the configurations
relevant to figures (b), (c) and (e). Note that not all of the magnetic
probes were functioning in some shots, for example the signals relevant
to the probes at $r=17$mm and $r=77$mm have been zeroed out in figure
\ref{fig:-for-four4configs}(e). Comparing figures \ref{fig:-for-four4configs}(a)
and (b), and noting, as outlined in section \ref{subsec::Levitation coil configurations},
that a 50\% increase in CT lifetime was expected with the switch to
the larger internal radius insulating tube, it can be seen how quartz
was significantly worse than ceramic as a plasma-facing material.
For these two shots, $|t_{lev}|$ was $300\upmu$s - as mentioned
in section \ref{subsec::Levitation coil configurations}, the strategy
of allowing the levitation field more time to soak into the steel
above and below the insulating wall led to slightly increased CT lifetimes
on the 6-coil configurations.

With CT lifetimes of up to $274\upmu$s, the longest-lived levitated
CTs were produced with the 25 turn coil configuration (figure \ref{fig:-for-four4configs}(c)).
The eleven coil configuration, with a field profile similar to that
of the 25-turn coil setup, also enabled the production of relatively
high-flux CTs with correspondingly increased lifetimes (figure \ref{fig:-for-four4configs}(d)).
In general, the recurrence rate of good shots in the 25-turn coil
configuration was poor compared with that in the 11-coil configuration.
However, it remains unclear why the longest-lived CTs produced with
the 25-turn configuration outlived those produced in the 11-coil configuration.
The most likely explanation  is connected to the fact that the 25-turn
coil extended farther above and below the insulating wall than the
stack of eleven coils. It may be that the increased levitation field,
relative to that for the 11-coil configuration, at the top and bottom
of the insulating wall, played a key role. At low formation settings,
without addition levitation circuit series resistance, levitated CT
lifetimes in the 25-turn configuration were comparable to those in
both the 11-coil and 6-coil configurations. It is clear that the feature
shared by the 25-turn and 11-coil configurations, of closing the gaps
that remained above and below the coil stack in the 6-coil configurations,
was responsible for enabling the formation of high flux CTs with correspondingly
increased lifetimes, and that the unconfirmed mechanism that enabled
(occasional) even better performance in the 25-turn configuration
was also effective only at high formation settings. Another possible
explanation concerns the ratio of the coil inductance to the levitation
circuit holding inductance, which was increased from $L_{coil}/L_{main}=600\mbox{nH}/6\upmu\mbox{H}=0.1$
for the 11-coil configuration to $L_{coil}/L{}_{main}=116\upmu\mbox{H}/6\upmu\mbox{H}\,\sim\,20$
for the 25-turn coil. When conductive plasma enters the pot (confinement
region) it reduces the inductance of the part of the levitation circuit
that includes the levitation/compression coil and the material that
the coil encompasses. The levitation current increases when the inductance
is reduced as plasma enters the pot. If the percentage rise of the
levitation current is increased, by increasing the ratio $L_{coil}/L{}_{main}$,
it means that levitation current prior to plasma bubble-in can be
minimised. This reduction in $I_{lev}$ reduces the likelihood that
the levitation field will be strong enough to partially block plasma
entry to the pot, while still allowing the field that is present,
when the plasma does enter, to be strong enough to levitate the plasma
away from the insulating wall. Comparing figures \ref{fig:-for-four4configs}(c)
and (d), it can be seen how the levitation current increases significantly
at bubble-in for the 25-turn coil only. FEMM models indicate that
the levitation fluxes found to be optimal at moderately high formation
settings for the 25-turn and 11-coil configurations were approximately
the same prior to plasma entry to the containment region. It may be
that the increased levitation flux at CT entry in the 25-turn configuration
was more efficient at keeping plasma off the wall. The optimal settings
for $|t_{lev}|$ in the various configurations were limited by $t_{rise}$,
the rise time of the levitation current for the particular configuration.
While the strategy of increasing $|t_{lev}|$ to allow the levitation
field more time to soak into the steel above and below the insulating
wall led to slightly increased CT lifetimes on the 6-coil configurations,
it was found that $|t_{lev}|$ should be reduced to as low a value
as possible on the 25-turn coil and 11-coil configurations for best
performance. Reducing $|t_{lev}|$ reduces the likelihood that the
levitation field will impede, through the line tying effect, plasma
entry to the containment region at formation. The benefit of slightly
reducing plasma-wall interaction by \emph{increasing} $|t_{lev}|$,
and the line-tying effect, outweighed the detrimental effect of pot-entry
blocking in the 6-coil configuration only. With the high inductance
25-turn coil, optimal $|t_{lev}|$ was equal to $t_{rise}\sim150\upmu$s,
while for the 11-coil configuration, optimal $|t_{lev}|$ was set
to $t_{rise}\sim50\upmu$s. It may be that allowing the level of levitation
flux that was present in the containment region upon plasma entry
in the 25-turn configuration to soak into the steel above and below
the wall, even for 50$\upmu$s in the 11-coil configuration, degraded
performance by impeding plasma entry to the containment region. The
requirement for increased $|t_{lev}|$, and consequent pot-entry blocking
may have been the cause of the poor repeatability of good shots in
the 25-turn configuration. 

Some tests were done to see the effect of allowing levitation field
to interact with a CT that was supported with a conducting wall. This
investigation was largely driven by concern over the absence, as discussed
in section \ref{subsec::Levitation coil configurations}, of the $n=2$
fluctuations, commonly observed with $\mbox{MRT}$ injector-produced
CTs, on levitated CT $B_{\theta}$ signals. Compared with the aluminum
flux conserver (figure \ref{fig:-for-four4configs}(f)), the resistivity
of the stainless steel flux conserver (figure \ref{fig:-for-four4configs}(e))
is increased by a factor of ten, leading to more magnetic field soakage,
and consequent impurity sputtering, radiative cooling, and reduced
CT lifetimes. $n=2$ magnetic fluctuations are apparent in both configurations
with metal walls, and remained even when a levitation field was allowed
to soak through the resistive stainless steel wall, but disappeared
when the levitation field was increased enough to push the CT significantly
off the stainless steel wall. It has been confirmed that $n=2$ fluctuations
are a sign of internal MHD activity associated with increased electron
temperature, as discussed in section \ref{subsec::Levitation coil configurations}.
It was thought that this correlation, and the absence of the fluctuations
on levitated CTs, was a sign that levitated CTs were colder than flux-conserved
CTs, and the problems encountered with plasma wall interaction in
the levitation configurations made that scenario more likely. However,
the CTs produced with the 25-turn configuration are longer-lived (by
up to 10\%) than, and may therefore be assumed to be hotter than the
CTs produced in the configuration with the stainless steel flux conserver.
It may be that the levitation field acts to damp out helically propagating
magnetic fluctuations at the outboard CT edge and that internal MHD
activity is relatively unchanged. The $n=1$ magnetic fluctuations
(not shown here), observed when 80kA additional crowbarred shaft current
was applied to the machine in the eleven coil configuration, confirmed
coherent toroidal CT rotation, and may have been a result of more
vigorous MHD activity that remained apparent despite damping. \\

\newpage{}

\section{\label{sec:Magnetic-compression}Magnetic compression }

\subsection{Overview of magnetic compression results\label{sec:Overview_compresults}\protect \\
}

With a compression coil current rise time of around $20\upmu$s, peak
CT compression is achieved at $t\sim t_{comp}+20\upmu$s. If the CT
remains stable during compression, it expands to its pre-compression
state (apart from resistive flux losses and thermal losses) between
$t\sim t_{comp}+20\upmu$s and $t\sim t_{comp}+40\upmu$s, when the
compression current falls to zero and changes direction. At this time,
the CT poloidal field reconnects with the compression field, and a
new CT with polarity opposite to that of the previous CT is induced
in the containment region, compressed, and then allowed to expand.
The process repeats itself at each change in polarity of the compression
current until either the plasma loses too much heat, or the compression
current is sufficiently damped. MHD simulations \cite{thesis-1,SIMpaper}
model this effect while closely reproducing experimental measurements
for $B_{\theta}$, line-averaged $n_{e}$, and $T_{i}$ (from the
ion-Doppler diagnostic), and Xray-phosphor imaging indicates the compressional
heating of up to three distinct plasmoids on many compression shots.
\begin{figure}[H]
\centering{}\includegraphics[scale=0.6]{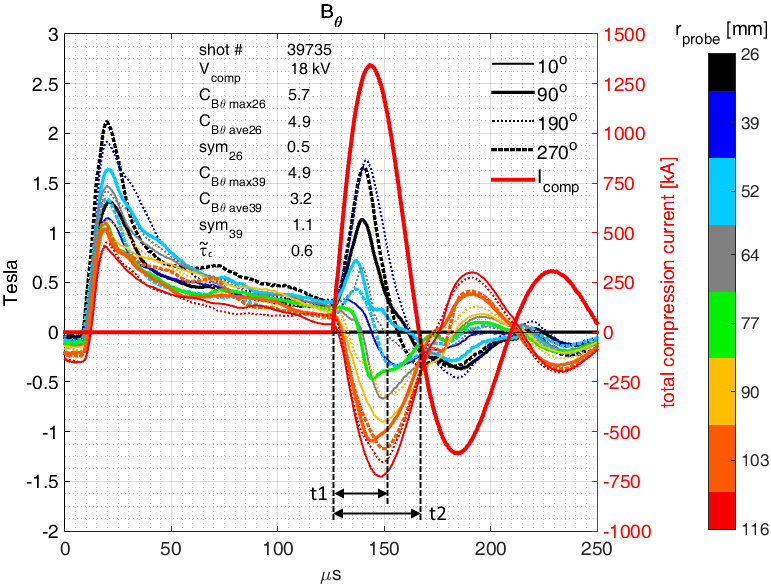}\caption{\label{fig:39735BpMagnetic-compression-shot}$B_{\theta}$ for shot
 39735 (11-coil configuration)}
\end{figure}
 Figure \ref{fig:39735BpMagnetic-compression-shot} shows $B_{\theta}$
traces for shot  39735, with $V_{comp}=18$kV (close to the maximum
setting), and $t_{comp}=130\upmu$s. The total peak compression current
(right axis), divided between the 11 coils, was $\sim1.3\mbox{MA}$,
and the total levitation current, on which the compression current
is superimposed, had a peak value of around $200$kA, and is not shown
here. In this shot, the CT is compressed inwards beyond the magnetic
probes at $r=77\mbox{mm}$, so, for example at $t\sim140\upmu$s,
$B_{\theta}$ recorded at the probes at $r\geq77\mbox{mm}$ is a measurement
of the external field ($i.e.,$ the combined compression and levitation
field), while the CT poloidal field is measured at $r<77\mbox{mm}$.
In this shot, the CT is being compressed more at $\phi=190^{o}$ than
at $\phi=10^{o}$, so that, for example, between $t=t_{comp}$ and
$t\sim150\upmu$s, the probes at $r=64\mbox{mm}$ measure CT field
at $190^{o}$ and external field at $10^{o}$. Some of the compression
parameters calculated for the shot are displayed on the graph. $C_{B\theta max}(r)$
and $C_{B\theta ave}(r)$ are the maximum and average of the two poloidal
magnetic compression ratios, obtained at the two probes located at
radius $r$ mm, $180^{o}$ apart toroidally, $e.g.,$ $C_{B\theta max26}=max(C_{B\theta_{90^{\circ}r26}},\,C_{B\theta_{270^{\circ}r26}})$,
where, for example, $C_{B\theta_{270^{\circ}r26}}=B_{\theta CTpeak}/B_{\theta CTpre}$,
where $B_{\theta CTpeak}$ and $B_{\theta CTpre}$ are the values
of $B_{\theta},$ measured with the probe at $r=26\mbox{\mbox{mm}},\,\phi=270^{\circ}$,
at the peak of compression and just before compression respectively.
The values of $C_{B\theta max26}=5.7\,\mbox{and }C{}_{B\theta ave26}=4.9$
obtained for this shot are particularly high, partly because the CT
was compressed late, with high $V_{comp},$ when most of its poloidal
flux had resistively decayed away.

Parameters $sym_{r}$ give an indication of the toroidal asymmetry
of the magnetic compression at the probes located at radius $r\,\mbox{mm}$.
Shots with $sym_{r}$ close to zero have toroidally symmetric compression
at radius $r\,\mbox{mm}$. With parameters $sym_{26}=0.5$, and $sym_{39}=1.1$,
shot  39735 had quite asymmetric compression at $r=26$mm, and very
asymmetric compression at $r=39$mm.

The parameter $\widetilde{\tau}_{c}$ indicates the level of compressional
flux conservation, and is calculated as $\widetilde{\tau}_{c}=t1/t2$,
where $t1$ and $t2$ are indicated in figure \ref{fig:39735BpMagnetic-compression-shot}.
$t2\sim50\upmu$s is the half-period of the compression current, and
$t1$ is the time from $t_{comp}$ to the average of the times when
$B_{\theta}$ at the two $r=26\mbox{mm}$ probes fall to their pre-compression
values (at $t=t_{comp}$). If the CT doesn't lose flux during compression,
the measured $B_{\theta}$ at the inner probes rises and falls approximately
in proportion to the compression current, and $t1\sim t2$. Shots
for which most of the CT's poloidal flux is conserved over compression
are characterised by $\widetilde{\tau}_{c}\sim1$. As shown in \cite{thesis-1},
MHD simulations support the idea that loss of CT poloidal flux at
compression leads to the collapse in poloidal field that is characterised
by having parameter $\widetilde{\tau}_{c}$ less than one. This characterisation
method assumes that the CT is not being compressed to a radius less
than $26\mbox{\mbox{mm}}$. If that did happen, the indication of
$B_{\theta}$ increase at $26\mbox{\mbox{mm}}$ should disappear early
($\widetilde{\tau}_{c}\ll1$), and then there would be no data whatsoever
available to assess the compression beyond $26\mbox{\mbox{mm}}$.
If the CT is being compressed beyond $26\mbox{\mbox{mm}}$, and stays
stable, it may expand back to $r>26\mbox{\mbox{mm}}$ after the peak
in compression field, but there are no examples of that occurrence
in the data. Shot 39735 has parameter $\widetilde{\tau}_{c}=0.6$,
which classifies it as a shot that lost a significant proportion of
its flux during compression.\\
\\
\\
\begin{figure}[H]
\centering{}\subfloat[]{\includegraphics[width=8.5cm,height=5cm]{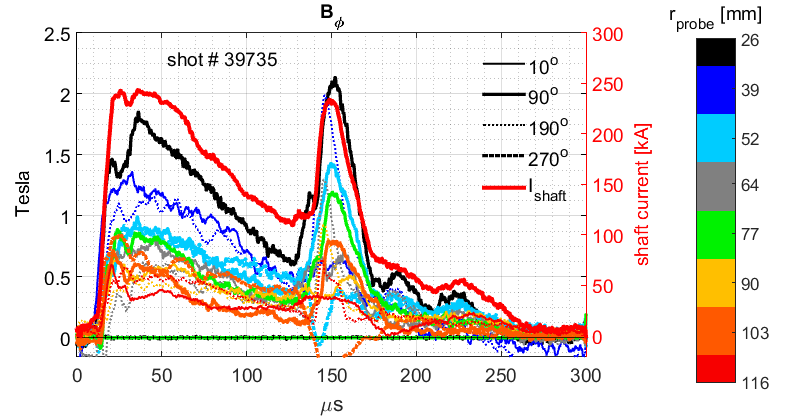}}\hfill{}\subfloat[]{\includegraphics[width=6.7cm,height=5cm]{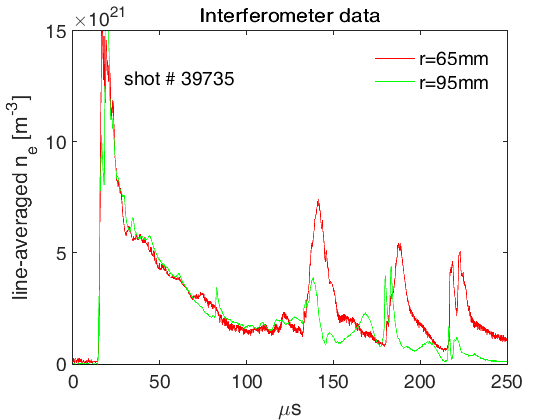}
\raggedleft{}}\caption{\label{fig:Bpand--traces}$B_{\phi}$ (figure (a)) and $n_{e}$ (b)
traces for shot  39735 (11-coil configuration)}
\end{figure}
Figure \ref{fig:Bpand--traces} shows measured $B_{\phi}$ and $n_{e}$
for shot  39735. As discussed earlier in section \ref{subsec:Magnetic-field-mesurements},
$B_{\phi}$ rises at compression as shaft current increases when it
is able to divert from the aluminum bars outside the insulating wall
to a lower inductance path through ambient plasma outside the CT.
For the 1st, 2nd and 3rd compressions, this is particularly evident
from the rise of the $r=26\mbox{mm}$ probe signal. An obvious exception
is during the 1st compression at $\sim150\upmu$s, when the toroidal
field at $\phi=270^{\circ}$ drops off - this is an indication of
the compressional instability that is discussed later in section \ref{subsec:Compressional-instability}.
The measured electron densities shown in figure \ref{fig:Bpand--traces}(b)
are line-averaged quantities obtained with He-Ne laser interferometers
looking down the vertical chords at $r=65\mbox{\mbox{mm}}$ and $r=95\mbox{\mbox{mm}}$
that are indicated in figure \ref{fig:Diagnostics-overview}(b). The
three distinct density peaks correspond to the three CT compressions.
From the time difference between the peaks at compression of the two
$n_{e}$ signals, the electron density front at the main compression
is found to move inwards at $\sim10\mbox{km/s}$.\\
\\
\\
\\
\begin{figure}[H]
\subfloat[shot  28426 (6-coil configuration)]{\raggedright{}\includegraphics[width=8cm,height=5cm]{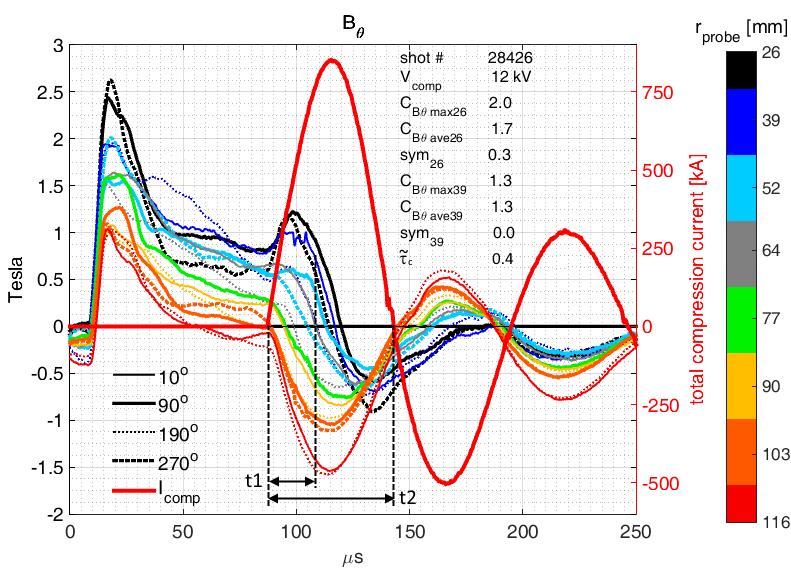}}\hfill{}\subfloat[shot  39475 (11-coil configuration)]{\raggedleft{}\includegraphics[width=8cm,height=5cm]{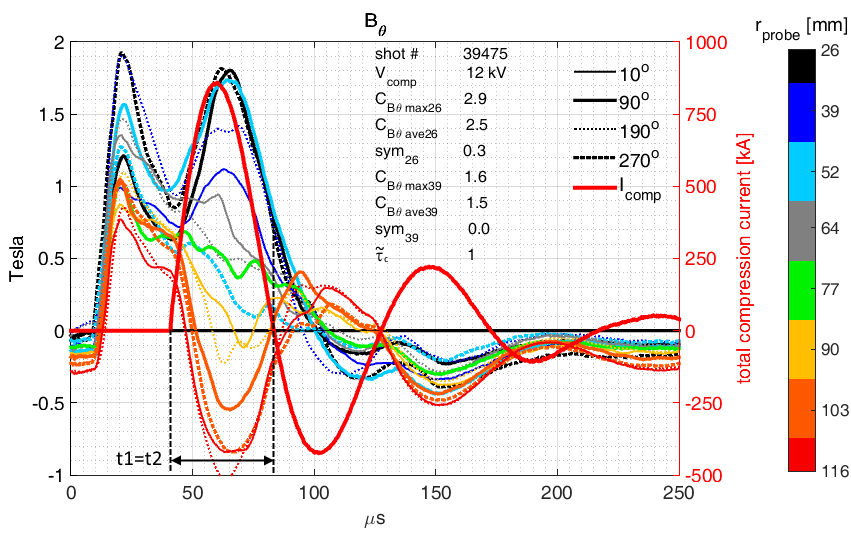}}

\caption{\label{fig:Bp_28426cf39475}$B_{\theta}$ traces - comparison of compressional
flux conservation }
\end{figure}
Poloidal field measured for two shots at moderate $V_{comp}=12$kV,
with peak total compression currents of around 800kA, is shown in
figure \ref{fig:Bp_28426cf39475}.  With parameter $\widetilde{\tau}_{c}=0.4$,
shot 28426 went unstable and lost most of its poloidal flux early
during compression. This was typical of compression shots in the 6-coil
configuration. In contrast, shot 39475 held onto its flux over the
main compression cycle, as was more usual in the 11-coil configuration.
As a consequence the magnetic compression ratios, indicated on these
graphs, are considerably higher in shot 39475, and in shots taken
in the 11-coil configuration in general. Both shots here, with low
values of $sym_{r}$, exhibited quite symmetric compression.\\
\\
\begin{figure}[H]
\subfloat[$B_{\theta}$, shot  39475]{\raggedright{}\includegraphics[width=7cm,height=5cm]{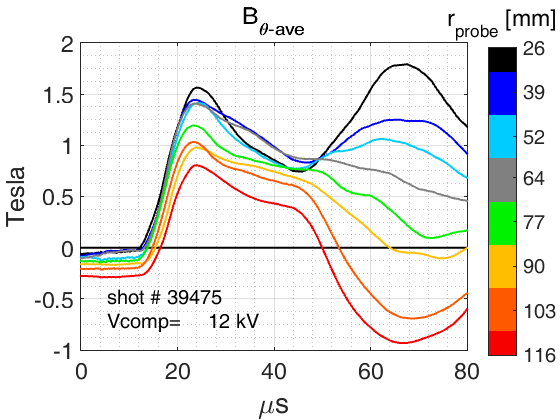}}\hfill{}\subfloat[$B_{\theta}$, simulation  2350]{\raggedleft{}\includegraphics[width=7.5cm,height=5cm]{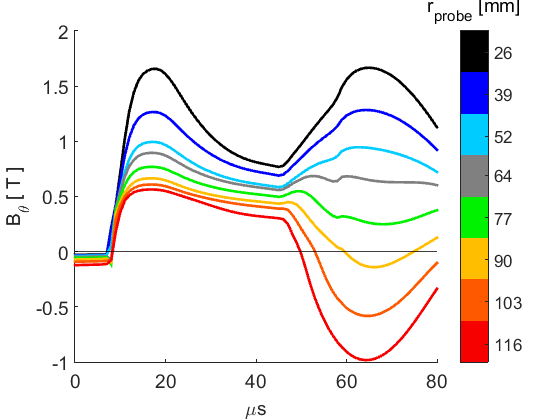}}

\caption{\label{fig:Bpol_meas_cf_sim39475}Comparison of measured (a) and simulated
(b) $B_{\theta}$ (compression - 11 coils) }
\end{figure}
 The good match observed between experimentally measured and simulated
$B_{\theta}$ when magnetic compression is included in the simulation
is evident in figure \ref{fig:Bpol_meas_cf_sim39475}.  For this shot
(and simulation), $V_{comp}=12$kV and $t_{comp}=45\upmu$s. For ease
of comparison, the toroidal averages of the poloidal field traces
measured at the two probes 180$^{o}$ apart at each of the eight radii,
at which the inner flux conserver magnetic probes are located, are
shown in figure \ref{fig:Bpol_meas_cf_sim39475}(a). These axisymmetric
MHD simulations allow for only resistive loss of flux and do not capture
inherently three-dimensional plasma instabilities that can lead to
poloidal flux loss. Shot $39475$ was a flux-conserving shot, and
a good match is found between experimentally inferred and MHD-simulated
poloidal field over the main compression cycle. 

\begin{figure}[H]
\subfloat[]{\raggedright{}\includegraphics[width=7cm,height=5cm]{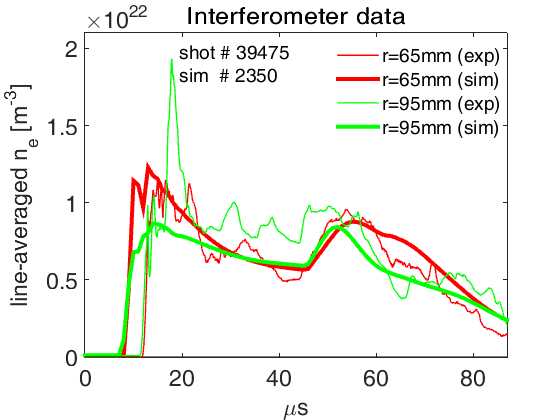}}\hfill{}\subfloat[]{\raggedright{}\includegraphics[width=7cm,height=5cm]{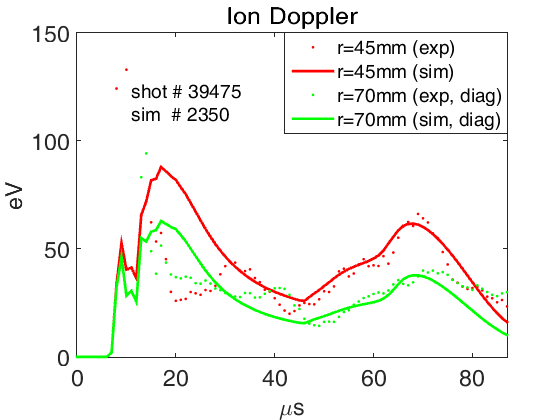}}
\begin{centering}
\subfloat[]{\raggedright{}\includegraphics[width=7cm,height=5cm]{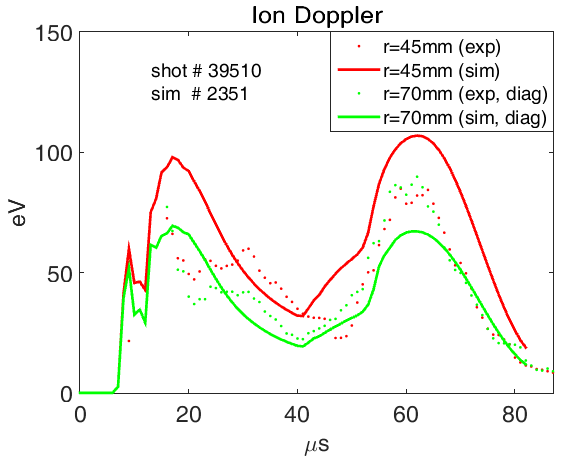}}
\par\end{centering}
\caption{\label{fig:ne_IDcomp}Comparison of measured and simulated $n_{e}$
(figure (a)) and $T_{i}$ ((b), (c)), (compression - 11 coils)}
\end{figure}
The agreement between experimentally measured and simulated electron
density and ion temperature in the case with magnetic compression
for shot 39475 is apparent from figures \ref{fig:ne_IDcomp}(a) and
\ref{fig:ne_IDcomp}(b).  The simulated line averaged electron density
along the interferometer chord at $r=35$mm hasn't been included in
figure \ref{fig:ne_IDcomp}(a) because the experimental data for that
chord is not available. Figure \ref{fig:ne_IDcomp}(c) shows the comparison
of experimental and simulated ion-Doppler measurement for shot  39510,
which was also a flux conserving shot, but with $t_{comp}=40\upmu$s,
and increased compressional energy, with $V_{comp}=18$kV. For this
shot, an increase in ion temperature by a factor of around four, from
$\sim25$eV to $\sim100$eV, is indicated in the region of the ion
Doppler chords. A maximum error in the temperature measurement due
to density broadening has been evaluated as $\sim$12eV for density
levels associated with shot 39510 at peak compression \cite{Kunze,Pittman}.
Careful analysis was undertaken to confirm that temperature broadening
rather than density broadening was the dominant broadening mechanism
for the compressed shots presented here. 
\begin{figure}[H]
\subfloat[]{\raggedright{}\includegraphics[width=7cm,height=5cm]{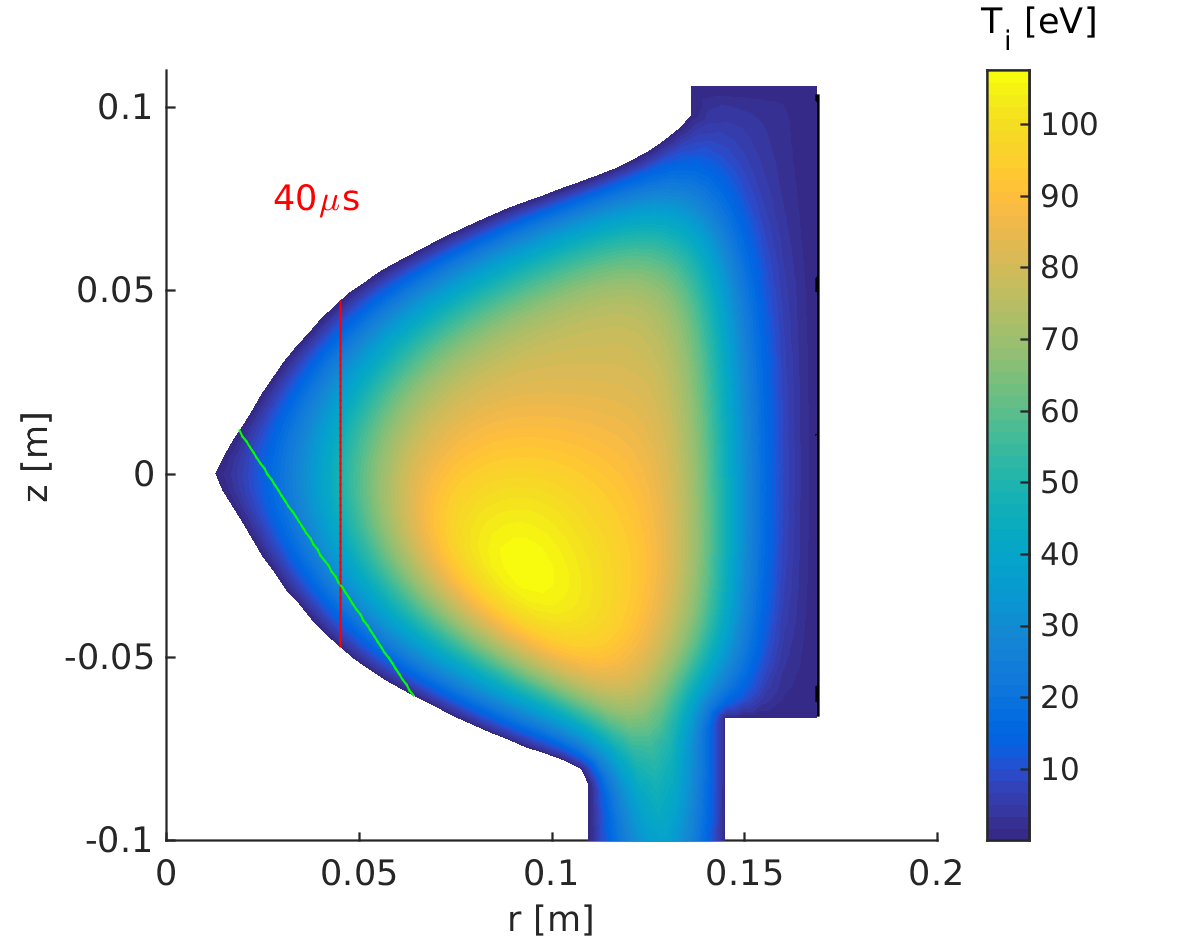}}\hfill{}\subfloat[]{\raggedright{}\includegraphics[width=7cm,height=5cm]{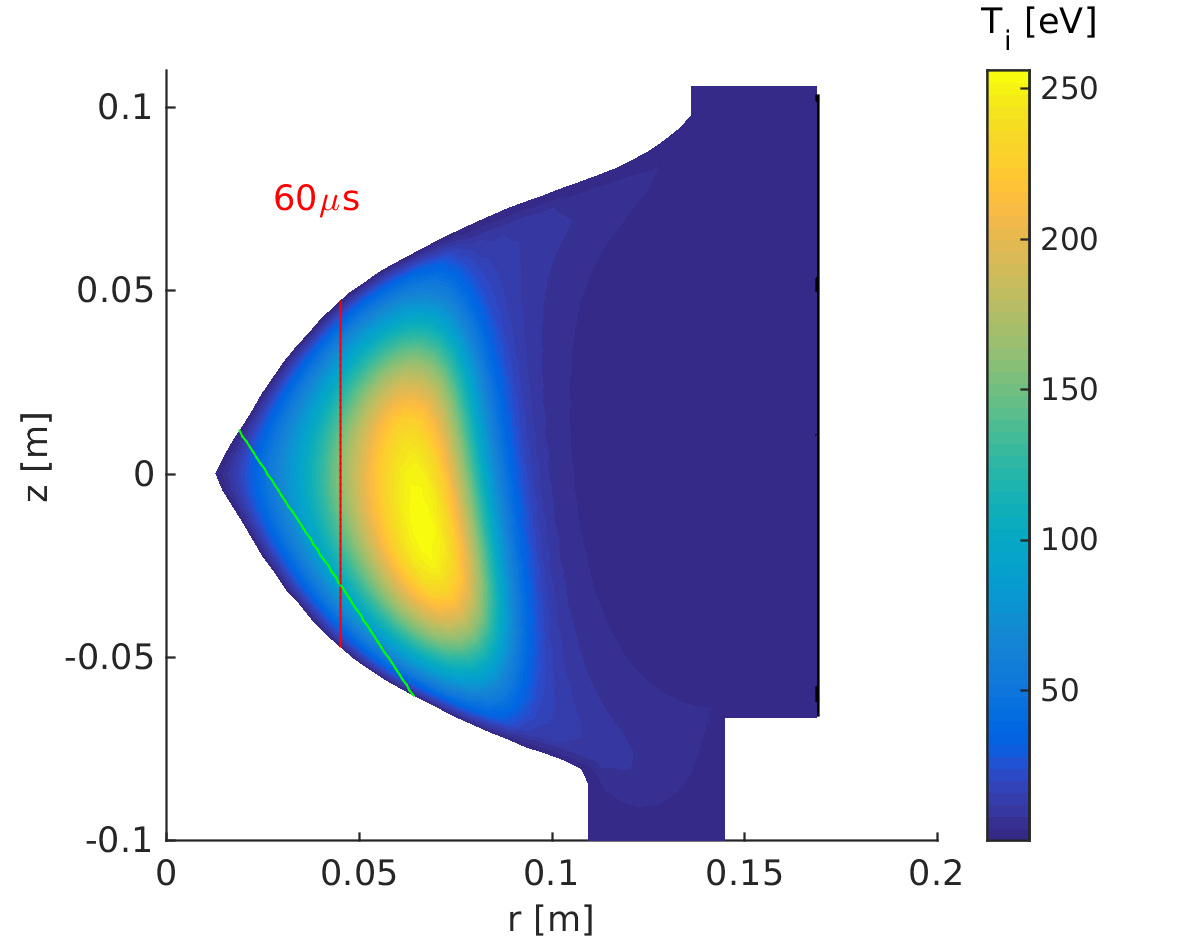}}

\caption{\label{fig:Ti}$T_{i}$ contours before compression (a) and at peak
compression (b), simulation 2351 }
\end{figure}
When the experimental ion-Doppler measurement is matched by simulations,
simulated core ion temperature increases by a factor of around 2.5
over the main compression cycle, as indicated in figure \ref{fig:Ti},
in which contours of ion temperature, for a simulation of shot 39510,
are shown just prior to compression ($t=40\upmu$s) and at around
peak compression ($t=60\upmu$s). As seen from figure \ref{fig:Chalice},
the ion-Doppler chords are located well away from the CT core. Note
that ion-Doppler temperature increases at compression were significant
on the 11-coil configuration only. 

\subsubsection{Compression field reversal}

As described at the beginning of section \ref{sec:Magnetic-compression},
when the compression current in the coils changes direction, the CT
poloidal field magnetically reconnects with the compression field,
and a new CT with polarity opposite to that of the previous CT is
induced in the containment region, compressed, and then allowed to
expand. The process repeats itself at each change in polarity of the
compression current until either the plasma loses too much heat, or
the compression current is sufficiently damped.

\begin{figure}[H]
\subfloat[]{\raggedright{}\includegraphics[width=5.5cm,height=5cm]{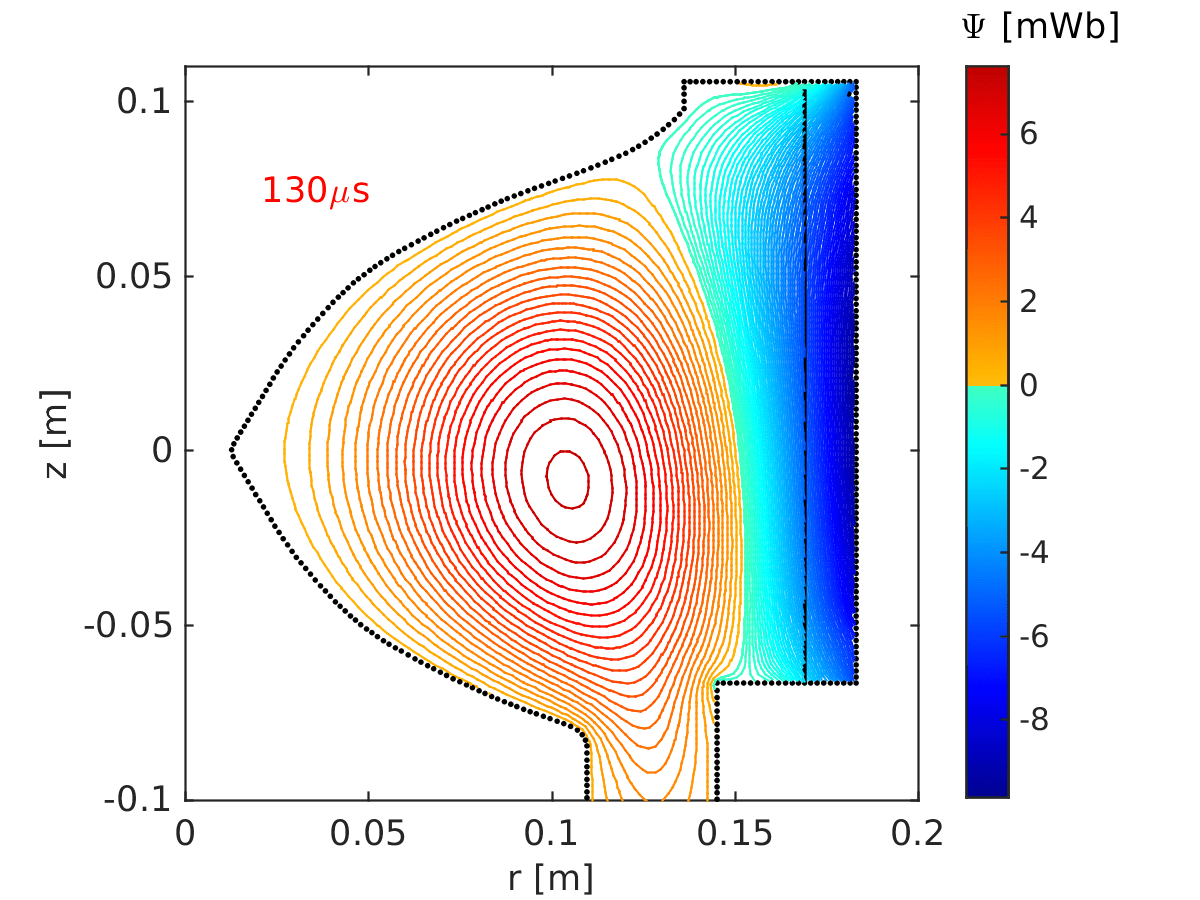}}\hfill{}\subfloat[]{\raggedleft{}\includegraphics[width=5.5cm,height=5cm]{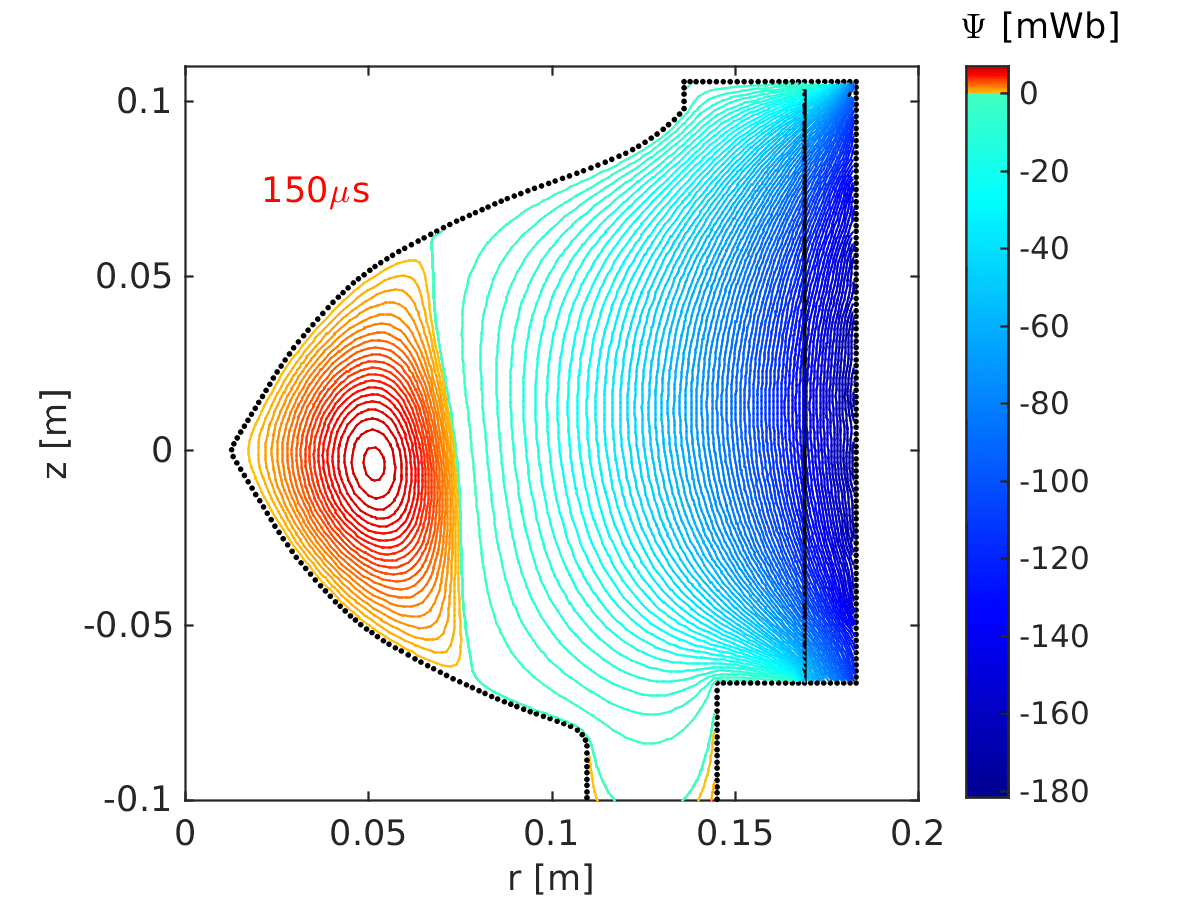}}\hfill{}\subfloat[]{\raggedright{}\includegraphics[width=5.5cm,height=5cm]{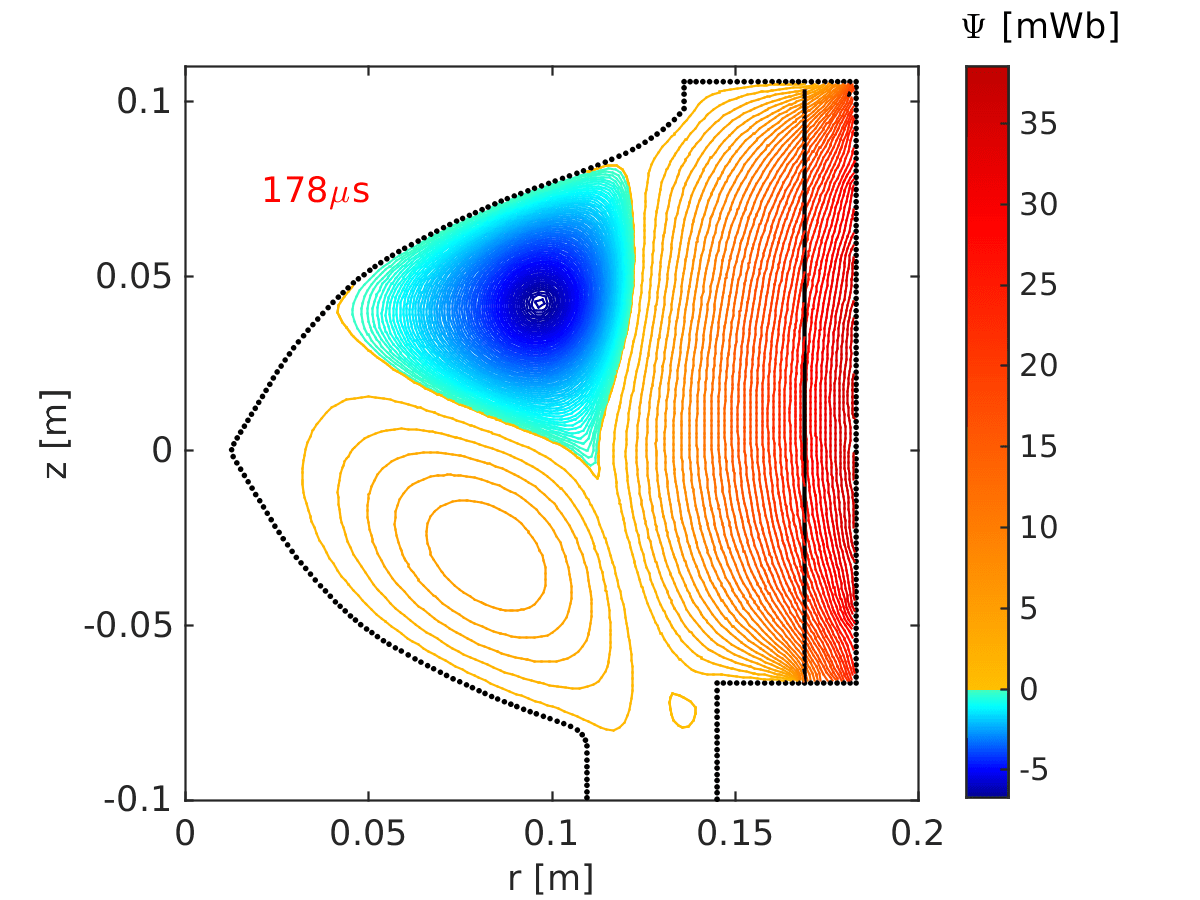}}

\subfloat[]{\raggedright{}\includegraphics[width=5.5cm,height=5cm]{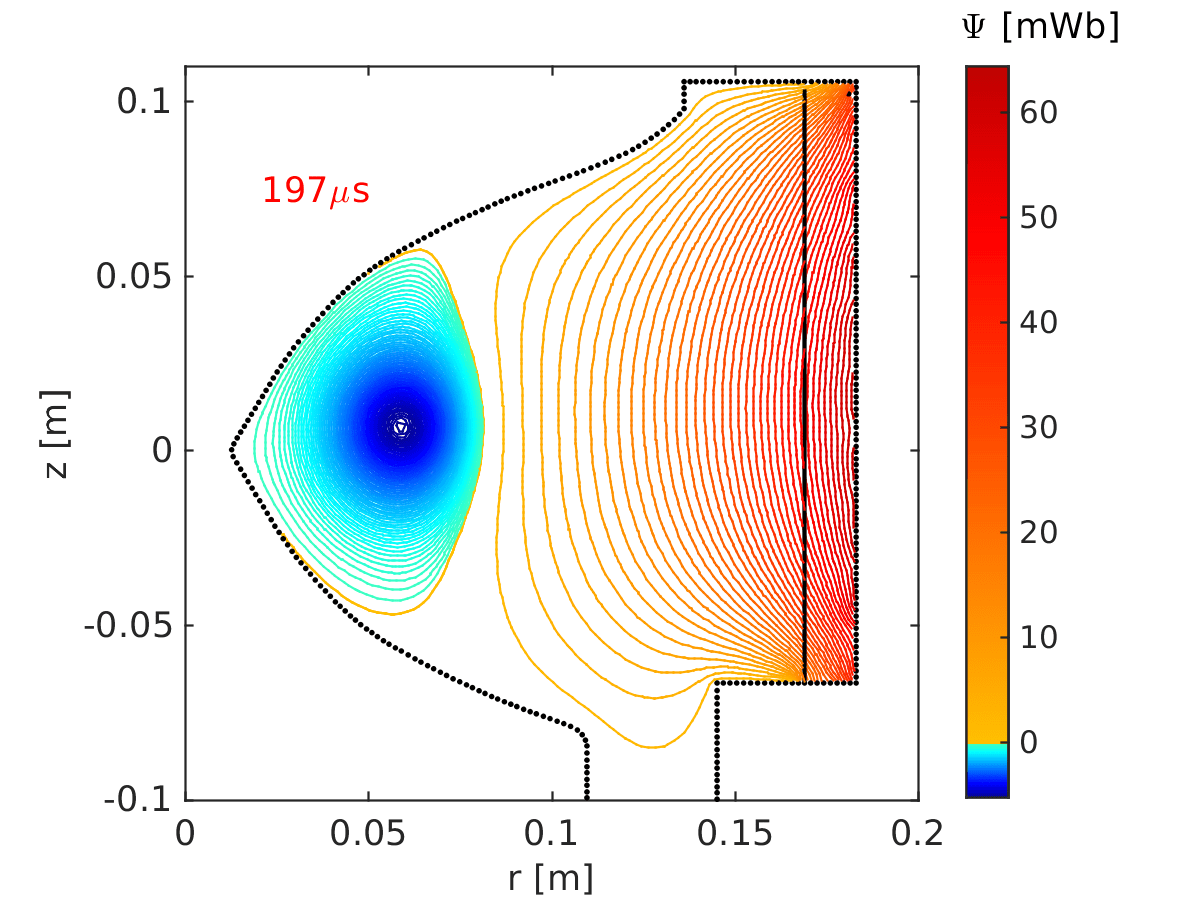}}\hfill{}\subfloat[]{\raggedleft{}\includegraphics[width=5.5cm,height=5cm]{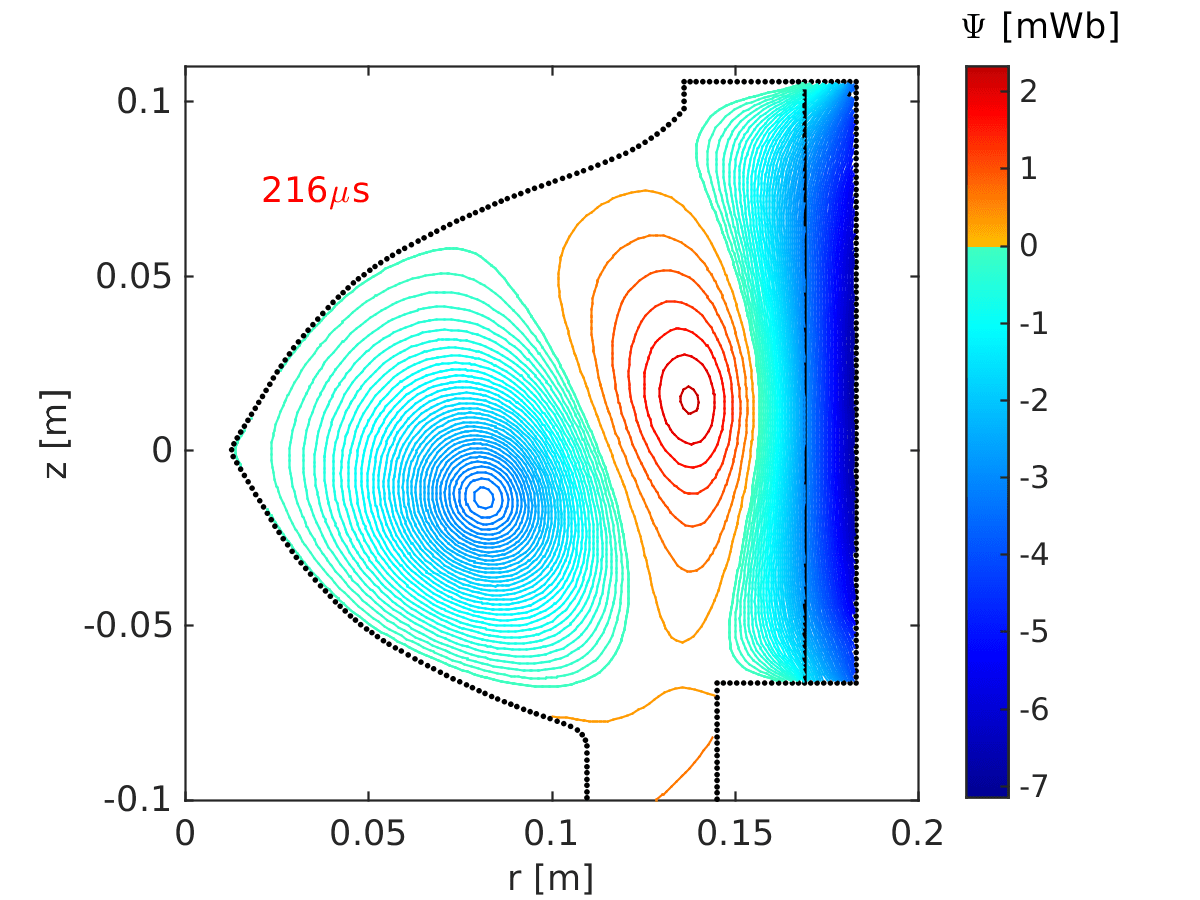}}\hfill{}\subfloat[]{\raggedright{}\includegraphics[width=5.5cm,height=5cm]{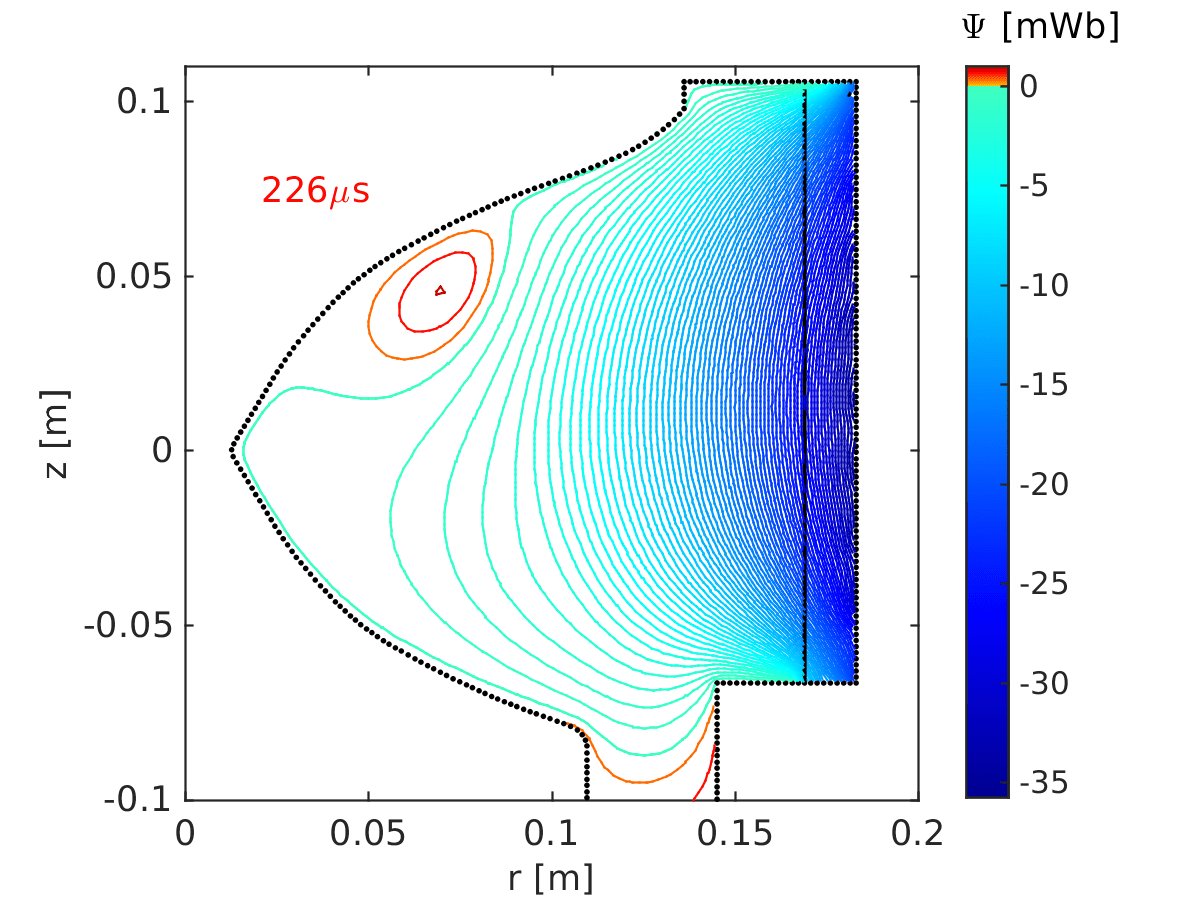}}

\caption{\label{fig:psi_2287}$\psi$ contours at various times ($130\upmu$s
(a), $150\upmu$s (b), $178\upmu$s (c), $197\upmu$s (d), $216\upmu$s
(e), $226\upmu$s (f)), simulation  2287}
\end{figure}
 Poloidal flux contours at various times from simulation  2287 are
shown in figure \ref{fig:psi_2287}.  Magnetic compression begins
at $t_{comp}=130\upmu$s, and peak compression is at $150\upmu$s
($c.f.$ shot 39735 (figure \ref{fig:39735BpMagnetic-compression-shot})).
By $178\upmu$s, the external compression field has changed polarity
and starts to reconnect with the CT poloidal field. Toroidal currents
are induced to flow in the ambient plasma initially located outboard
of the original CT, enabling the formation of a new CT (blue closed
contours) with polarity opposite to that of the original CT. The new
induced CT is magnetically compressed inwards by the increasing reversed
polarity compression field, with peak compression at around $197\upmu$s
(figure \ref{fig:psi_2287}(d)). The compression field polarity rings
back to its original state by $216\upmu$s, when a third CT is induced,
with the same polarity as the original CT. By $226\upmu$s, the poloidal
field of the second CT has reconnected with the compression field,
and the poloidal flux of the third CT, which is being compressed inwards
during the third compression cycle, has almost decayed away (figure
\ref{fig:psi_2287}(f)). 
\begin{figure}[H]
\begin{centering}
\subfloat{\raggedleft{}\includegraphics[width=8cm,height=5.7cm]{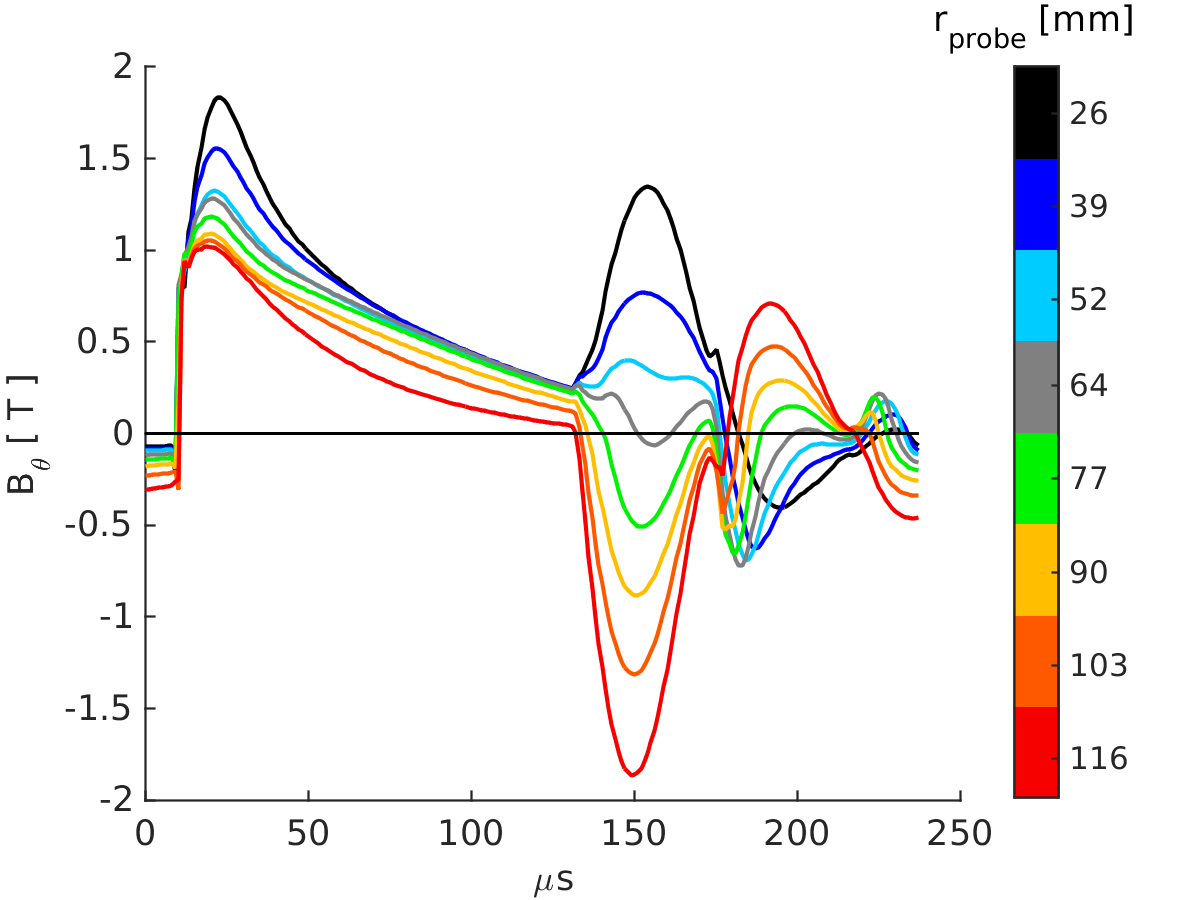}}
\par\end{centering}
\caption{\label{fig:Bpolexpcf_sim39735_2287}$B_{\theta}$, simulation 2287}
\end{figure}
Poloidal field from the same simulation, pertaining to shot 39735
(figure \ref{fig:39735BpMagnetic-compression-shot}) is shown in figure
\ref{fig:Bpolexpcf_sim39735_2287}.  In shot  39735, the poloidal
field measured at the inner probes collapses at $\sim145\upmu$s,
while the compression coil current peaks at $\sim150\upmu$s. Because
of this, as outlined previously, shot  39735 had parameter $\widetilde{\tau}_{c}=0.6$,
implying that poloidal flux was not well conserved during compression.
Apart from resistive losses, CT poloidal flux is conserved in the
simulation, so the poloidal field at the inner probes continues to
rise until the compression coil current peaks. 

\subsubsection{Experimentally measured separatrix radius ($r_{s}$) for compression
shots\label{sec:Rsep_comp}}

Using the method outlined in appendix \ref{sec:rsep}, it is possible
to determine $r_{s}$, the outboard CT separatrix at the equator ($z\sim0\mbox{mm}$),
and compare with simulations for compression shots. 
\begin{figure}[H]
\subfloat[averaged $B_{z}$ (from side probes), shot 39738]{\raggedright{}\includegraphics[width=8cm,height=5cm]{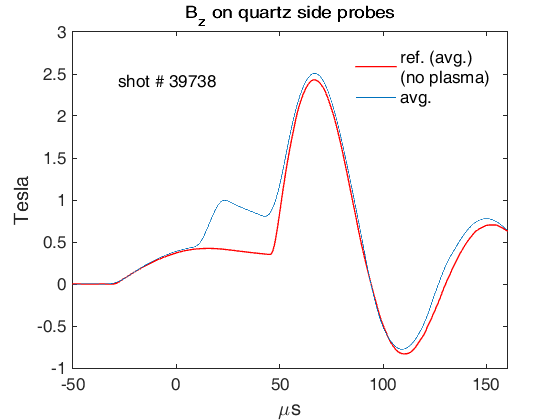}}\hfill{}\subfloat[Comparison of $r_{s}$ - experiment and simulation ]{\raggedleft{}\includegraphics[width=8cm,height=5cm]{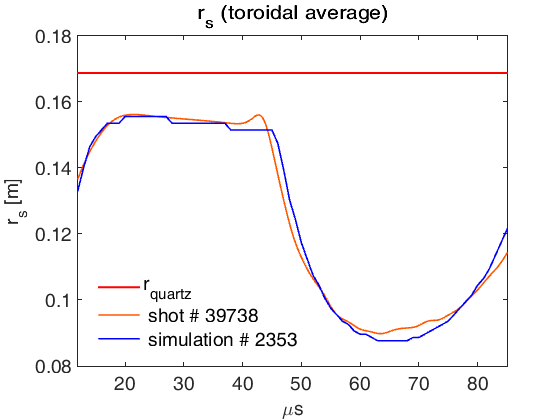}}

\caption{Experimental data (a) and comparison of measured and simulated $r_{s}$
(b) for shot $39378$\label{fig:rsepCOMP39738} }
\end{figure}
The averages of the eight $B_{z}(\phi,\,t)$ and $B_{zref}(\phi,\,t)$
signals for shot $39738$ are depicted in figure \ref{fig:rsepCOMP39738}(a).
 Note that, as described in appendix \ref{sec:rsep}, $B_{z}(\phi,\,t)$
refers to the axial field mesured at the probes located outside the
insulating wall, and $B_{zref}(\phi,\,t)$ the reference signals,
are the averages of the signals measured at the same probes during
three levitation-only shots taken without charging or firing the formation
banks. Shot 39738 had $V_{comp}=18$kV, $t_{lev}=30\upmu$s, $t_{comp}=45\upmu$s,
and was taken in the 11-coil configuration, so that the functional
fit indicated in figure \ref{Bz_rsep39650_1}(b) was used to find
$r_{s}(t)$. For compression shots, it is convenient to find the toroidally
averaged $r_{s}$ using toroidally averaged probe data. As seen in
figure \ref{fig:rsepCOMP39738}(a), at compression, the reference
$B_{z}$ is very close to $B_{z}$ with the CT present, so that errors
in probe signal response can lead to instances when $B_{zref}(\phi,\,t)>B_{z}(\phi,\,t)$,
and consequent complex-valued $r_{s}(\phi,\,t)$ solutions. Using
the toroidally averaged signals reduces the likelihood of this error.
The MHD simulations allow for only resistive loss of flux and do not
capture the mechanisms that led to flux loss in many compression shots.
Shot $39738$ was a flux-conserving shot, and a good match is found,
as indicated in \ref{fig:rsepCOMP39738}(b), between experimentally
inferred and MHD-simulated $r_{s}$, indicating a radial compression
factor, in terms of equatorial outboard CT separatrix, of $C_{s}=1.7$.
Note that $r_{s}\sim9$cm at peak compression. As noted in section
\ref{subsec:Rsep_Lev_70mOhm}, when $r_{s}\lesssim9$cm, the slope
of the functional fit in \ref{Bz_rsep39650_1}(b) is too flat to be
successfully inverted with good accuracy. For this reason, $C_{s}$
cannot be evaluated if the CT is compressed more than in shot  39378.
An example of a shot in which compression is too strong for successful
evaluation of $C_{s}$ is shot  39735 (figure \ref{fig:39735BpMagnetic-compression-shot})
which also has $V_{comp}=18$kV, but is compressed later ($t_{comp}=130\upmu$s),
when pre-compression CT flux has decayed to lower levels and therefore
compression is more extreme.

\subsection{Compressional instability\label{subsec:Compressional-instability}}

\begin{figure}[H]
\begin{centering}
\subfloat{\raggedright{}\includegraphics[width=8.5cm,height=5cm]{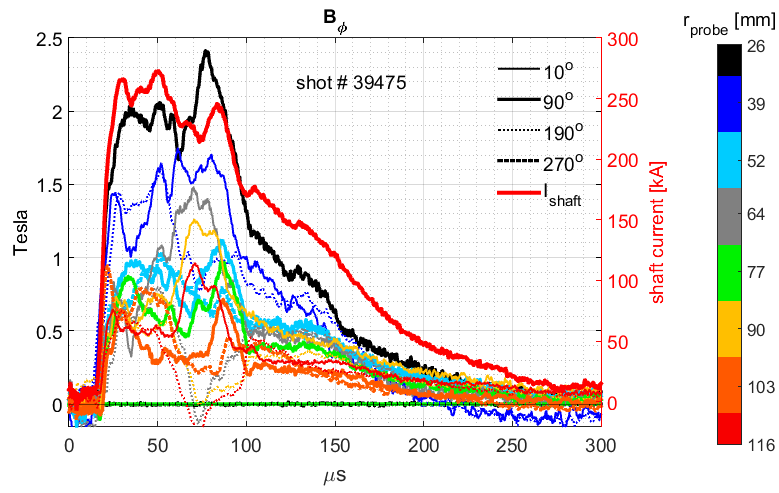}}
\par\end{centering}
\caption{\label{fig:Bt_39475}$B_{\phi}$ for shot  39475}
\end{figure}
Measured $B_{\phi}$ traces for shot  39475 are shown in figure \ref{fig:Bt_39475}.
 The $B_{\phi}$ signals for this shot are a good exemplification
of the indication of the instability that was observed on most compression
shots. It can be seen how $B_{\phi}$ at all four probes at $190^{\circ}$
drops during compression, while the field increases at the other toroidal
angles. The angle at which the signal drops varies, apparently randomly,
from shot to shot, but shots were generally quite consistent in displaying
this behaviour. \\
\\
\begin{figure}[H]
\centering{}\includegraphics[scale=0.6]{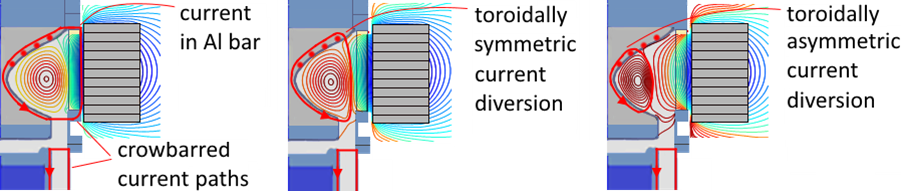}\caption{\label{fig:divertedcurrent}Asymmetric current diversion}
\end{figure}
Figure \ref{fig:divertedcurrent} shows a possible explanation for
this instability. After the $50\upmu$s formation capacitor-driven
pulse, toroidal-flux conserving crowbarred current continues to flow,
primarily along two separate paths as indicated. In addition, it is
likely that there is a third current path, consisting of the merger
of the two paths indicated here. Referring to the upper path, initially
most of the outboard part of this current is in the aluminum bars
depicted in figure \ref{fig:Schematic-of-6-1}(a). Shaft current,
and $B_{\phi}$ at probes, rise at compression as the current path
shifts symmetrically to a lower inductance path (central subfigure);
now the outboard part of the current loop travels through the ambient
plasma outboard of the CT. The asymmetric current diversion depicted
in the right subfigure will be discussed after outlining how the symmetric
shifting current path mechanism is reproduced in MHD simulations:
\begin{figure}[H]
\subfloat[]{\raggedright{}\includegraphics[width=6.7cm,height=5cm]{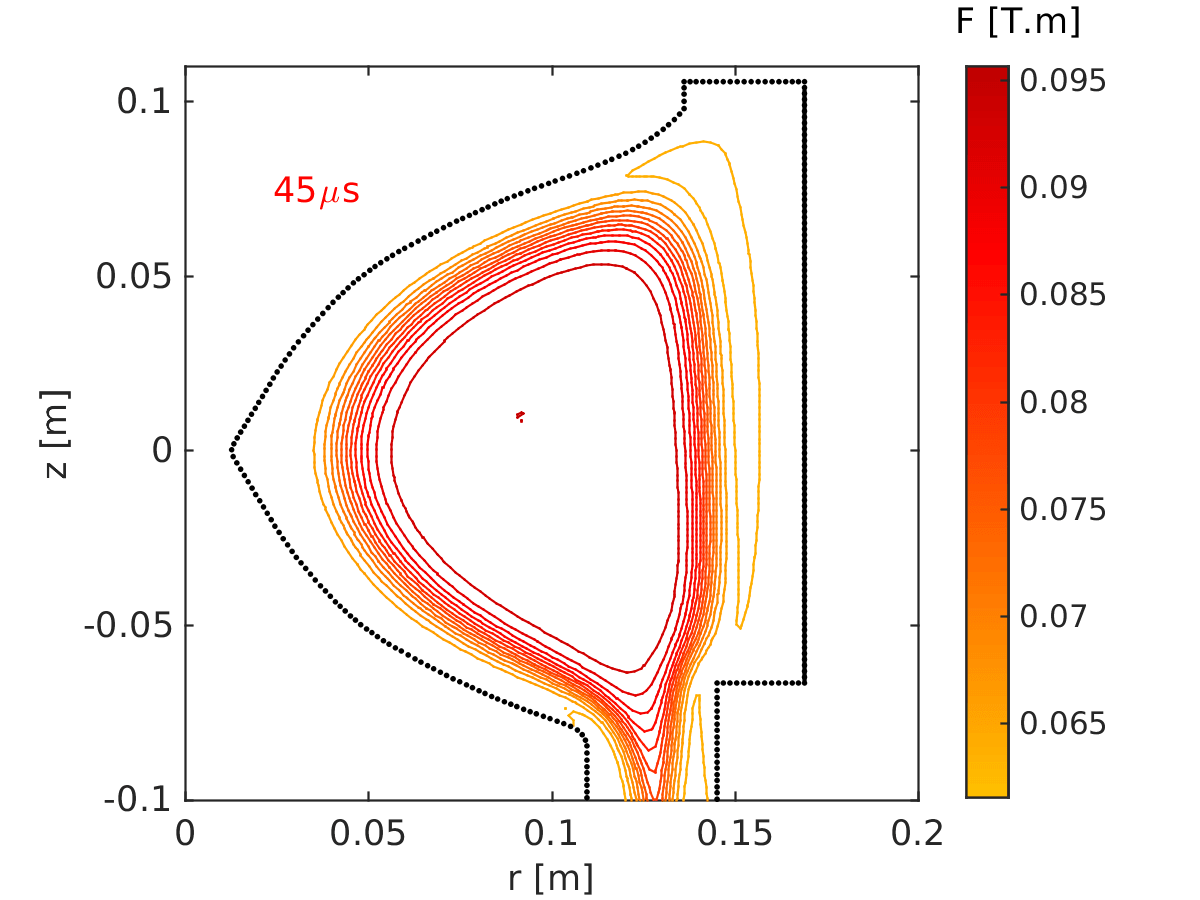}}\hfill{}\subfloat[]{\raggedleft{}\includegraphics[width=6.7cm,height=5cm]{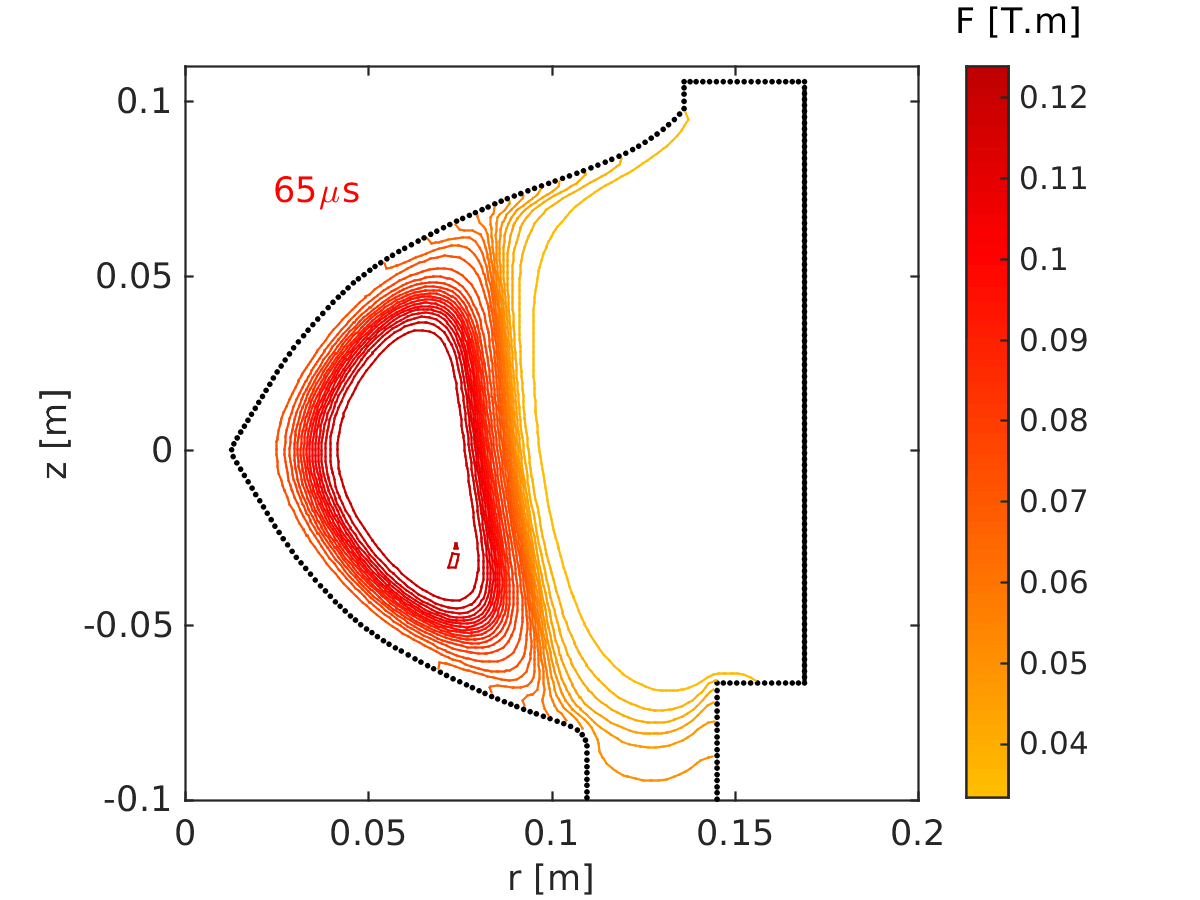}}

\caption{\label{fig:CompInstab_F-1}$f$ contours before compression (a) and
at peak compression (b), simulation  2350}
\end{figure}
Contours of $f$ at $45\upmu$s just prior to magnetic compression,
and at $65\upmu$s, at peak magnetic compression, are depicted in
figure \ref{fig:CompInstab_F-1}.  Note that in figure \ref{fig:CompInstab_F-1},
$f$ includes the contributions from poloidal plasma current \emph{and}
from poloidal current flowing in the external boundary. In the simulation,
as described in \cite{SIMpaper,thesis-1}, the boundary (including
the part of the boundary representing the aluminum bars indicated
in figure \ref{fig:Schematic-of-6-1}(a)) is modelled as being perfectly
electrically conducting. Contours of $f=rB_{\phi}$ represent paths
of poloidal current. Closely spaced contours indicate regions of high
gradients of $f$, which in turn are regions of high currents. The
MHD equations implemented to code are formulated such that the code
has various conservation properties \cite{SIMpaper}, including conservation
of toroidal flux. It can be seen in figure \ref{fig:CompInstab_F-1}(b)
how the imposition of toroidal flux conservation leads to the induction,
at magnetic compression, of poloidal currents flowing from wall-to-wall
through the ambient plasma just external to the outboard boundary
of the CT.

If some mechanism causes the CT to be compressed more at a particular
toroidal angle (an effect which the axisymmetric MHD code cannot reproduce),
the inductance of the current path at that angle will be reduced further
and more current will flow there (right subfigure in figure \ref{fig:divertedcurrent}),
enhancing the instability. This is analogous to the mechanisms behind
external kink and toroidal sausage type instabilities. As the current
path moves inwards past the probes at a particular toroidal angle,
$B_{\phi}$ at the probes will change polarity at that toroidal angle,
as is observed on most compression shots. As the CT decompresses as
$I_{comp}$ decreases, the current path returns towards its pre-compression
path. It is noteworthy that although the magnetically compressed CTs
generally exhibit this instability, there is a noticeable correlation
in that the compression shots that have a high value of $\widetilde{\tau}_{c}$
($i.e.,$ apparent flux conservation during compression) seem to exhibit
the clearest manifestation of the instability, through the behaviour
of the $B_{\phi}$ signals - shot $39475$ above is a good example
of this. As mentioned earlier, even levitated, but non-compressed
shots, exhibited this behaviour to some degree ($e.g.,$ figure \ref{fig:Poloidal-field-for}(c)),
in cases where the levitation currents were not optimised to decay
at near the rate of the CT currents. 

\begin{figure}[H]
\subfloat[]{\raggedright{}\includegraphics[width=5.5cm,height=5cm]{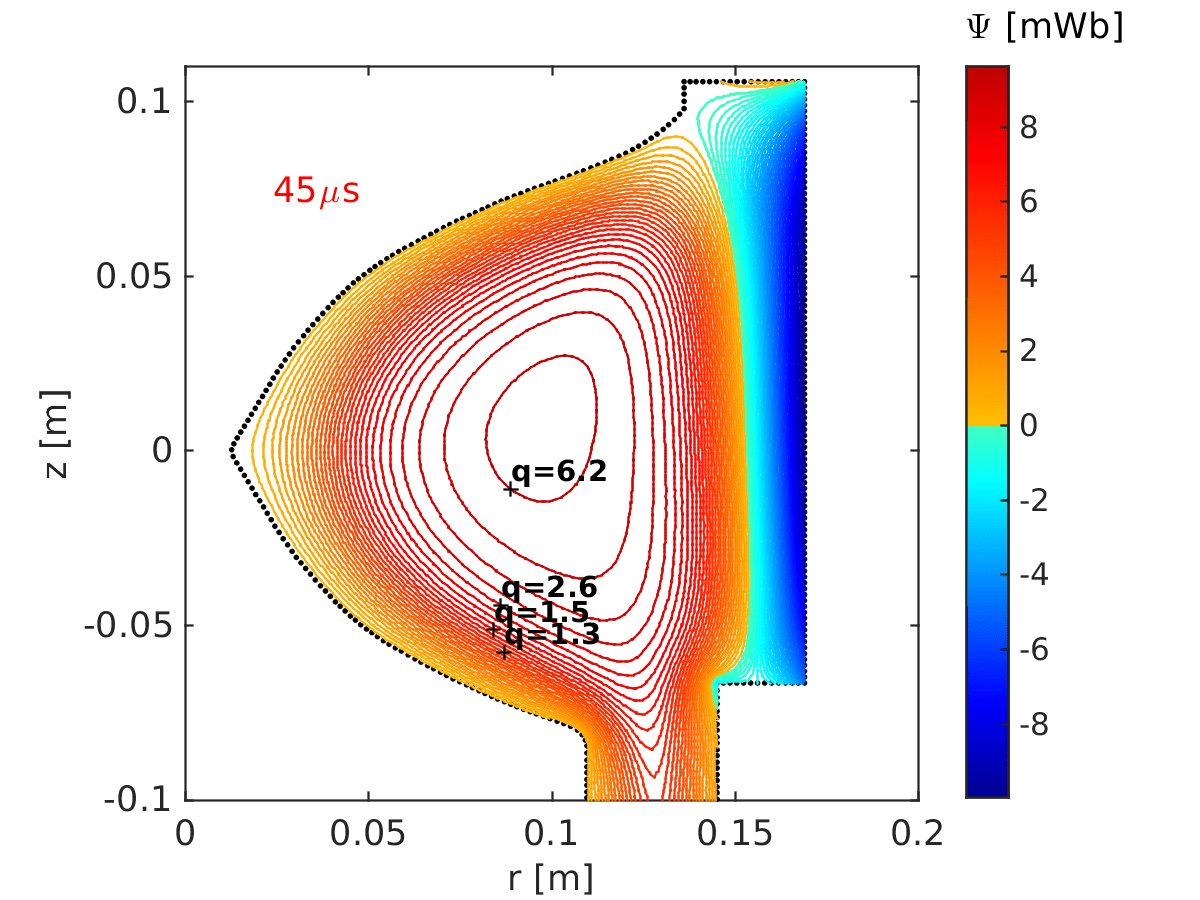}}\hfill{}\subfloat[]{\raggedleft{}\includegraphics[width=5.5cm,height=5cm]{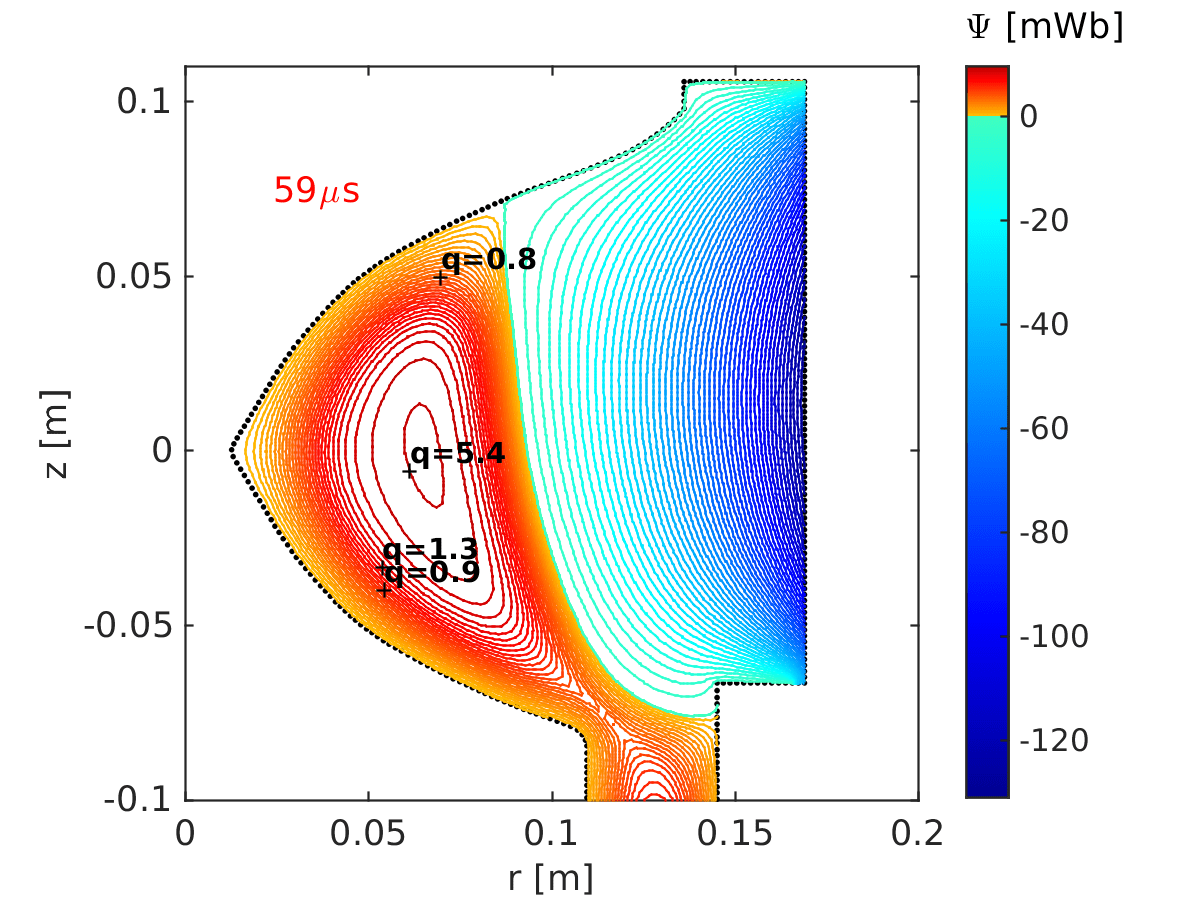}}\hfill{}\subfloat[]{\raggedright{}\includegraphics[width=5.5cm,height=5cm]{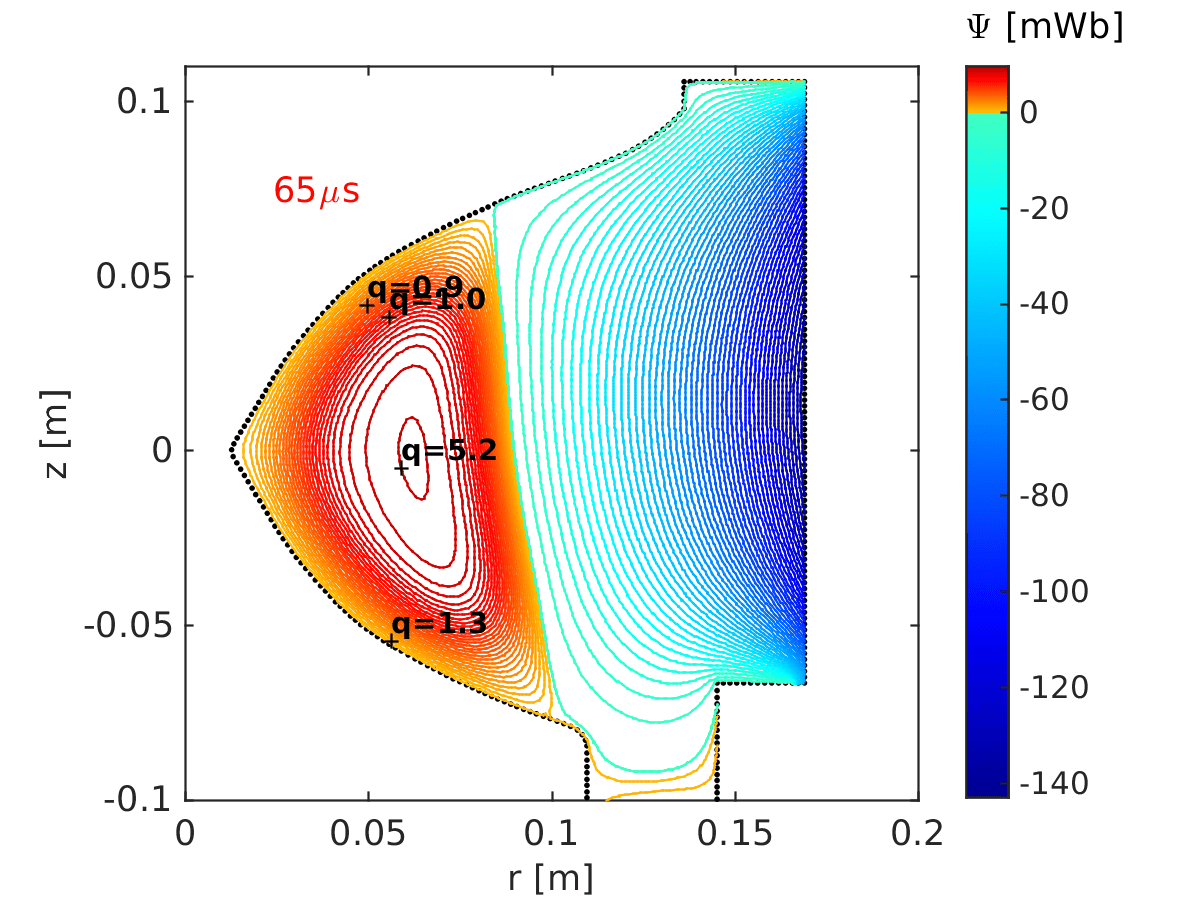}}

\subfloat[]{\raggedright{}\includegraphics[width=7cm,height=5cm]{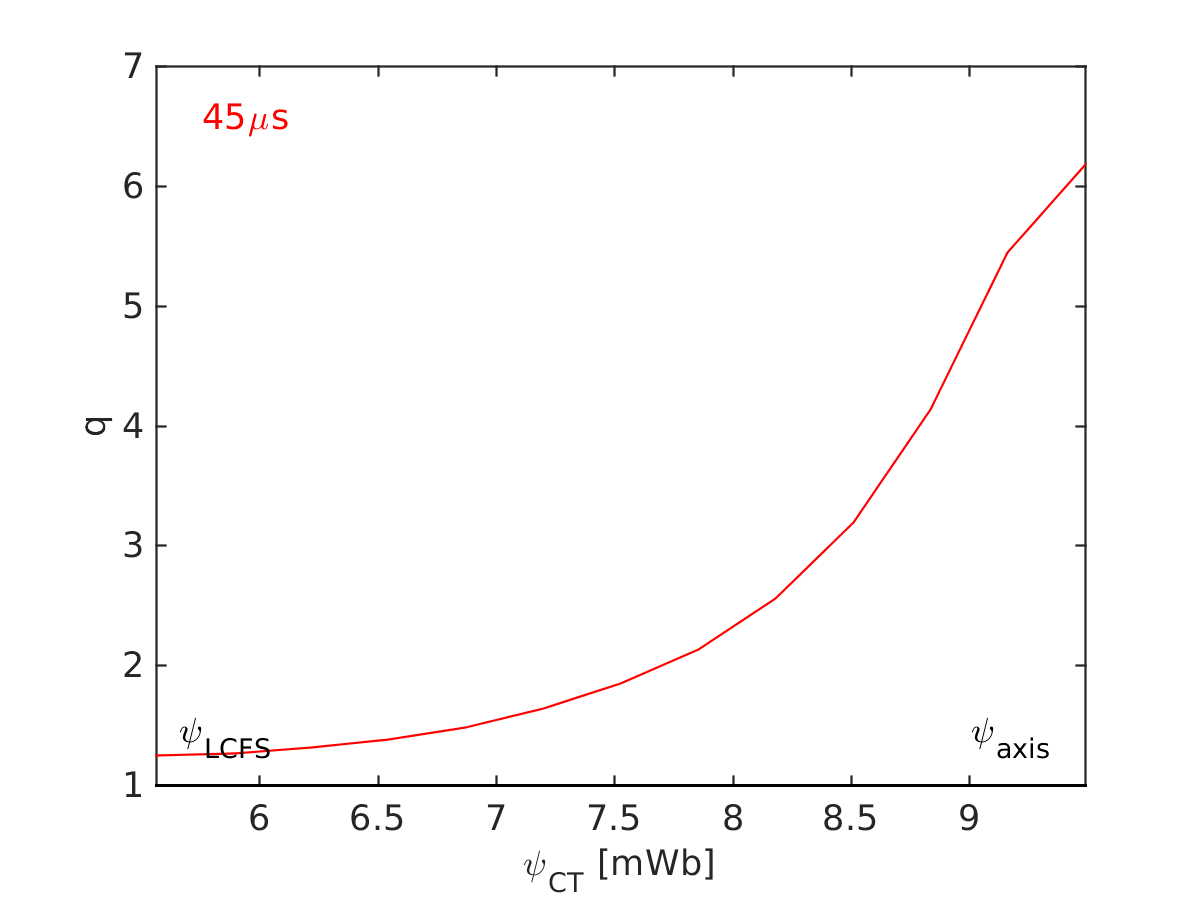}}\hfill{}\subfloat[]{\raggedleft{}\includegraphics[width=7cm,height=5cm]{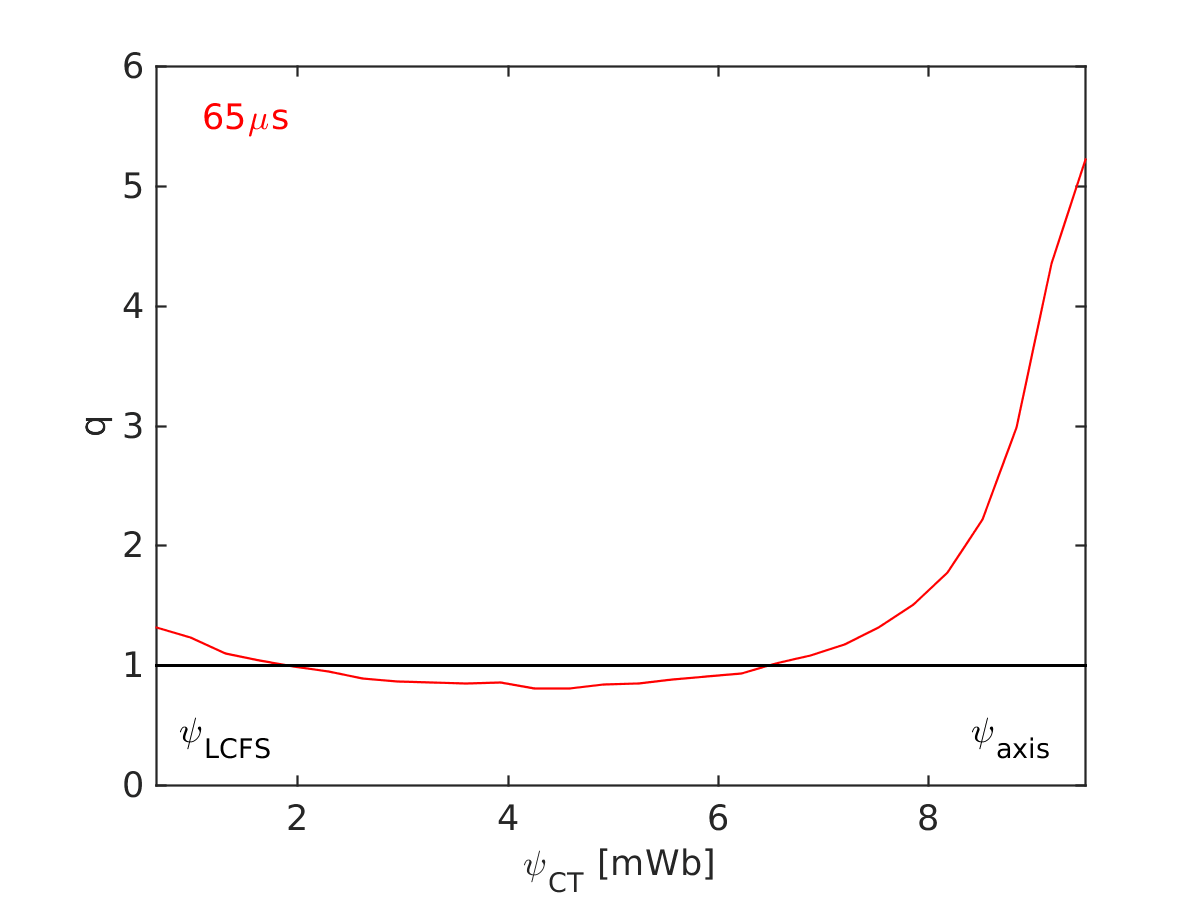}}

\caption{\label{fig:Qpsi_2350}$q$ profiles (before compression (figures (a),
(d)), mid compression (b), peak compression ((c), (e)), simulation
 2350}
\end{figure}
The profile of safety factor $q(\psi)$ obtained for simulation  2350
is shown at various times in figure \ref{fig:Qpsi_2350}.  Simulated
$q(\psi)$ shows two trends at compression, depending on the value
of $t_{comp}$. When compression banks are fired early in the CT's
life, for example at $t_{comp}=45\upmu$s as in simulation  2350,
the CT, defined by regions of closed $\psi$ contours, is still, at
$t=t_{comp}$, surrounded by open field lines that are pinned to the
inner and outer electrodes (figure \ref{fig:Qpsi_2350}(a)), and $q(\psi)$
ranges from $q\sim6.2$ near the magnetic axis (at $\sim9.5\mbox{mWb}$)
to $q\sim1.2$ at the last closed flux surface ($\mbox{LCFS}$) at
$\sim5.5\mbox{mWb}$ (figure \ref{fig:Qpsi_2350}(d)). During magnetic
compression, the open field lines surrounding the CT are pinched off
and reconnect to form additional closed field lines, as depicted in
figure \ref{fig:Qpsi_2350}(b), that are then associated with the
exterior of the CT, as indicated in figure \ref{fig:Qpsi_2350}(c).
High levels of toroidal current flowing along the originally open
field lines results in these field lines being associated with low
$q$ when they are pinched off. At $65\upmu$s, $q(\psi)$ ranges
from $q\sim5.3$ near the magnetic axis (at $\sim9.3\mbox{mWb}$)
to $q\sim1.3$ at the $\mbox{LCFS}$ at $\sim0.7\mbox{mWb}$ (figure
\ref{fig:Qpsi_2350}(e)), while dipping below $q=1$ over a large
extent between the magnetic axis and the $\mbox{LCFS}$. When compression
is started late in the CT's life, for example at $t_{comp}=130\upmu$s
in shot 39735 (figure \ref{fig:39735BpMagnetic-compression-shot}),
simulations indicate that most of the open poloidal field lines that
previously surrounded the CT have already reconnected because $T_{e}$
has dropped and $\eta$ has increased. Then, MHD simulations typically
show $q>1$ at the $\mbox{LCFS}$, while the region with $q<1$ extends
all the way to the magnetic axis, both prior to and during compression.

For both early and late magnetic compression, simulations indicate
that the $q$ profile is not contingent to magnetohydrodynamic stability,
as $q<2$ at the $\mbox{LCFS}$ in both cases \cite{WessonTokamaks}.
Also, for both early and late compression, $q$ drops below one over
extensive spans between the magnetic axis and the LCFS. Note that
the 2D simulations, which neglect inherently three dimensional turbulent
transport and flux conversion, are likely to overestimate the level
of hollowness of the current profiles, and lead to an underestimation
of $q$ towards the CT edge, but without further internal experimental
diagnostics or 3D simulations, the level of underestimation remains
uncertain. The Kruskal-Shafranov limit determines that magnetically
confined plasma are unstable to external kink modes when $q<1$. An
obvious solution towards mitigating the instability would be to drive
more shaft current around the machine. This would lead to increased
CT toroidal field (higher $q)$ which can stabilise the external kink
and toroidal sausage modes. This was attempted briefly, shortly before
the machine was decommissioned, when one of the levitation coil circuits
was used to drive additional crowbarred shaft current with an $RC$
decay time of around 200$\upmu$s and a peak of up to $80\mbox{kA}$.
From the data obtained from 30-40 compression shots, this had no apparent
effect on improving stability during compression. Its likely that
insufficient shaft current was driven. More recent SPECTOR plasma
injectors at GF drive up to 1MA crowbarred shaft current, largely
to improve CT stability.

\subsection{Comparison of compression parameters between configurations\label{sec:Comparison-of-compression}}

The impedances (effective resistance to alternating current, due to
combined effects of reactance and ohmic resistance) of the coil arrays
are slightly different for the 6-coil and 11-coil configurations,
leading to a few percent variation in peak compression current at
the same $V_{comp}$. At $V_{comp}=18$kV, measured peak $I_{comp}$
was $\sim200\mbox{kA}$ per coil in the 6-coil configuration, compared
with $\sim210\mbox{kA}$ per coil-pair (and $\sim210\mbox{kA}$ in
the single coil 3rd from the bottom of the stack) in the 11-coil configuration.
At the same $V_{comp}$, compressional flux ($\psi_{comp}$), as estimated
from $\mbox{FEMM}$ model outputs, is around $1.7$ times higher in
the 6-coil configuration, relative to the 11-coil configuration, due
to the large gaps (see figure \ref{fig:Schematic-of-6-1}(b) $cf.$
figure \ref{fig:11-coil-configuration}(b)), above and below the coil
stack, that ease entry of compressional flux into the containment
region for the 6-coil configuration. 

In general, as external compressional flux is increased, the level
of CT flux conservation at compression is reduced, while, for the
same level of flux conservation, magnetic compression ratios are increased.
A fair comparison of compression performance metrics across builds
can be obtained by comparing the metrics for shots with the same level
of external compressional flux. For around the same external compressional
flux, we would ideally compare shots with $V_{comp}=$14kV in the
11-coil configuration against shots with $V_{comp}=$8.2kV in the
6-coil configuration. With limited available data, a reasonable comparison
of compression parameter trending can be made looking at shots with
$V_{comp}=14\mbox{kV}$ for the 11-coil configuration, and $V_{comp}=7\mbox{kV}-9\mbox{kV}$
in the 6-coil configuration. 
\begin{figure}[H]
\centering{}\includegraphics[scale=0.3]{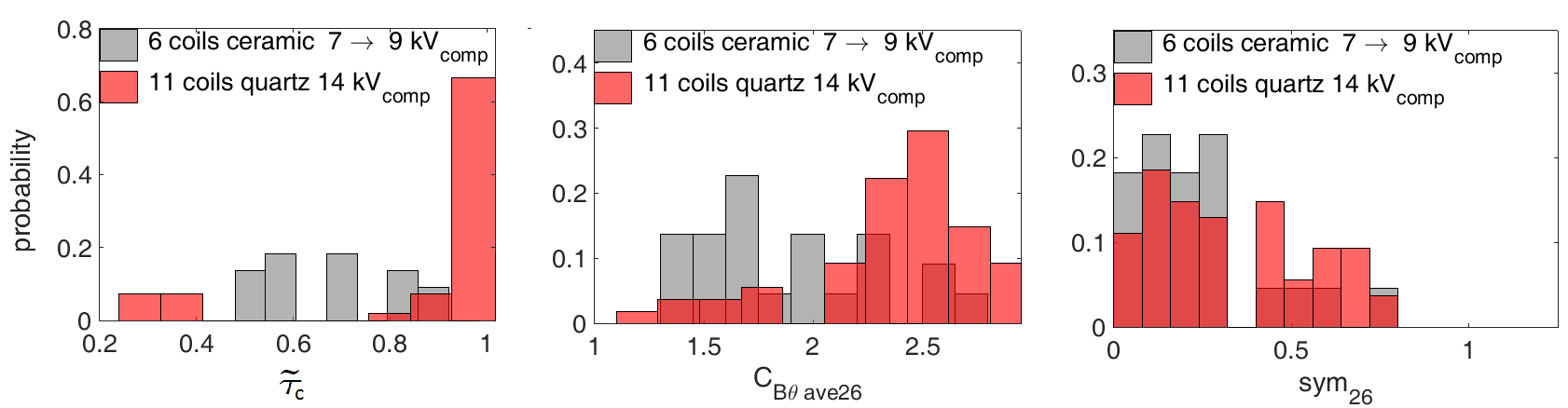}\caption{\label{fig:Comparison-of-compression_3}Comparison of compression
parameters}
\end{figure}
Normalised histograms of the key compression parameters (defined below
figure \ref{fig:39735BpMagnetic-compression-shot}) are shown in figure
\ref{fig:Comparison-of-compression_3}.  The recurrence rate of shots
that conserved CT flux at compression was significantly improved in
the 11-coil configuration. Around $70\%$ of shots had good CT flux
conservation ($i.e.,\,\widetilde{\tau}_{c}\sim1$) in the 11-coil
configuration, while only $\sim10\%$ of shots conserved $\sim80\%$
of flux ($i.e.,\,\widetilde{\tau}_{c}\sim0.8$) in the 6-coil configuration.
Poloidal magnetic compression ratios (characterised by $C_{B\theta ave26}$)
would be expected to be low when CT flux is lost, and it can be seen
how the ratios are nearly doubled on average in the 11-coil configuration.
Compression asymmetry (characterised by $sym_{26}$) remains poor
in both configurations.

While reduced plasma wall interaction at formation and consequent
reduced impurity radiation cooling in the 11-coil configuration was
certainly behind the huge improvement in lifetimes of levitated CTs
(figure \ref{fig:Effect-of-lithium}), it seems likely, but can't
be confirmed without further experiment or 3D simulation, that a different
mechanism was responsible for the orders of magnitude improvement
in the rates of shots with good CT flux conservation at compression.
Supporting this, shots taken in the 11-coil configuration with compression
fired late when plasma has had time for significant diffusive cooling
($e.g.,$ figure \ref{fig:39735BpMagnetic-compression-shot}) generally
conserved \emph{more} flux than those fired early in time in the 6-coil
configuration ($e.g.,$ figure \ref{fig:Bp_28426cf39475}(a)). The
improvement is likely to be largely due to the compression field profile
itself, which led to more uniform outboard compression, as opposed
to largely equatorial outboard compression with the six coil configuration.
Equatorially-focused outboard compression may have caused the CT to
bulge outwards and upwards/downwards above and below the equator,
leading to poloidal field reconnection, CT depressurisation, and possible
disruption.

\section{Conclusions\label{sec:Conclusion}}

In the study of CT formation into a levitation field, interaction
between plasma and the outer insulating wall during the CT formation
process led to high levels of plasma impurities and consequent radiative
cooling. The longest levitated CT lifetimes were up to $\sim270\upmu$s
with the 25-turn coil configuration, despite the presence of the quartz
wall. This was almost double the maximum of $\sim150\upmu$s lifetimes
seen with six coils around quartz wall, but still less than the $\sim340\upmu$s
lifetimes observed without levitation with an aluminum flux conserver.
The ceramic alumina wall was far less contaminating than the quartz
wall. In the six coil configuration, best levitated CT lifetimes decreased
significantly when the ceramic wall was replaced with a quartz wall,
despite the larger inner radius of the quartz wall, which should have
allowed for a 50\% increase in lifetime if the material had not also
been changed. A revised wall design, such as one implementing a thin-walled
tube of pyrolytic boron nitride located inside an alumina tube for
vacuum support, would likely be beneficial. Future designs may ideally
use levitation/compression coils internal to the vacuum vessel, but
that would introduce further complications.

In the original six-coil configurations, plasma being rapidly advected
into the containment region during the formation process was able
to displace the levitation field into the large gaps above the coil
stack, and come into contact with the insulating wall. Some mitigation
of this effect was achieved by firing the levitation banks earlier,
allowing the levitation field to soak through and become resistively
pinned in the steel above and below the insulating wall. As supported
by MHD simulation, this line-tying effect reduced the level of penetration
of pre-CT magnetic field in the insulating wall during bubble-in to
the containment region, resulting in up to $20\upmu$s increase in
CT lifetime. Also supported by MHD simulation, plasma/material interaction
during formation was reduced with the modified levitation field profiles
of the 25-turn coil and 11-coil configurations, in which current carrying
coils extended along the entirety of the outer surface of the insulating
wall. Spectrometer data and observations of CT lifetime confirm that
the improved design led to reduced levels of plasma impurities and
radiative cooling. Consistent with this explanation for the improvement,
at the same initial CT poloidal flux, as determined by the voltage
on the formation capacitors and the current in the main coil, CT lifetimes
were around the same for the six-coil, 25-turn coil, and eleven-coil
configurations. However the setups with the 25-turn coil and with
eleven coils allowed for the successful formation of higher flux,
physically larger, CTs - formation voltage could be increased from
$12$ to $16\mbox{kV}$ and main coil current could be increased from
45 to 70A, corresponding to an increase in $\Psi_{gun}$ from around
8 to 12 mWb. In contrast, the benefit of increased initial CT flux
was surpassed by the performance degradation due to increased wall
interaction in the six coil setup. Although the recurrence rate of
good shots in the 25-turn coil configuration was significantly worse
than that in the 11-coil configuration, the longest lived CTs produced
in the former configuration endured for noticeably longer than those
produced in the latter configuration. The stronger levitation field
at the top and bottom of the insulating wall in the 25-turn coil configuration,
and consequent reduction in plasma-wall interaction and radiative
cooling may partially account for this. In addition, the increased
ratio of the coil to levitation circuit holding inductance associated
with the 25-turn coil configuration, which led to more levitation
flux increase upon plasma entry to the containment region, may have
played a role. The longer rise time of the levitation current associated
with the 25-turn configuration required an increase in the delay between
the firing of levitation and formation banks, which can lead to impediment,
through the line-tying effect, of plasma entry to the CT containment
region. This is likely to have been the cause of the poor repeatability
of good shots in the 25-turn configuration. Future designs should
optimise between the ideals of low coil inductances and high coil
to levitation circuit inductance ratios.

Compared with the aluminum flux conserver, the stainless steel wall
led to more impurities and shorter CT lifetimes, likely due to more
CT field-diffusion in the material, leading to enhanced impurity sputtering.
Magnetic perturbations with toroidal mode number $n=2$ were observed
on CTs produced with both stainless steel and aluminum outer flux
conservers, and remained even when a moderate levitation field was
allowed to soak through the stainless steel wall, but were absent
in all configurations tested in which a CT was held off an outer insulating
wall by a levitation field. It is known that $n=2$ fluctuations are
a sign of internal MHD activity associated with increased electron
temperature. However, the longest-lived CTs produced with the 25-turn
configuration endured for up to 10\% longer than, and may therefore
be reasonably assumed to be hotter than the CTs produced with the
stainless steel outer flux conserver. It is possible that the levitation
field acts to damp out helically propagating magnetic fluctuations
at the outboard CT edge and that internal MHD activity is relatively
unchanged. 

Indications of an instability, thought to be an external kink, occurred
very frequently during magnetic compression and during under-damped
magnetic levitation. Levitation circuit modification to match the
decay rates of the levitation and plasma currents led to more stable,
longer lived plasmas, and a greatly increased recurrence rate of good
shots, by avoiding unintentional magnetic compression during CT levitation.
MHD simulation results, which closely match the available experimental
measurements, indicate that $q<1$ over extensive regions between
the CT magnetic axis and LCFS. An obvious improvement to the experiment
design would be to drive additional shaft current and raise the $q$
profile to MHD stable regimes. 

The recurrence rate of shots in which the CT poloidal flux was conserved
during magnetic compression increased from around 10\% to 70\% with
the transition to the levitation/compression field profile of the
eleven-coil configuration. The improvement is likely to be largely
due to the compression field profile itself, which led to more uniform
outboard compression, as opposed to the largely equatorial outboard
compression associated with the six coil configuration. The effect
of having a reduced impurity concentration and increased CT plasma
temperature prior to compression initiation, as a consequence of the
improved levitation field profile, may also have played a role. Due
to improved flux conservation at compression, magnetic compression
ratios increased significantly with the eleven coil configuration.
Magnetic compression usually did not exhibit good toroidal symmetry.

In the eleven coil configuration, poloidal field at the CT edge, at
fixed $r=26$mm, increased by a factor of up to six at compression,
while line averaged electron density at fixed $r=65$mm was observed
to increase by 400\%, with the electron density front moving inwards
at up to 10km/s. Ion Doppler measurements, at fixed $r=45$mm indicated
ion temperature increases at magnetic compression by a factor of up
to four. Increases in poloidal field, density, and ion temperature
at compression were significant only in the eleven coil configuration.
The experimental technique developed to measure the CT outboard separatrix
confirmed that increasing the damping of the levitation field over
time led to CTs that remained physically larger over extended times.
Separatrix radii trajectories from MHD simulations matched those obtained
experimentally for various magnetic levitation and compression scenarios,
and indicated a radial compression factor, in terms of equatorial
outboard CT separatrix, of up to $1.7$. As shown in appendix \ref{subsec:SimDiagCompression-scalings},
MHD simulation results indicate that CT aspect ratio is approximately
constant over compression, and that internal CT poloidal and toroidal
fields, and CT toroidal current, scale approximately adiabatically,
increasing over the main compression cycle by factors of approximately
four, three and two respectively. 

\section{Acknowledgments}

Funding was provided in part by General Fusion Inc., Mitacs, University
of Saskatchewan, and NSERC. Particular thanks to Ivan Khalzov, Meritt
Reynolds, Aaron Froese, Charlson Kim, and Masayoshi Nagata for useful
discussions. Thanks also to Mike Donaldson, Blake Rablah, Curtis Gutjahr
and James Wilkie for assistance and advice relating to operation scheduling
and hardware, especially during the final stages of the experiment.
Thanks to the entire General Fusion Team for making the work possible.
We acknowledge the University of Saskatchewan ICT Research Computing
Facility for computing time.

\appendix

\section{Appendix: Using side-probe data to find CT outer separatrix radius\label{sec:rsep} }

A set of eight magnetic probes with windings to measure $B_{r},\,B_{\phi}$,
and $B_{z}$, were attached to the outside of the insulating wall
at $\phi=10^{\circ},\,55{}^{\circ},\,100^{\circ},\,145{}^{\circ},\,190^{\circ},\,235{}^{\circ},\,280^{\circ},\,\mbox{ and\,\ensuremath{\,}}325{}^{\circ}$.
The probes were installed at $z=0\,\mbox{mm}$ ($i.e.,$ around the
equator of the plasma torus) on the earlier configuration with six
coils around the ceramic (alumina) wall (with stainless steel extension
in place), and at $z=6\,\mbox{mm}$ on the configuration with eleven
coils around the quartz wall. The probes measured the levitation field
which is compressed when the plasma enters the confinement region.
A bigger CT will displace a greater proportion of the levitation flux,
so that $B_{\theta}(\phi,\,t)$$=B_{\theta}(r_{s}(\phi,\,t))$, where
$B_{\theta}(\phi,\,t)$ is the poloidal field measured at the side
probe, and $r_{s}(\phi,\,t)$ is the radius of the CT's separatrix
at the z-coordinate of the probe. By definition of the separatrix,
$\psi_{CT}(\phi,\,t)+\psi_{lev}(\phi,\,t)=0$ at $r_{s}(\phi,\,t)$,
where $\psi_{CT}$ and $\psi_{lev}$ are the contributions to $\psi$
that arise due to CT currents and external coil currents respectively.
We expect that $\psi_{lev}(\phi,\,t)\approx\psi_{lev}(t),$ with any
deviation from toroidal symmetry being due either to coil misalignment,
or asymmetry associated with discrete coils. However, the radial distribution
of CT poloidal flux, and therefore $r_{s},$ can vary with toroidal
angle, depending on MHD activity in the CT. The $r$ component of
the experimentally measured field at the probes proved to be negligible,
so we made the approximation $B_{z}\approx B_{\theta}.$ We used a
set of FEMM models to estimate the value of $B_{z}$ that would be
measured at the probes for varying $r_{s}$.
\begin{figure}[H]
\subfloat[$r_{sF}=168\mbox{mm}$.]{\raggedright{}\includegraphics[width=7cm,height=4cm]{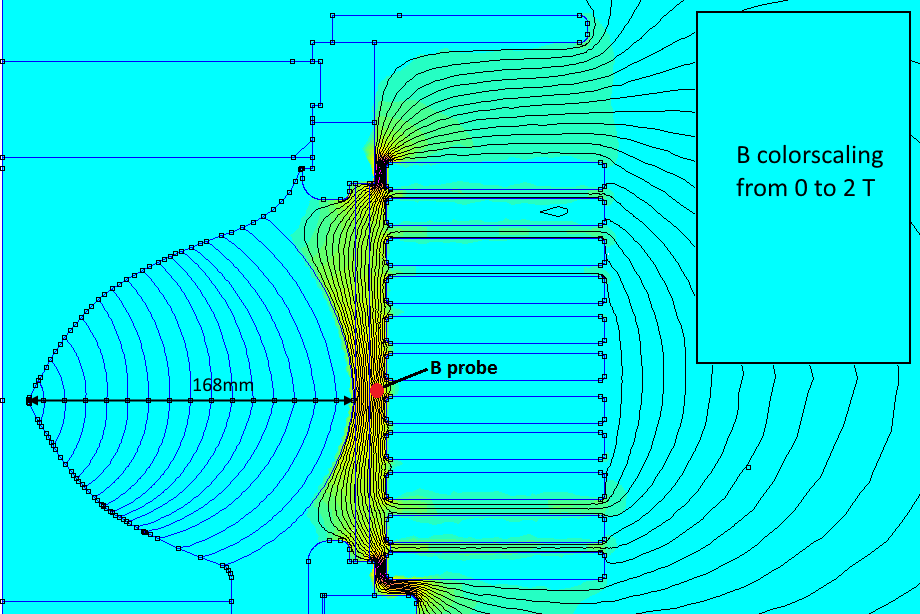}}\hfill{}\subfloat[$r_{sF}=60\mbox{mm}$.]{\raggedleft{}\includegraphics[width=7cm,height=4cm]{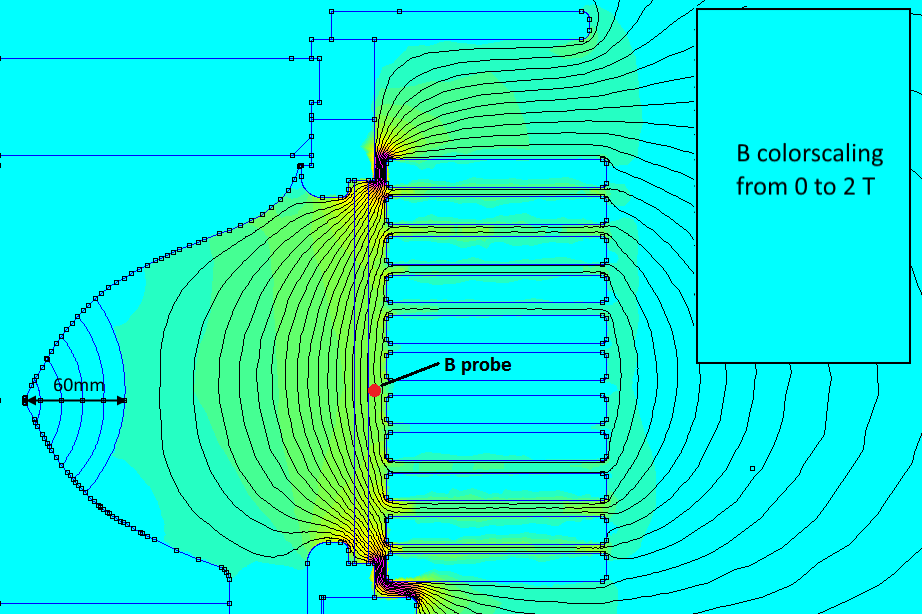}}

\caption{\label{rsepFEMM-1}FEMM models for finding $r_{s}$}
\end{figure}
Two of the seventeen FEMM models used to find $r_{s}$ from $B_{z}$
recorded in the experiment at the side probes, for the 11-coil configuration,
are shown in figure \ref{rsepFEMM-1}.  A material with artificially
high conductivity ($\sigma=10^{12}$S/m) was assigned to the areas
representing the nested \textquotedbl plasmas\textquotedbl{} in FEMM.
The seventeen models are identical except that, starting with the
model with $r_{sF}=168\mbox{mm}$ (note $r_{quartz}$, the inner radius
of the insulating wall is at 170mm), the outermost of the set of nested
shells representing \textquotedbl plasma\textquotedbl{} material
is removed for each successive model. Figures \ref{rsepFEMM-1}(a)
and (b) show the FEMM solutions (contours of $\psi,$ coloured by
$|B|$) for models with $r_{sF}=168\mbox{mm}$ and 60mm respectively,
where $r_{sF}$ is the radius, at the same z coordinate as the probes,
of the outermost layer of \textquotedbl plasma material\textquotedbl{}
in the FEMM model.

With the levitation coil currents in the models determined from experimental
measurements, and the coil current frequencies set to a high value
($\sim1$MHz), FEMM was run for each model. The high conductivity
of the material representing plasma, and the high current frequency,
ensure minimal penetration of levitation field into the \textquotedbl plasma\textquotedbl{}
region, so that the true separatrix radius is modelled. A LUA script
was written so that each of the models can be loaded successively
and run automatically through FEMM, and the required data for each
solution can be written to file for processing. The required data
consists, for each model, of $r_{sF}$ and $B_{zF}(r_{sF})$, where
$B_{zF}(r_{sF})$ is the FEMM solution for $B_{z}$ at the probe location
$((r,\,z)=(177\mbox{mm},\,6\mbox{mm}))$ for a given $r_{sF}$. The
process was repeated for another set of FEMM models based on the six
coil configuration with the reduced insulating wall inner radius ($r_{ceramic}=144\mbox{mm})$,
with the probe location being $(r,\,z)=(161\mbox{mm},\,0\mbox{mm}$). 

\subsection{Levitation only shots with 70$\mbox{m}\Omega$ cables\label{subsec:Rsep_Lev_70mOhm}}

\begin{figure}[H]
\subfloat[$B_{z}$ on side probes, shot 39650]{\raggedright{}\includegraphics[width=8cm,height=5cm]{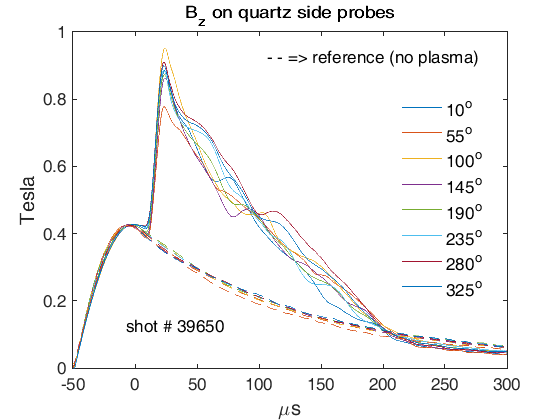}}\hfill{}\subfloat[FEMM outputs and functional fit]{\raggedleft{}\includegraphics[width=8cm,height=5cm]{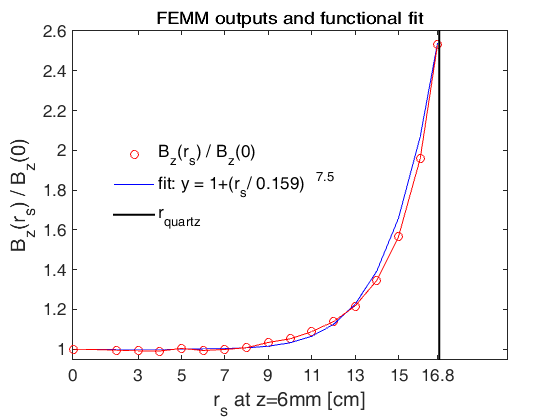}}

\caption{Experimental data and functional fit to FEMM data (70m$\Omega$ cables)\label{Bz_rsep39650_1}}
\end{figure}
 The $B_{z}(\phi,\,t)$ signals measured at the eight probes for shot
$39650$, along with $B_{zref}(\phi,\,t)$, the reference signals,
which are the averages of the signals measured at the same probes
during three levitation-only shots taken without charging or firing
the formation banks, are shown in figure \ref{Bz_rsep39650_1}(a).
 The measured $B_{z}(\phi,\,t)$ and $B_{zref}(\phi,\,t)$ signals
were calibrated using $B_{zF}(0)$, which is determined from a similar
FEMM model without any superconducting \textquotedbl plasma\textquotedbl{}
material, to determine the peak field amplitude at $\sim0\upmu$s
(before CT entry to the containment region). Shot $39650$ was taken
in the 11-coil configuration with $70\mbox{m}\Omega$ cables in place
between each main levitation inductor and coil-pair/coil, with $t_{lev}=50\upmu$s.
Figure \ref{Bz_rsep39650_1}(b) shows $B_{zF}(r_{sF})/B_{zF}(0)$
plotted against $r_{sF}$ using data from the set of FEMM models relevant
to the 11-coil configuration. A function of the form $y=1+(r_{sF}/0.159)^{7.5}$
was found to be a good fit to the data. Using the data from each of
the eight probes, this functional fit is inverted to find $r_{s}(\phi,\,t)$
at each of the toroidal angles associated with the probes. At each
probe, we have recorded $B_{zref}(\phi,\,t)$, and $B_{z}(\phi,\,t)$,
so $r_{s}(\phi,\,t)$ can be found using the formula 
\begin{equation}
r_{s}(\phi,\,t)=0.159\,(B_{z}(\phi,\,t)\,/\,B_{zref}(\phi,\,t)\,-1)^{\frac{1}{7.5}}\label{eq:103-1}
\end{equation}
Note that $r_{s}(\phi,\,t)$ becomes complex-valued if $B_{zref}(\phi,\,t)>B_{z}(\phi,\,t)$
- care has to be taken to ensure that probe signals are properly calibrated,
and signals from any probes that have unusual responses must be ignored,
in order for the method to work. Note also that for $r_{s}\lesssim9$cm,
the slope of the function fit in \ref{Bz_rsep39650_1}(b) is too flat
to be successfully inverted with good accuracy.
\begin{figure}[H]
\begin{centering}
\subfloat[]{\raggedright{}\includegraphics[scale=0.5]{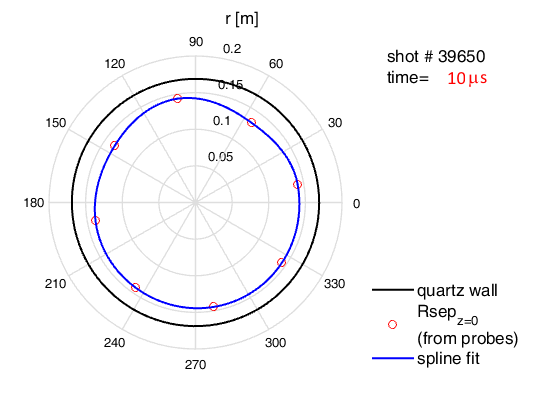}}\hspace{1.5cm}\subfloat[]{\raggedleft{}\includegraphics[scale=0.5]{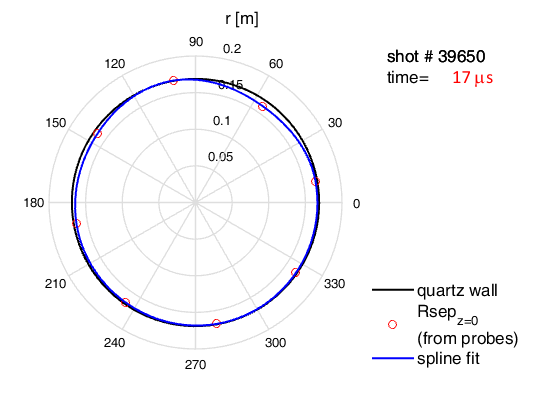}}
\par\end{centering}
\begin{centering}
\subfloat[]{\raggedright{}\includegraphics[scale=0.5]{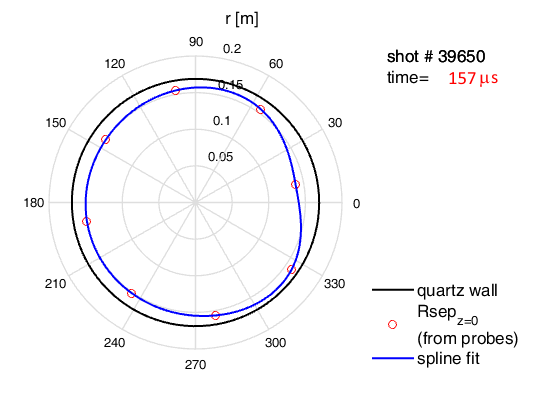}}\hspace{1.5cm}\subfloat[]{\raggedleft{}\includegraphics[scale=0.5]{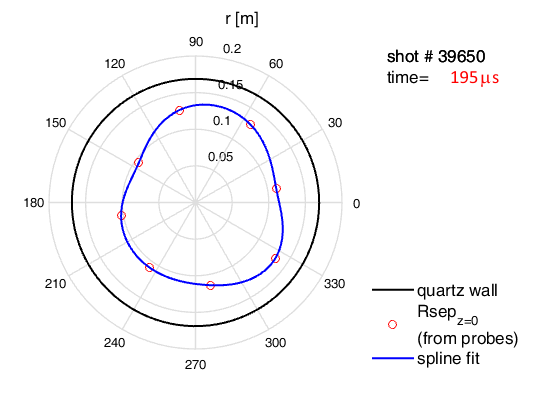}}
\par\end{centering}
\caption{\label{fig:rsep39650VID}$r_{s}(\phi,t)$ for shot $39650\:(70\mbox{m}\Omega\:\mbox{cables}),$at
10$\upmu$s (a), 17$\upmu$s (b), 157$\upmu$s (c), 195$\upmu$s (d).}
\end{figure}
Images from a movie that is the output of a code that finds $r_{s}(\phi,\,t)$,
based on $B_{z}(\phi,\,t)$ recorded at the side probes during shot
$39650$, according to the functional fit in equation \ref{eq:103-1},
are shown in figure \ref{fig:rsep39650VID}.  It can be seen that
the plasma enters the confinement region at $t=10\upmu$s, and that
at $t=17\upmu$s the CT fills the space right up to the inner radius
of the quartz wall, at $z=6\,\mbox{mm}$ (z coordinate of the side
probes). It remains at around this size and then starts to shrink
at around $157\upmu$s. At this time it looks like it is being pushed
in more at around $\phi=10^{\circ}$. At $195\upmu$s, there are signs
of an $n=3$ mode - the CT is being pushed in more at around $\phi=10^{\circ}$
and $150^{\circ}$, and is reacting by starting to bulge outwards
at $\phi=80{}^{\circ},210^{\circ}\,\mbox{and}\:330^{\circ}$. The
CT is gone shortly after $195\upmu$s.
\begin{figure}[H]
\subfloat[]{\raggedright{}\includegraphics[width=8cm,height=5cm]{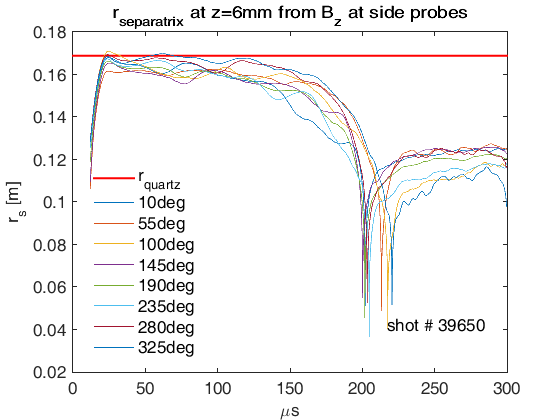}}\hfill{}\subfloat[]{\raggedleft{}\includegraphics[width=8cm,height=5cm]{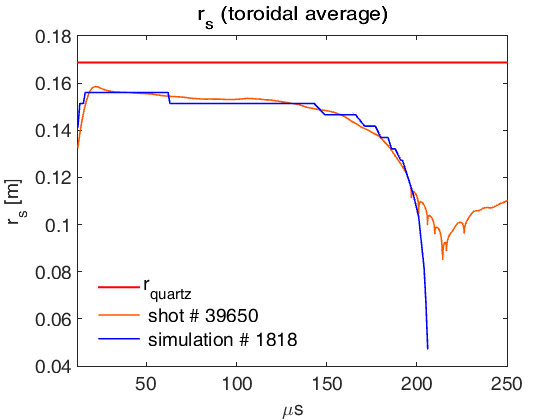}}

\caption{\label{fig:rsep39650_2}$r_{s}(\phi,t)$ for shot $39650$ (a), and
comparison with simulation (b)}
\end{figure}
Figure \ref{fig:rsep39650_2}(a) is a plot of the modelled $r_{s}(\phi,\,t)$
against time. As also indicated in figure \ref{fig:rsep39650VID}(c),
the CT starts to shrink in from the inner radius of the wall at around
$150\upmu$s. As mentioned, calculated $r_{s}(\phi,\,t)$ is not valid
when $B_{zref}(\phi,\,t)>B_{z}(\phi,\,t)$. It can be seen in figure
\ref{Bz_rsep39650_1}(a) that (due to inaccuracies in probe responses
etc.) $B_{zref}(\phi,\,t)>B_{z}(\phi,\,t)$ after around $200\upmu$s.
Figure \ref{fig:rsep39650_2}(b) shows the close match obtained between
the toroidally-averaged experimentally inferred $r_{s}(t)$ and $r_{s}(t)$
as determined by MHD simulation. \\
\\

\subsection{Levitation only shots with 2.5$\mbox{m}\Omega$ cables}

\begin{figure}[H]
\subfloat[]{\raggedright{}\includegraphics[width=8cm,height=5cm]{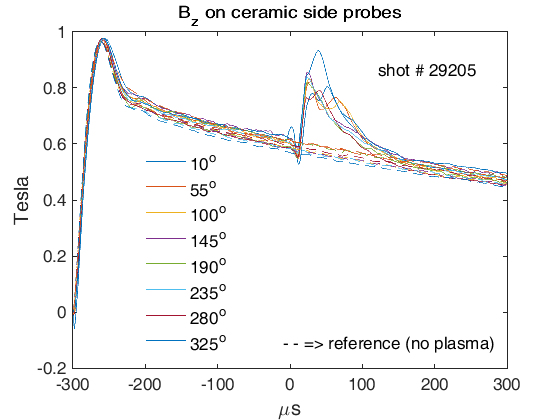}}\hfill{}\subfloat[]{\raggedleft{}\includegraphics[width=8cm,height=5cm]{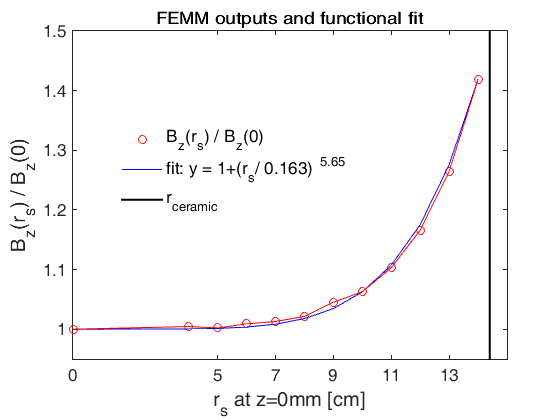}}

\caption{Experimental data (a) and functional fit to FEMM data (b) (2.5m$\Omega$
cables)\label{Bz_rsep29205_1}}
\end{figure}
The $B_{z}(\phi,\,t)$ and $B_{zref}(\phi,\,t)$ signals for shot
$29205$ are shown in figure \ref{Bz_rsep29205_1}(a).  Shot $29205$
was taken in the configuration with six coils surrounding the shortened
alumina insulating wall, with a $2.5\mbox{m}\Omega$ cable between
the main levitation inductor and the coil in each levitation/compression
circuit, and with $t_{lev}=300\upmu$s to allow for enhanced field
line pinning and reduced plasma-wall interaction in the 6 coil configuration.
Figure \ref{Bz_rsep29205_1}(b) shows $B_{zF}(r_{sF})/B_{zF}(0)$
plotted against $r_{sF}$ for the 6 coil configuration. A function
of the form $y=1+(r_{s}/0.163)^{5.65}$ was found to be a good fit
to the FEMM data and the procedure followed to get $r_{s}(\phi,\,t)$
from the experiment data is as described above.
\begin{figure}[H]
\begin{centering}
\subfloat[]{\raggedright{}\includegraphics[scale=0.5]{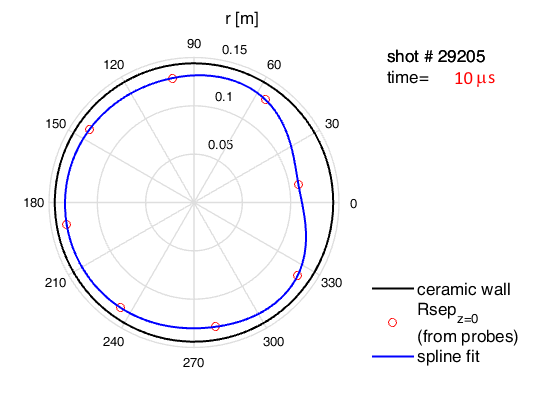}}\hspace{1.5cm}\subfloat[]{\raggedleft{}\includegraphics[scale=0.5]{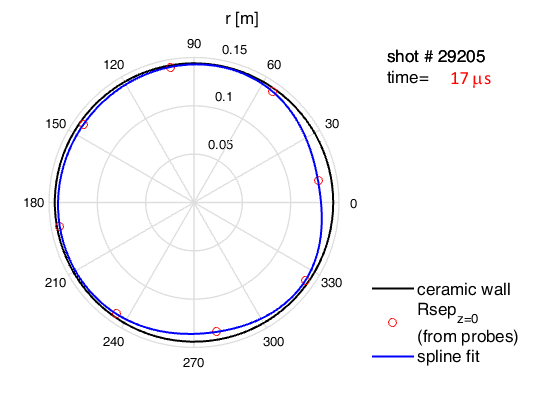}}
\par\end{centering}
\begin{centering}
\subfloat[]{\raggedright{}\includegraphics[scale=0.5]{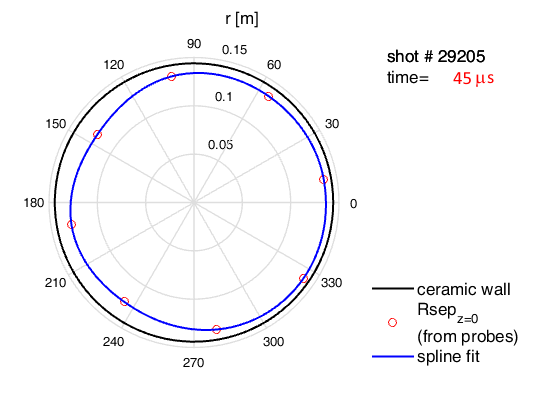}}\hspace{1.5cm}\subfloat[]{\raggedleft{}\includegraphics[scale=0.5]{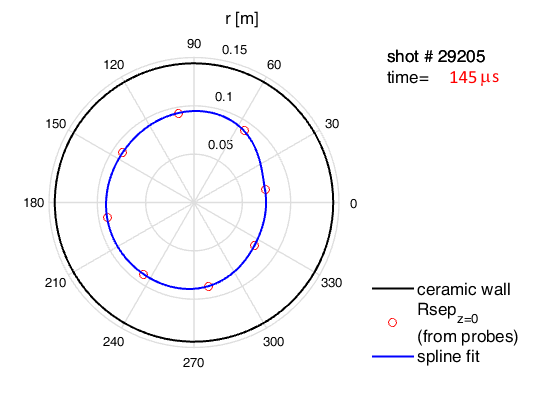}}
\par\end{centering}
\caption{\label{resp29205_2}$r_{s}(\phi,t)$ for shot $29205\ (2.5\mbox{m}\Omega\:\mbox{cables})$,
at 10$\upmu$s (a), 17$\upmu$s (b), 45$\upmu$s (c), 145$\upmu$s
(d). }
\end{figure}
Images at four times, indicating $r_{s}(\phi,\,t)$ based on the FEMM
model outputs are shown in figure \ref{resp29205_2}. As in figure
\ref{fig:rsep39650VID}, it can be seen that the plasma enters the
confinement region at $t=10\upmu$s, and that at $t=17\upmu$s the
CT fills the space right up to the inner radius of the insulating
wall. With the low resistance cables, for a shot on the 6-coils with
ceramic wall configuration, the levitation field is constantly compressing
the CT. It can be seen (figure \ref{resp29205_2} (c)) how the CT
has already started to shrink at $45\upmu$s, whereas the CT retains
its maximum volume up until around $157\upmu$s when the levitation
field decay rate is optimized (figure \ref{fig:rsep39650VID}(c)).
The CT continues to be pushed inwards rapidly and is extinguished
shortly after $145\upmu$s.
\begin{figure}[H]
\subfloat[]{\raggedright{}\includegraphics[width=8cm,height=5cm]{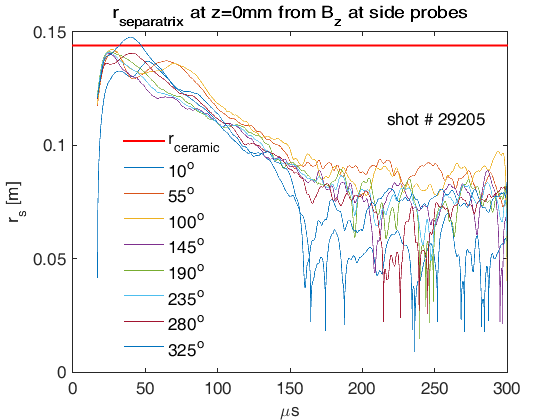}}\hfill{}\subfloat{\centering{}\includegraphics[width=8cm,height=5cm]{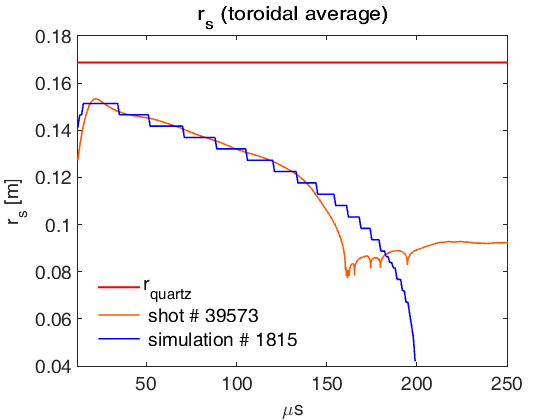}}

\caption{$r_{s}(\phi,\,t)$ for shot $29205\:\mbox{ (a), and comparison of measured and simulated \ensuremath{r_{s}} for shot 39573 (b) }(2.5\mbox{m}\Omega$
cables)\label{fig:rsep29205_39573}}
\end{figure}
A plot of experimentally determined $r_{s}(\phi,\,t)$ for the same
shot 29205 is shown in figure \ref{fig:rsep29205_39573}(a). As also
indicated in figure \ref{resp29205_2}(c), the CT starts to shrink
in from the inner radius of the wall at around $50\upmu$s. A comparison
between toroidally-averaged experimentally inferred and simulated
$r_{s}(t)$ for shot 39573 is shown in figure \ref{fig:rsep29205_39573}(b).
Shot 39573 was also taken in the configuration with $2.5\mbox{m}\Omega$
cables in parallel between the main levitation inductors and the coils,
but the 11-coil setup was used, and therefore the functional fit indicated
in figure \ref{Bz_rsep39650_1}(b) was used to extract $r_{s}(\phi,\,t)$.
The CT in shot 39573 ($V_{form}=16\mbox{kV}$), lives longer than
that in 29205 ($V_{form}=12\mbox{kV}$). However, the CTs in shots
29205 and 39573 are similar in that they both shrink rapidly in comparison
with shot 39650 (figure \ref{fig:rsep39650VID}), in which the decay-rate
matching strategy was used. In simulation  1815 (figure \ref{fig:rsep29205_39573}(b)),
the boundary conditions for $\psi_{lev}$ were scaled over time according
to the levitation current measured with the $2.5\mbox{m}\Omega$ cables
in place. The compressional instability, which is not captured by
2D MHD, causes the CT to shrink rapidly and be extinguished at $\sim150\upmu$s
in the case without decay rate matching. 

\section{Appendix: Adiabatic compression scalings\label{subsec:SimDiagCompression-scalings}}

As discussed in \cite{Furth}, if a magnetically confined plasma is
compressed on a time-scale that is short compared with the resistive
magnetic decay time and thermal and particle confinement times of
the plasma, ideal adiabatic compression scaling laws should apply.
It is worth noting that, as described in \cite{Kaur}, experimentally
determined adiabatic compression scalings may be assessed and used
to identify the equation of state. In contrast to the experiments
described in \cite{ATCconf,ATCpaper,S1spheromak,S1_compression,Kaur},
diagnostics internal to the CT that would enable assessment of the
scalings are not available for the SMRT experiment, but it is possible
to estimate them using outputs from simulations that match the available
fixed-point external diagnostics for magnetic field, and internal
line-averaged diagnostics for density and ion temperature along fixed
chords. The CT cross-sectional area in the poloidal plane is non-circular
- the scalings relevant to this case are collected in table \ref{tab:Scaling-parameters-for}.
\begin{table}[H]
\centering{}%
\begin{tabular}{|c|c|c|c|c|c|c|c|}
\hline 
parameter & $B_{\theta}$ & $B_{\phi}$ & $n$ & $T$ & $I_{p}$ & $\beta_{\phi}$ & $\beta_{\theta}$\tabularnewline
\hline 
 &  &  &  &  &  &  & \tabularnewline
scaling & $a^{-1}R^{-1}$ & $S^{-1}$ & $V^{-1}$ & $V^{-\frac{2}{3}}$ & $L\,a^{-1}R^{-1}$ & $V^{-\frac{5}{3}}S^{2}$ & $V^{-\frac{5}{3}}a^{2}\,R^{2}$\tabularnewline
\hline 
\end{tabular}\caption{\label{tab:Scaling-parameters-for}Parameter scalings for adiabatic
compression}
\end{table}
Here, $R(t)$ is the major axis, and $a(t)$ is the distance from
the CT magnetic axis radially outwards at poloidal angle $\theta=0$
to the last closed flux surface (LCFS). $L(t)$ and $S(t)$, the perimeter-length
and area of the poloidal CT cross section, and $V(t)$, the CT volume,
can be calculated using the coordinates of the points that define
the $\psi$ contour pertaining to the LCFS. As discussed previously,
and illustrated in figure \ref{fig:MHDform}, poloidal field lines
that remain open surrounding closed CT flux surfaces are pinched off
during magnetic compression, and reconnect to form additional closed
flux surfaces. This affects the definition of $\psi_{LCFS}$, and
therefore of the values of the geometric parameters $a(t),\,L(t),\,S(t)\mbox{ and }V(t)$
that are defined by the location of the LCFS and are required to determine
the predicted adiabatic scalings. Hence, compression scalings are
best assessed from simulations in which compression is initiated relatively
late in time when $\psi_{LCFS}$ is close to zero, and few open poloidal
field lines surround the CT. In addition, as also depicted in figure
\ref{fig:MHDform}, simulations indicate that closed poloidal CT field
lines that extend partially down into the gun barrel entrance can
be pinched off, and reconnect at compression, which also affects the
geometric parameters. A solution is to assess the parameters of interest,
including the geometric parameters, relevant to a $\psi$ contour,
defined by a fixed value of $\psi=\psi^{0}$, that is internal to
the pre-compression LCFS, and doesn't extend partially into the gun
barrel, a strategy that naturally does not affect the compression
scalings. 
\begin{figure}[H]
\subfloat[]{\raggedright{}\includegraphics[width=6cm,height=4cm]{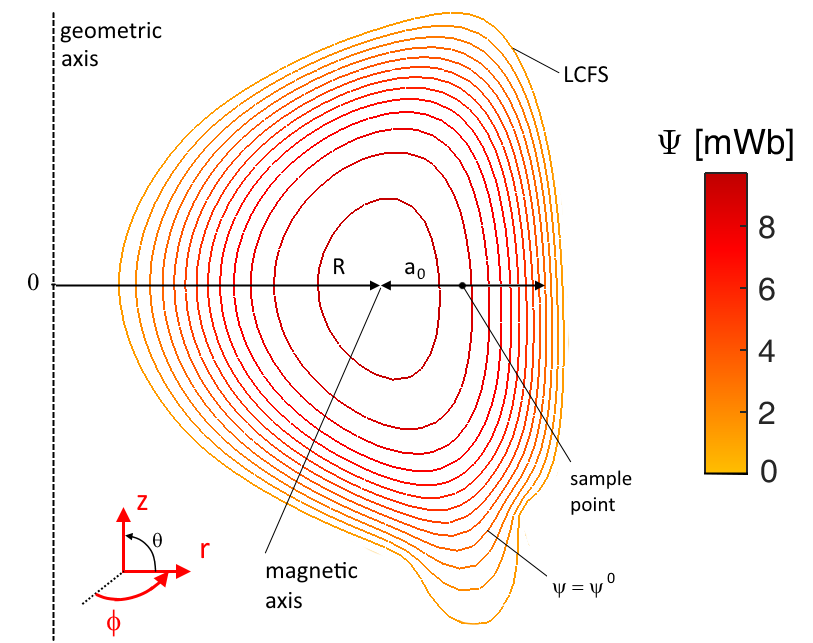}}\hfill{}\subfloat[]{\raggedleft{}\includegraphics[width=6cm,height=4cm]{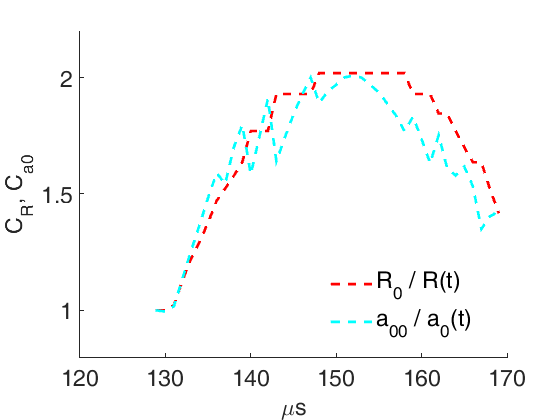}}

\subfloat[]{\raggedright{}\includegraphics[width=6cm,height=4cm]{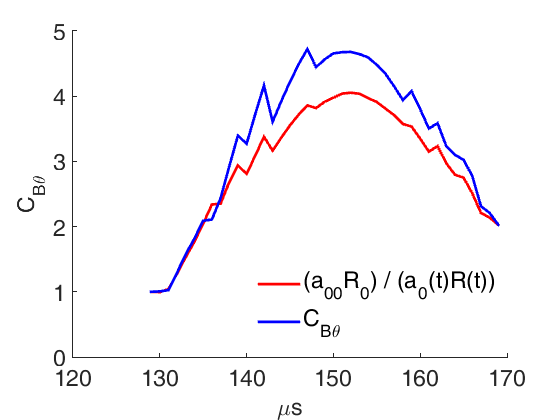}}\hfill{}\subfloat[]{\raggedleft{}\includegraphics[width=6cm,height=4cm]{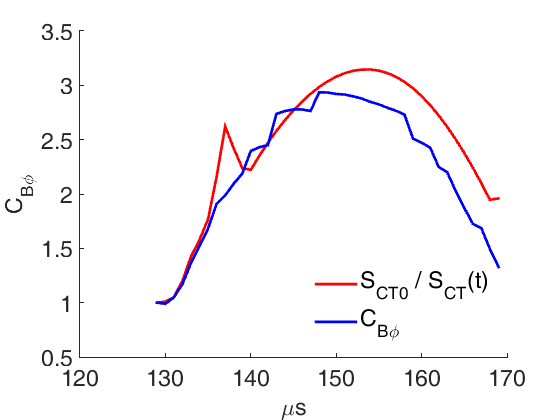}}

\subfloat[]{\raggedright{}\includegraphics[width=6cm,height=4cm]{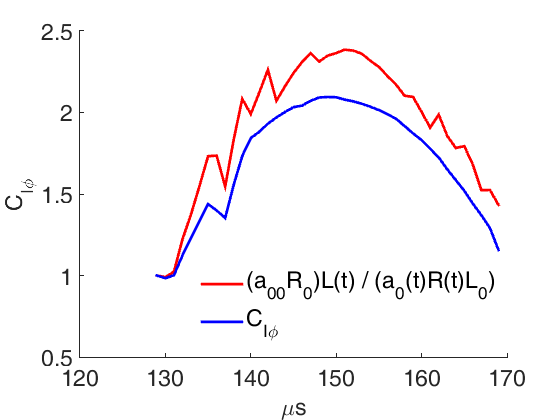}}\hfill{}\subfloat[]{\raggedright{}\includegraphics[width=6cm,height=4cm]{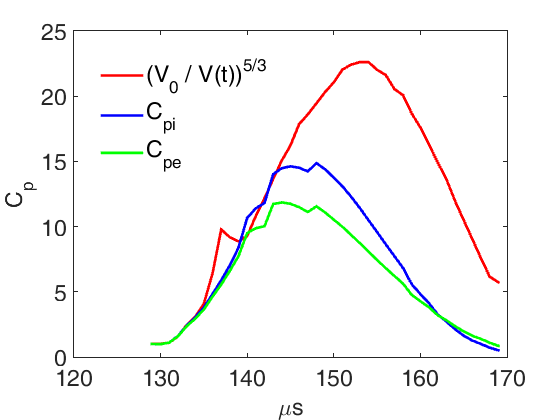}}

\caption{\label{fig:Compscalings2275}Geometry definitions (a), with indications
of simulated compression scalings for major and minor radius (b),
poloidal field (c), toroidal field (d), toroidal plasma current (e)
and ion and electron pressure (f), for simulation 2287}
\end{figure}
Pre-compression CT $\psi$ contours, from an MHD simulation at relatively
late time, when most open poloidal field lines have reconnected to
form closed CT field lines, and $\psi_{LCFS}\sim0$, are shown in
figure \ref{fig:Compscalings2275}(a).  A flux contour with $\Psi=\Psi^{0}=2$mWb,
that is suitable for assessing compression scaling parameters, and
definitions of $R(t)$ and $a_{0}(t)$ are also indicated. Note that
$a_{0}(t)$ is defined as the distance from the CT magnetic axis radially
outwards at poloidal angle $\theta=0$ to the closed flux surface
defined by $\Psi=\Psi^{0}$. Simulated geometric compression scalings
for $R(t)$ and $a_{0}(t)$ from simulation 2287 are shown in figure
\ref{fig:Compscalings2275}(b), where the subscript 0 denotes pre-compression
values. This indicates approximately constant aspect ratio (in irregular
geometry), and that the compression is close to the $Type\,A$ compression
regime defined in \cite{Furth}. With constant aspect ratio, this
indicates a geometric compression factor, in terms of equatorial outboard
CT separatrix, of $C_{s}\sim C_{R}\sim C_{a0}\sim2$. As described
in section \ref{sec:Rsep_comp}, a geometric compression factor $C_{s}\sim1.7$
was determined experimentally, and confirmed by MHD simulation, for
shot 39738 ($V_{comp}=18$kV and $t_{comp}=45\upmu$s), and that more
extreme compression in $r_{s}$ cannot be experimentally evaluated
due to limitations on the technique. More extreme compression would
be expected for shots at comparable $V_{comp}$, with $t_{comp}$
delayed to when pre-compression CT flux has decayed to lower levels.
Simulation 2287 pertains to shot 39735 ($V_{comp}=18$kV and $t_{comp}=130\upmu$s),
so the increased estimate for $C_{s}$ is consistent with the shot
parameters. Note that, as outlined in section \ref{sec:Magnetic-compression},
with $\widetilde{\tau}_{c}=0.6$, shot 39735 is not classified as
a flux-conserving shot, so the adiabatic compression scalings evaluated
here pertain to the shot only up until the time when flux started
to be lost, just before peak compression. As outlined above, the technique
described here cannot practically be applied to simulations with compression
initiated early, and flux-conserving compressed shots were generally
taken with $t_{comp}=45\upmu$s. Only a few shots were taken with
late compression, none of which conserved flux very well over compression,
as determined by the $\widetilde{\tau}_{c}$ metric.

Figure \ref{fig:Compscalings2275}(c) shows how, for simulation  2287,
poloidal field scales approximately adiabatically as $B_{\theta}\rightarrow a_{0}^{-1}R^{-1}$,
where the sample point used to determine the scalings, indicated in
figure \ref{fig:Compscalings2275}(a), is located halfway between
the magnetic axis and the outboard point where $\psi=\psi^{0}$ at
the same axial coordinate as the magnetic axis. The notation $C_{B\theta}$
denotes the scaling of poloidal field as $C_{B\theta}(t)=\frac{B_{\theta}(t)}{B_{\theta0}}$
where $B_{\theta0}$ is the pre-compression magnetic field at the
sample point. Similarly, figure \ref{fig:Compscalings2275}(d) shows
how toroidal field at the same sample point also scales adiabatically,
as $B_{\phi}\rightarrow S^{-1}$. Figure \ref{fig:Compscalings2275}(e)
shows how plasma current, calculated as the integral of toroidal current
density over the area inside the closed flux surface at $\psi=\psi^{0}$,
evolves approximately according to the adiabatic scaling for plasma
current. 

As indicated in \ref{fig:Compscalings2275}(f), the scaling for pressure
(and hence also for the $\beta$ scalings) does not follow the adiabatic
prediction of $p\rightarrow V^{-\frac{5}{3}}$, due to the presence
of artificial density diffusion, which effectively relocates particles
from high density to low density regions. For this simulation, density
diffusion was $\zeta=50$m$^{2}$/s, which is close to the minimum
value required for numerical stability at moderate timestep and mesh
resolution for simulations including magnetic compression, and $n_{e}$
follows the adiabatic scaling $n_{e}\rightarrow V^{-1}$ for only
the first $5\upmu$s after compression initiation. Ion and electron
pressures follow the adiabatic predictions for $15\upmu$s - the extension
is due to approximate internal force balance during this portion of
the compression cycle, which leads to increased temperature in regions
of low density. Temperatures at compression increase more, while density
increases less, than the predicted increases based on the adiabatic
scalings. When $\zeta$ is increased to $150$m$^{2}$/s, the duration
over which $n_{e}$ follows the adiabatic scaling is reduced further
to around $2\upmu$s. 

This simulation, which produces results that closely match the available
experimental measurements for shot 39735 over most of the compression
cycle, indicates that CT aspect ratio is approximately constant over
compression, with $C_{s}\sim C_{R}\sim C_{a0}\sim2$, and that internal
CT poloidal and toroidal fields, and CT toroidal current, scale approximately
adiabatically, increasing over the main compression cycle by factors
of approximately four, three and two respectively. \newpage{}

\end{document}